\documentclass[apj]{emulateapj-rtx4}
\usepackage{graphicx}
\usepackage{natbib}
\usepackage{lscape}
\bibpunct{(}{)}{;}{a}{}{,}
\newcommand{\mw}{MW}

\newcommand\smm{submillimeter}
\newcommand\fir{far-infrared}
\newcommand\mir{mid-infrared}
\newcommand\nir{near-infrared}

\newcommand\av{\mbox{$A_V$}}% visual extinction
\newcommand\ak{\mbox{$A_K$}}% extinction at K
\newcommand{\lsun}{\mbox{L$_\odot$}}% Lsun
\newcommand{\msun}{\mbox{M$_\odot$}}% Msun
\newcommand{\rsun}{\mbox{R$_\odot$}}% Rsun
\newcommand{\lbol}{\mbox{$L_{bol}$}} % bolometric luminosity
\newcommand{\lfir}{\mbox{$L_{FIR}$}} % far-infrared luminosity
\newcommand{\lir}{\mbox{$L_{IR}$}} % infrared luminosity
\newcommand{\lhcn}{\mbox{$L_{HCN}$}} % HCN 1-0 line luminosity
\newcommand{\ee}[1]{\mbox{${} \times 10^{#1}$}}% scientific number format
\newcommand{\eten}[1]{\mbox{$10^{#1}$}}% power of ten
\newcommand\cmv{\mbox{cm$^{-3}$}}
\newcommand\cmc{\mbox{cm$^{-2}$}}
\newcommand{\sfr }{\mbox{$\dot M_{*}$}}% Star formation rate
\newcommand{\fdense }{\mbox{$f_{dense}$}}% dense gas fraction
\newcommand{\rgal }{\mbox{$R_{gal}$}}% distance from gal center
\newcommand{\sigmasfr}{\mbox{$\Sigma({\rm SFR})$}}% Star formation rate sig
\newcommand{\msunmyr}{\mbox{M$_\odot$ Myr$^{-1}$}}% Msun per million years
\newcommand{\msunyr}{\mbox{M$_\odot$ yr$^{-1}$}}% Msun per year
\newcommand{\msunkpc}{\mbox{\msunyr kpc$^{-2}$}}% Msun per yr per sq kpc
\newcommand{\tk}{\mbox{$T_K$}}
\newcommand{\td}{\mbox{$T_d$}}
\newcommand{\tex}{\mbox{$T_{ex}$}}
\newcommand{\kms}{\mbox{km s$^{-1}$}}% km/s
\newcommand{\mstar}{\mbox{$M_{\star}$}}
\newcommand{\mcloud}{\mbox{$M_{cloud}$}}
\newcommand{\mgas}{\mbox{$M_{gas}$}}
\newcommand{\hi}{\mbox{\ion{H}{1}}}
\newcommand{\hii}{\mbox{\ion{H}{2}}}
\newcommand{\cii}{\mbox{\ion{C}{2}}}
\newcommand{\sigmagas}{\mbox{$\Sigma_{gas}$}}% surface density of gas 
\newcommand{\sigmamol}{\mbox{$\Sigma_{mol}$}}% surface density of molecular gas 
\newcommand{\sigmahtwo}{\mbox{$\Sigma_{\hh}$}}% surface density of H2, no He 
\newcommand{\sigmahi}{\mbox{$\Sigma_{\rm{HI}}$}}% surface density of HI, no He
\newcommand{\tff}{\mbox{$t_{ff}$}} % free-fall time
\newcommand{\tdyn}{\mbox{$t_{dyn}$}} % dynamical time
\newcommand{\tcross}{\mbox{$t_{cross}$}} % crossing time
\newcommand{\torb}{\mbox{$t_{orb}$}} % orbital time
\newcommand{\tdep}{\mbox{$t_{dep}$}} % depletion time
\newcommand{\msunpc}{\mbox{M$_\odot$ pc$^{-2}$}}% Msun per square pc

\newcommand{\mean}[1]{\mbox{$\langle#1\rangle$}} %generic mean for defined qu.
\newcommand{\halpha}{\mbox{H$\alpha$}}
\newcommand{\hh}{\mbox{H$_2$}}
\newcommand{\coo}{\mbox{$^{13}$CO}}
\newcommand{\cooo}{\mbox{C$^{18}$O}}
\newcommand{\hcop}{\mbox{HCO$^+$}}
\newcommand{\jj}[2]{\mbox{$J = #1\rightarrow#2$}}
\newcommand{\ico}{\mbox{$I({\rm CO})$}}% I(CO)
\newcommand{\xco}{\mbox{$X({\rm CO})$}}% X(CO)
\newcommand{\lco}{\mbox{$L({\rm CO})$}}% L(CO)
\newcommand{\aco}{\mbox{$\alpha_{\rm CO}$}}% alpha(CO)
\slugcomment{To Appear in ARAA, vol. 50}
\shorttitle{Star Formation}
\shortauthors{Kennicutt and Evans}

\begin{document}

%\input epsf.tex    %<-If you need EPS figures to be
                   %  called in {figure} environment for PC
%\input epsf.def   %<-If you need EPS figures to be
                   %  called in {figure} environment for Macintosh

%\input AR2e_ps/psfig.sty

%\jname{Annu. Rev. Astron. Astrophys.}
%\jyear{2012}
%\jvol{50}
%\ARinfo{1056-8700/97/0610-00}

\title{Star Formation in the Milky Way and Nearby Galaxies}

%\markboth{Verso Running Head}{Recto Running Head}
%\markboth{Kennicutt \& Evans }{Star Formation}

\author{ Robert C. Kennicutt, Jr.}
\affil{Institute of Astronomy, University of Cambridge, Madingley Road,
Cambridge, CB3 0HA, United Kingdom}
\email{robk@ast.cam.as.uk} 
\author{Neal J. Evans II}
\affil{The University of Texas at Austin, Department of Astronomy, 
2515 Speedway, Stop C1400 Austin, TX 78712-1205, USA \\  and \\
European Southern Observatory, Casilla 19001, Santiago 19, Chile}
\email{nje@astro.as.utexas.edu}

%\begin{keywords}
%star formation, galaxies, Milky Way
%\end{keywords}

\begin{abstract}

We review progress over the past decade in observations of
large-scale star formation, with a focus on the interface between
extragalactic and Galactic studies.  
Methods of measuring gas contents and star formation rates are discussed,
and updated prescriptions for calculating star formation rates are provided. 
We review relations between star formation and gas  on
scales ranging from entire galaxies to individual molecular clouds.

\end{abstract}

%\maketitle

\section{Overview}\label{overview}
% overview
% sec1.tex
% RCK 05/02/2011
% 06/10/11 NJE
% 090611 NJE mild edits and added a few refs from localintro
% localintro was then eliminated
% 100311 NJE work in questions.tex reference so it will be subsec(box) 
% 100511 NJE added subsection and label
% 100711 NJE minor cleanup
% 110411 NJE minor edits
% 110611 RCK minor edits
% 121211 NJE added some caveats about scope
% 121611 NJE minor edits
% 012412 NJE edits after comments
% 020212 NJE minor cleanup
% 031212 RCK minor corrections (via NJE)
% 031212 NJE more minor corrections
% 031612 NJE adjust for emulateapj, refer to section 3, not inset box

\subsection{Introduction}\label{introduction}

Star formation encompasses the origins of 
stars and planetary systems, but it is also
a principal agent of galaxy formation and evolution, and
hence a subject at the roots of astrophysics on its largest
scales.   

The past decade has witnessed an unprecedented 
stream of new observational information on star formation
on all scales, thanks in no small part to 
%the impact of
new facilities such as the Galaxy Evolution Explorer (GALEX),
the Spitzer Space Telescope, the Herschel Space Observatory,
the introduction of powerful new instruments on the
Hubble Space Telescope (HST), and a host of groundbased
optical, infrared, submillimeter, and radio telescopes.
These new observations are providing a detailed reconstruction
of the key evolutionary phases and physical processes
that lead to the formation of individual stars in 
interstellar clouds, while at the same time extending the
reach of integrated measurements of star formation rates
(SFRs) to the most distant galaxies known.  The new data
have also stimulated a parallel renaissance in theoretical
investigation and numerical modelling of the star formation
process, on scales ranging from individual protostellar
and protoplanetary systems to the scales of molecular
clouds and star clusters, entire galaxies and ensembles
of galaxies, even to the first objects, which are thought
to have reionized the Universe and seeded today's stellar
populations and Hubble sequence of galaxies.  

This immense expansion of the subject, both in terms of
the volume of results and the range of physical scales explored,
may help to explain one of its idiosyncracies, namely the
relative isolation between the community studying individual
star-forming regions and stars in the Milky Way (usually abbreviated
hereafter as \mw, but sometimes referred to as ``the Galaxy"), and the largely
extragalactic community that attempts to characterise the star formation
process on galactic and cosmological scales.  Some aspects
of this separation have been understandable.  The key 
physical processes that determine how molecular clouds
contract and fragment into clumps and cores and finally
clusters and individual
stars can be probed up close only in the Galaxy, and much
of the progress in this subject has come from in-depth case
studies of individual star-forming regions.  Such detailed
observations have been impossible to obtain for even relatively
nearby galaxies.  Instead the extragalactic branch of the 
subject has focused on the collective effects of star formation,
integrated over entire star-forming regions, or often
over entire galaxies.  It is this collective conversion of
baryons from interstellar gas to stars and the emergent
radiation and mechanical energy from the stellar populations that
is most relevant to the formation and evolution of galaxies. 
As a result, much of our empirical knowledge of star
formation on these scales consists of scaling laws and
other parametric descriptions, in place of a rigorous,
physically-based characterization.  Improving our knowledge
of large-scale star formation and its attendant feedback
processes is essential to understanding the 
birth and evolution of galaxies.

Over the past several years, it has become increasingly 
clear that many of the key processes influencing star
formation on all of these scales lie at the interface
between the scales within individual molecular clouds
and those of galactic disks. 
It is now clear that the large scale SFR is determined
by a hierarchy of physical processes spanning a
vast range of physical scales: the accretion of gas
onto disks from satellite objects and the intergalactic medium
(Mpc); the cooling of this gas to form a cool neutral phase (kpc); 
the formation of molecular clouds ($\sim 10- 100$ pc); 
the fragmentation and accretion of this molecular gas to form 
progressively denser structures such as clumps ($\sim 1$ pc) and 
cores ($\sim 0.1$ pc); and the subsequent
contraction of the cores to form stars (\rsun) and planets ($\sim$ AU).
% NJE put in scales re GB 2 comment; Rob should check
% NJE also left out "formation of bound clouds", not clear where it belongs
The first and last of these processes operate on
%in distinctly
galactic (or extragalactic) and local cloud scales, 
respectively, but the others occur at the boundaries
between these scales, and the
coupling between processes is not yet well understood.
Indeed it is possible that different physical processes
%serve as 
provide the ``critical path" to star formation in different
interstellar and galactic environments.  Whatever the 
answer to these challenging questions, however, Nature
is strongly signalling that we need a unified approach to understanding
star formation, one which incorporates observational and 
astrophysical constraints on star formation efficiencies,
mass functions, etc. from small-scale studies along
with a much deeper understanding of the processes that
trigger and regulate the formation of star-forming clouds
on galactic scales, which, in turn, set the boundary conditions for
star formation within clouds.

This review makes a modest attempt to present a consolidated
view of large-scale star formation, one which incorporates
our new understanding of the star formation within clouds
and the ensemble properties of star formation in our own
\mw\ into our more limited but broader understanding
of star formation in external galaxies.  As will become
clear, the subject itself is growing and transforming rapidly,
so our goal is to present a progress report in what remains
an exciting but relatively immature field.  Nevertheless
a number of factors make this a timely occasion for such a
review.  The advent of powerful multi-wavelength observations
has transformed this subject in fundamental ways since the
last large observational review of the galaxy-scale aspects 
\citep{1998ARA&A..36..189K},
%Kennicutt 1998a ARAA
hereafter denoted K98. 
Reviews have covered
various observational aspects of star formation within the \mw\
(\citealt{1999ARA&A..37..311E, 2007ARA&A..45..339B, 2007ARA&A..45..481Z}),
% Evans et al. , Bergin and Tafalla, Zinnecke and Yorke
but the most comprehensive reviews have been theoretically based
(\citealt{1987ARA&A..25...23S, 2007ARA&A..45..565M}).
% Shu et al., McKee-Ostriker, 
We will take an observational perspective, with emphasis on 
the interface between local and galactic scales. 
We focus on nearby galaxies and the \mw, bringing in results from
more distant galaxies only as they bear on the issues under discussion.

The remainder of this article is organized as follows. 
In the next subsection, we list some definitions and conventions
that will be used throughout the paper.
All observations in this subject rest on quantitative diagnostics
of gas properties and star formation rates on
various physical scales, and we review the current state of
these diagnostics (\S \ref{ismdiagnostics}, \S \ref{sfrdiagnostics}).  
In \S \ref{local},  we review
those properties of Galactic star-forming regions which are most
relevant for comparison to other galaxy-wide studies.  
In \S \ref{outside},
we review the star-forming properties of galaxies on the large
scale, including the \mw; much of this section is effectively
an update of the more extended review presented in K98.
Section \ref{sflobs}
updates our knowledge of relations between star formation
and gas (e.g., the Schmidt law) and local tests of these relations.
The review concludes in \S \ref{synthesis}  with our attempted synthesis 
of what has been learned from the confluence of local and global studies
and a look ahead to future prospects.

We 
%show in the inset box a list of 
list in \S \ref{questions} some
key questions in the field.
Some have been or will be addressed by other reviews.
Some are best answered by observations of other galaxies; some
can only be addressed by observations of local star-forming regions
in the \mw.  For each question we list sections in this
review where we review the progress to date in answering them,
but many remain largely unanswered and are referred
to the last section of this review on future prospects.

Space limitations prevent us from citing even a fair fraction
of the important papers in this subject, so we instead cite
useful examples and refer the reader to the richer lists of
papers that are cited there.

% introduction
% definitions.tex
% 100411 Initialized by NJE
% Try to collect some definitions.
% 100611 NJE, added and smoothed out interface with massfunctions.tex
% 100711 NJE minor cleanup
% 110411 NJE minor edits
% 110611 RCK minor edits
% 120711 NJE minor edits
% 121211 NJE added some definitions and clarified some
% 121611 NJE minor edits
% 012412 NJE minor edits
% 020212 NJE minor clean up and move mass functions to end
% 021212 RCK add Schechter function stuff
% 021312 NJE edited re RCK comments
% 021712 NJE fixed various details
% 031212 NJE minor edits
% 051412 NJE minor edits

\subsection{Definitions and Conventions}\label{definitions}

Here we define some terms and symbols that will be used throughout
the review.

The term ``cloud" refers to a structure in the interstellar medium
(ISM) separated from its surroundings by the rapid change of some
property, such as pressure, surface density, or chemical state.
Clouds have complex structure, but theorists have identified two
relevant structures: clumps are the birthplaces of clusters; cores
are the birthplaces of individual or binary stars 
(e.g., \citealt{2007ARA&A..45..565M}). The observational
equivalents have been discussed 
(e.g., Table 1 in \citealt{2007ARA&A..45..339B}), and cores
reviewed  (\citealt{2007prpl.conf...17D, 2007prpl.conf...33W}),
and we discuss them further in \S \ref{names}.

We will generally explicitly use ``surface density" or ``volume density",
but symbols with $\Sigma$ will refer to surface density, $N$ will
refer to column density, and symbols with $n$ or $\rho$ will refer to 
number or mass volume density, respectively.
The surface density of \hi\ gas is represented by 
\sigmahi\ if He is not included and by
$\Sigma$(atomic) if He is included.
The surface density of molecular gas is symbolized by 
\sigmahtwo, which generally does not include He, or \sigmamol, which
generally does include He.
For \mw\ observations, He is almost always included in mass and 
density estimates, and \sigmamol\ may be determined by extinction or emission 
by dust (\S \ref{dusttracers}) or by observations of CO isotopologues
(\S \ref{gastracers}).
When applied to an individual cloud or clump, we refer to it as
$\Sigma$(cloud) or $\Sigma$(clump).
For extragalactic work, \sigmahtwo\ or \sigmamol\
are almost always derived from CO observations, and inclusion of
He is less universal.
In the extragalactic context, \sigmamol\ is best interpreted as related 
to a filling factor
of molecular clouds until $\sigmamol \sim 100$ \msunpc, where the 
area filling factor of molecular gas may become unity (\S \ref{gastracers}). 
Above this point, the meaning may change substantially. 
The total gas surface density, with He, is 
$\sigmagas = \sigmamol+ \mu \sigmahi$, where $\mu$ is the mean
molecular weight per H atom; $\mu = 1.41$ for \mw\ abundances
\citep{2008A&A...487..993K}, 
% Kauffmann et al. 2008, see appendix
though a value of 1.36 is commonly used.

The conversion from CO intensity (usually in the \jj10\ rotational
transition) to column density of \hh, not including He, is denoted
\xco, and the conversion from CO luminosity to mass, which usually
includes the mass of He, is denoted $\alpha_{\rm CO}$. 

The term ``dense gas" refers generally to gas above some threshold
of surface density, as determined by extinction or dust continuum
emission, or of volume density,
as indicated by emission from molecular lines that trace densities
higher than does CO. While not precisely defined, suggestions
include criteria of a threshold surface density, such as
$\sigmamol > 125$ \msunpc\
(\citealt{2008ApJ...680..428G, 2010ApJ...724..687L, 2010ApJ...723.1019H}),
% Lada et al, Heiderman et al.
a volume density criterion (typically $n > \eten4$ \cmv)
(e.g., \citealt{1992ApJ...393L..25L}),
%Lada, E. 1992 L1630 study
and detection of a line from certain molecules, such as HCN.
The surface and volume density criteria roughly agree in nearby
clouds \citep{2012ApJ...745..190L} but may not in other environments.
% Lada 2012 paper

The star formation rate is symbolized by SFR or $\sfr$, often with units of
$\msun$ yr$^{-1}$ or \msunmyr, and its meaning depends on how much 
averaging over time and space is involved.  The surface density of
star formation rate, \sigmasfr, has the same averaging issues.
The efficiency of star formation is symbolized by $\epsilon$ when it
means $\mstar/(\mstar + \mcloud)$, as it usually does in Galactic studies.
For extragalactic studies, one usually means $\sfr/\mgas$, which we
symbolize by $\epsilon^{\prime}$. The depletion time, $t_{dep} = 
1/\epsilon^{\prime}$. Another time often used is the ``dynamical" time,
$\tdyn$, which can refer to the free-fall time (\tff), the crossing
time (\tcross), or the galaxy orbital time (\torb).  In recent years, it also
has become common to compare galaxies in terms of their SFR per unit
galaxy mass, or ``specific star formation rate" (SSFR).  The
SSFR scales directly with
the stellar birthrate parameter $b$, the ratio of the SFR today to 
the average past SFR over the age of the galaxy, and thus provides
a useful means for characterizing the star formation history of a galaxy. 

The far-infrared luminosity has a number of definitions, and
consistency is important in converting them to \sfr.  A commonly 
used definition integrates the dust emission over the wavelength
range 3--1100 \micron\ \citep{2002ApJ...576..159D},
%Dale & Helou (2002)
and following those authors we refer to this as the total-infrared
or TIR luminosity.  Note however that other definitions based
on a narrower wavelength band or even single-band infrared measurements
are often used in the literature.

We refer to mass functions with the following conventions.
Mass functions are often approximated by power laws in log$M$:

\begin{equation}
dN/d{\rm log} M \propto M^{-\gamma},
\label{massfunceq1}
\end{equation}
In particular, the initial mass function (IMF) of stars is often
expressed by equation \ref{massfunceq1} with a subscript to indicate
stars (e.g., $M_*$).

Structures in gas are usually characterized by distributions
versus mass, rather than logarithmic mass:
\begin{equation}
dN/d M \propto M^{-\alpha},
\label{massfunceq2}
\end{equation}
for which $\alpha = \gamma + 1$. 
Terminology in this area is wildly inconsistent. To add
to the confusion, some references use cumulative distributions, which
decrease the index in a power law by 1.

The luminosities and masses of galaxies are well represented
by an exponentially truncated power-law function \citep{1976ApJ...203..297S}:
%Schecter 1976

\begin{equation}
\Phi(L) = (\Phi^*/L^*)(L/L^*)^\alpha e^{(-L/L^*)}
\end{equation}

\noindent
The parameter $\alpha$ denotes the slope of the power-law function
at low luminosities, and 
$L^*$ represents the luminosity above which
the number of galaxies declines sharply.  For the relatively
shallow slopes ($\alpha$) which are typical of present-day galaxies,
$L^*$ also coincides roughly with the peak contribution
to the total light of the galaxy population.  As discussed in 
\S \ref{demographics}, the Schechter function also provides a good fit to the
distribution function of total SFRs of galaxies, with the 
characteristic SFR in the exponential
designated as SFR$^*$, in analogy to $L^*$ above.

The term ``starburst galaxy" was introduced by \citet{1981ApJ...248..105W},
%Weedman et al. 1981
but nowadays the term is applied to a diverse array of galaxy
populations.  The common property of the present-day populations of starbursts
is a SFR out of equilibrium, much higher than the long-term average
SFR of the system.  No universally accepted quantitative definition
exists, however.  Some of the more commonly applied criteria are 
SFRs that cannot be sustained for
longer than a small fraction (e.g., $\le10$\%) of the Hubble time,
i.e., with gas consumption timescales of less than $\ll$1 Gyr,
or galaxies with disk-averaged SFR surface densities 
$\sigmasfr \ge 0.1$ \msunkpc (\S\ref{demographics}).  
Throughout this review we will use the term ``quiescent
star-forming galaxies" merely to characterize the non-starbursting 
galaxy population.  
Note that these local criteria for identifying starbursts are
not particularly useful for high-redshift galaxies; a young galaxy
forming stars at a constant SFR might more resemble a present-day starburst
galaxy than a normal galaxy today.

% definitions
%questions.tex
% 100511  RCK: major edits
% 100611 NJE edit
% 100711 NJE added refs to sections, needs more work
% 110411 NJE minor edits
% 120611 NJE try consolidating and re-ordering (and adding...)
% 121311 NJE incorporated Rob's suggestions
% 012412 NJE edits after comments
% 020212 NJE minor edits, correct section refs, etc.
% 031612 NJE adapt to emulateapj, assign subsec reference

\subsection{Questions}\label{questions}

%{\bf 1.  Astrophysical Questions}

\begin{enumerate}

% 1
\item How should we interpret observations of the main molecular
diagnostic lines (e.g., CO, HCN) and millimeter-wave dust emission? 
How does this interpretation
change as a function of metallicity, surface density, location
within a galaxy, and star formation environment?  
%Are the variations continuous or bimodal in nature?
(\S \ref{gastracers}, \S \ref{milkyway}, \S \ref{future})

% 2
\item How does the structure of the ISM, the structure of star-forming
clouds, and the star formation itself change as a function of metallicity,
surface density, location within a galaxy, and star formation environment? 
(\S \ref{gastracers}, \S \ref{outside}, \S \ref{future}).

% 3
\item How do the mass spectra of molecular clouds and dense clumps
in clouds vary between galaxies and within a galaxy? 
(\S \ref{massfunctions}, \S \ref{milkyway}, \S \ref{future})

% 4
\item How constant is the IMF, and how are star formation rate
measurements affected by possible changes in the IMF or by incomplete
sampling of the IMF?
(\S \ref{massfunctions}, \S \ref{halpha}, \S \ref{highmass},
\S \ref{localtests})

% 5
\item 
What are the limits of applicability of current star
formation rate tracers?  How are current measurements
biased by dust attenuation or the absence of dust, and
how accurately can the effects of dust be removed?
How do different tracers depend on metallicity, and what
stellar mass ranges and timescales do they probe?
(\S \ref{sfrdiagnostics})

% 6
\item How long do molecular clouds live and how can we best measure
lifetimes? Do these lifetimes
change systematically as functions of cloud mass, location in a
galaxy, or some other parameter? (\S \ref{sftheory}, \S \ref{future})

% 7
\item Do local observations provide any evidence for bimodality
in modes of star formation, for example distributed versus
clustered, low-mass versus high-mass star formation?
(\S \ref{local}, \S \ref{future})

% 8
\item On the scale of molecular clouds, what are the star formation 
efficiencies [$\epsilon = \mstar/(\mstar + \mcloud)$],  and
star formation rates per unit mass ($\epsilon^{\prime} = \sfr/\mcloud$), 
and do these efficiencies
vary systematically as functions of cloud mass or other parameters?
(\S \ref{local}, \S \ref{future})

% 9
\item How do spatial sampling and averaging affect the observed
form of the star formation relations expressed in terms of total,
molecular, or dense gas surface densities?
How well do \sfr\ diagnostics used in extragalactic studies work
on finer scales within a galaxy?  
(\S \ref{challenge}, \S \ref{sflobs}, \S \ref{localtests}, \S \ref{future})

% 10
\item Are there breakpoints or thresholds where either tracers or
star formation change their character?
(\S \ref{ismdiagnostics}, \S \ref{sfrdiagnostics}, \S \ref{synthesis})

\end{enumerate}

% questions

\section{ISM DIAGNOSTICS: GAS AND DUST}\label{ismdiagnostics}

%ismintro.tex
%060611
%070511 NJE
%072211 NJE added info on HI
%092011 NJE small fix.
%100711 NJE small fixes
%110411 NJE minor edits
%121211 NJE minor edits
%011712 NJE edits after comments
%020312 NJE added a ref to cold flows
%031212 NJE minor edits
%051412 NJE minor edits, used ion

\subsection{ISM Basics} \label{ismintro}

The interstellar medium (ISM) is a complex environment encompassing
a very wide range of properties (e.g., \citealt{1977ApJ...218..148M,
2005ARA&A..43..337C}). 
%McKee and Ostriker 1977, Cox, 2005
In the Milky Way, about half the volume
is filled by a hot ionized medium (HIM) with $n < 0.01$ \cmv\ and $\tk > 
\eten5$ K, and half is filled by
some combination of warm ionized medium (WIM) and warm neutral medium (WNM),
with $n \sim 0.1 - 1$ \cmv\ and $\tk$ of several thousand K
\citep{2005ARA&A..43..337C}.
The neutral gas is subject to a thermal instability that predicts
segregation into the warm neutral and cold neutral media (CNM),
the latter with $n > 10$ \cmv\ and $\tk < 100$ K
\citep{1969ApJ...155L.149F}.
%Field, Goldsmith, and Habing, 1969
The existence of the CNM is clearly revealed by emission in the hyperfine 
transition of \hi, with median temperature about 70 K, determined by
comparison of emission to absorption spectra
\citep{2003ApJ...586.1067H}.
% Heiles and Troland Millenium Arecibo Survey
However, the strict segregation into cold and warm stable phases
is not supported by observations suggesting that
about 48\% of the WNM may lie in the thermally unstable phase
\citep{2003ApJ...586.1067H}.
% Heiles and Troland Millenium Arecibo Survey
Finally, the molecular clouds represent the coldest ($\tk \sim 10$ K
in the absence of star formation), 
densest ($n > 30$ \cmv) part of the
ISM.  We focus on the cold ($\tk < 100$ K) phases here, as they are 
the only plausible sites of star formation. The warmer phases
provide the raw material for the cold phases, perhaps through
``cold" flows from the intergalactic medium \citep{2006MNRAS.368....2D}.
% Dekel and Birnboim
In the language of this review, such flows would be ``warm."

In the next section, we introduce some terminology used for the
molecular gas as a precursor to the following discussion 
of the methods for tracing it. We then discuss the use of dust
as a tracer (\S \ref{dusttracers}) before discussing the gas tracers 
themselves (\S \ref{gastracers}).

% ismintro
%names.tex
%updated 042211 NJE
% 060611 NJE moved mass functions to a separate section later in paper
%060711
%061011 NJE moving bits to other subsections
%070511
%072611 small fixes
%110411 NJE small fixes
%121211 NJE minor edits
%011712 NJE edits after comments
%020312 NJE very minor edit
%031212 NJE very minor edits

\subsection{Clouds and Their Structures}\label{names}

The CNM defined in the preceding section presents a wide array of
structures, with a rich nomenclature, but we focus here on the objects
referred to as clouds, invoking a kind of condensation process. This
name is perhaps even more appropriate to the molecular gas, which
corresponds to a change in chemical state.

The molecular cloud boundary is usually defined by the detection, above
some threshold, of emission from the lower rotational transitions of CO. 
Alternatively, a certain level of extinction of background stars is often used. 
In confused regions, velocity coherence can be used to separate clouds at 
different distances along the line of sight. However, molecular clouds are 
surrounded by atomic envelopes and a transition region in which the hydrogen
is primarily molecular, but the carbon is mostly atomic 
\citep{1988ApJ...334..771V}.
%(van Dishoeck and Black 1988). 
These regions are known as PDRs, which stands either for Photo-Dissociation
Regions, or Photon-Dominated Regions \citep{1997ARA&A..35..179H}.
%Hollenbach and Tielens 1997
More recently, they have been referred to as ``dark gas."

The structure of molecular clouds is complex, leading to considerable 
nomenclatural chaos. 
Cloud projected boundaries, defined by either dust or CO emission or 
extinction, can be characterized by a fractal dimension around 1.5 
(\citealt{1991ApJ...378..186F, 1990ASSL..162..151S, 1998A&A...336..697S}),
% Scalo 1990 in book, Falgarone, Phillips, and Walker 1991) 
suggesting an intrinsic 3-D dimension of 2.4.
When mapped with sufficient sensitivity and dynamic range, clouds are highly
structured, with filaments the dominant morphological theme; denser,
rounder cores are often found within the filaments
(e.g., \citealt{2010A&A...518L.102A, 2010A&A...518L.103M, 
2010A&A...518L.100M}).  
%Andre et al., Menshchikov et al. 2010, Molinari et al. 2010, all from Herschel 

The probability distribution function of surface densities, as mapped by 
extinction, can be fitted to a lognormal function for low extinctions, 
$\av \leq 2-5$ mag \citep{2010A&A...512A..67L},
%Lombardi et al. 2010
with a power law tail to higher extinctions, at least in some clouds. 
In fact, it is the clouds 
with the power-law tails that have active star formation 
\citep{2009A&A...508L..35K}.
% Kainulainen et al. 2009
Further studies indicate that the power-law tail begins at 
$\sigmamol \sim 40-80$ \msunpc, 
and that the material in the power-law tail lies
in unbound clumps with mean density about \eten3 \cmv\ \citep{2011A&A...530A..64K}.
% Kainulainen et al. 2011

Within large clouds, and especially within the filaments,
there may be small cores, which are much less massive
but much denser. Theoretically, these are the sites of formation of individual
stars, either single or binary \citep{2000prpl.conf...97W}.
%Wiliams, Blitz, McKee 2000
Their properties have been reviewed by \citet{2007prpl.conf...17D}
%Di Francesco et al. (2007) 
and \citet{2007prpl.conf...33W}.
%Ward-Thompson et al. (2007). 
Those cores without evidence of 
ongoing star formation are called starless cores, and the subset of those that
are gravitationally bound  are called prestellar cores. 
Distinguishing these two requires a good mass estimate, independent of
the virial theorem, and observations of lines that trace the kinematics.
A sharp decrease in turbulence may provide an alternative way to distinguish
a prestellar core from its surroundings \citep{2010ApJ...712L.116P}.
% Pineda et al. 2010, ammonia maps in Perseus

While clouds and cores are reasonably well defined, intermediate structures
are more problematic. Theoretically, one needs to name the entity that
forms a cluster. Since large clouds can form multiple (or no) clusters, 
another name is needed. \citet{2000prpl.conf...97W}
%Williams et al. (2000) 
established the name ``clump" for this structure.
The premise was that this region should also be gravitationally bound,
at least until the stars dissipate the remaining gas.
Finding an observable correlative has been a problem
(\S \ref{massfunctions}).

Finally, one must note that this hierarchy of structure is based on relatively
large clouds. There are also many small clouds \citep{1988ApJS...68..257C},
% Clemens and Barvainis 1988 
some of which are their own clumps and/or cores.
The suggestion by \citet{1947ApJ...105..255B}
% Bok and Reilly 1947
that these could be the sites of star formation was seminal.

% names
%dusttracers.tex
%060611
%061011 NJE
%070111 NJE
%070511 NJE
%070911 NJE
%072611 NJE
%092011 NJE condense a bit and fix refs
%100711 NJE minor fixes
%110411 NJE minor edits
%120511 NJE changed Planck refs to published ones
%121211 NJE minor edits
%011712 NJE edits after comments
%020312 NJE very minor edits
% 031612 NJE used url command for emulateapj version
% 051412 NJE minor edits, used ion

\subsection{Dust as an ISM Tracer}\label{dusttracers}

In the Milky Way, roughly 1\% of the ISM is found in solid form,
primarily silicates and carbonaceous material
\citep{2003ARA&A..41..241D}
% Draine (2003) ARAA 41, 241.
(see also a very useful website\footnote{
\url{http://www.astro.princeton.edu/$\sim$draine/dust/dust.html}}).
Dust grain sizes range from 0.35 nm to around 1 \micron, 
with evidence for grain growth in molecular clouds.
The column density of dust can be measured by studying either 
the reddening (using colors) or the extinction of 
starlight, using star counts referenced to an unobscured field.
The relation between extinction (e.g., \av) and reddening 
[e.g., $E(B-V)$] is generally parameterized by $R_V = A_V/E(B-V)$.
In the diffuse medium, the reddening 
has been calibrated against atomic and molecular gas to obtain
$N(\hi) + 2N(\hh) = 5.8\ee{21}$ cm$^{-2} E(B-V)$(mag), not including
helium
\citep{1978ApJ...224..132B}. 
%Bohlin et al. 1978
For $R_V = 3.1$, this relation translates to 
$N(\hi) + 2N(\hh) = 1.87\ee{21}$ cm$^{-2}$ \av(mag).
This relation depends on metallicity, with coefficients larger by
factors of about 4 for the LMC and 9 for the SMC
\citep{2001ApJ...548..296W}.
% Weingartner and Draine 2001

The relation derived for diffuse gas 
is often applied in molecular clouds, but grain growth
flattens the extinction curve, leading to larger values of $R_V$,
and a much flatter curve in the infrared
(\citealt{2007ApJ...663.1069F,2009ApJ...690..496C}).
%Flaherty07, Chapman 09 (clouds) 
More appropriate conversions can be found in
the curves for $R_V = 5.5$ in the website referenced above.
There is evidence of continued grain growth in denser regions, which
particularly affects the opacity at longer wavelengths, such as
the mm/submm regions \citep{2011ApJ...728..143S}.
% Yancy and Tracy  study of dust opacities).

Reddening and extinction of starlight provided early evidence
of the existence of interstellar clouds 
(\citealt{1908AN....177..231B}).
%Barnard note in Astronomische Nachrichten on bright and dark nebulae
\citet{1962ApJS....7....1L}
% Lynds catalog of dark nebulae
provided the mostly widely used catalog of ``dark clouds"
based on star counts, up to $\av \sim 6$ mag, at which point the clouds
were effectively opaque at visible wavelengths.
The availability of infrared photometry for large numbers of 
background stars from {\it 2MASS}, {\it ISO},  and {\it Spitzer} 
toward nearby clouds has allowed extinction measurements up to 
$\av \sim 40$ mag. These have proved very useful in characterizing
cloud structure and in checking the gas tracers discussed in \S
\ref{gastracers}.
Extinction mapping with sufficiently high spatial resolution
can identify cores (e.g., \citealt{2007A&A...462L..17A}),
%Alves et al. 2007
but many are not gravitationally bound
\citep{2008ApJ...672..410L},
%Lada et al.
and they are likely to  dissipate.

At wavelengths with strong, diffuse Galactic background emission, such
as the mid-infrared, dust absorption against the continuum can be
used in a way parallel to extinction maps 
(e.g., \citealt{2000A&A...361..555B, 2007ApJ...665..466S}).
% Bacmann, A. et al. 2000 using ISOCAM on prestellar cores
% Stutz et al. using Spitzer 24 micron
The main uncertainties in deriving the gas column density
are the dust opacity in the mid-infrared and the fraction of
foreground emission.

The energy absorbed by dust at short wavelengths is emitted at
longer wavelengths, from infrared to millimeter, with a spectrum
determined by the grain temperature and the opacity as a function of
wavelength. At wavelengths around 1 mm, the emission is
almost always optically thin and close to the Rayleigh-Jeans limit for
modest temperatures. (For $\td = 20$K, masses, densities, etc.
derived in the Rayleigh-Jeans limit at $\lambda = 1$ mm must be
multiplied by 1.5). 
Thus, the millimeter-wave continuum
emission provides a good tracer of dust (and gas, with knowledge of the
dust opacity) column density and mass.

As a by-product of studies of the CMB, {\it COBE}, {\it WMAP}, and
{\it Planck} produce large-scale maps of mm/submm emission from the
Milky Way. With {\it Planck}, the angular resolution reaches
$\theta = 4.7\arcmin$ at 1.4 mm in  
an all-sky map of dust temperature and optical depth
\citep{2011A&A...536A..19P}.
%  Ade et al. 2011 on dark gas constraints from all sky map
Mean dust temperature increases from $\td = 14 -15$ K in the outer
Galaxy to 19 K around the Galactic center.
A catalog of cold  clumps with $\td = 7 - 17$ K,
mostly within 2 kpc of the Sun, many with surprisingly low
column densities, also resulted from analysis of {\it Planck} data
\citep{2011A&A...536A..23P}.
%  Ade et al. 2011 on cold clump catalog
Herschel surveys will be providing higher resolution 
[25\arcsec\ at 350 \micron\ \citep{2010A&A...518L...3G}
% Griffin et al. 2010, SPIRE performance paper
and 6\arcsec\ at 70 \micron\ \citep{2010A&A...518L...2P}] 
% Poglitsch et al. 2010 PACS performance paper
images of the nearby clouds 
\citep{2010A&A...518L.102A},
%Andre et al, initial Herschel GB results
the Galactic Plane \citep{2010A&A...518L.100M},
%Molinari et al) high gal description
and many other galaxies
(e.g., \citealt{2011ApJ...738...89S, 2011ApJ...738..106W}).
% Skibba et al., Wuyts et al..
Still higher resolution is available at $\lambda \geq 350$ \micron\
from ground-based
telescopes at dry sites. Galactic plane surveys at wavelengths
near 1 mm have been carried out in both hemispheres 
(\citealt{2011ApJS..192....4A,2009A&A...504..415S}).
% Aguirre et al. bolocam, and Schuller et al. ATLASGAL
Once these are put on a common footing, we should have maps of the
Galactic plane with resolution of 20\arcsec\ to 30\arcsec.
Deeper surveys of nearby clouds \citep{2007PASP..119..855W},
%Ward-Thompson et al. 
the Galactic Plane, and extragalactic fields
are planned with SCUBA-2.

While studies from ground-based telescopes offer higher resolution,
the need to remove atmospheric fluctuations leads to loss of sensitivity
to structures larger than typically about half the array footprint on the sky
\citep{2011ApJS..192....4A}.
% Aguirre
While the large-scale emission is lost, the combination of sensitivity
to a minimum column density and a maximum size makes these surveys
effectively select structures of a certain mean volume density. In nearby
clouds, these correspond to dense cores \citep{2009ApJ...692..973E}.
%Enoch et al. 2009
For Galactic plane surveys, sources mainly correspond to clumps but
can be whole clouds at the largest distances 
\citep{2011ApJ...741..110D}.
%Dunham et al. in press 

% dusttracers
%gastracers.tex
%042611
%060611
%060711
%061011
%062211
%070511
%070711 NJE added ref to Reiter
%070811 NJE moved discussion about HI surveys of MW to milkyway
%070911 NJE
%072211 NJE updated based on Bolatto comments and other fixes
%082411 NJE further updates based on more exchange with Bolatto
%090911 NJE added some stuff on SMC
%092111 NJE clean up
%100311 NJE minor edits, added warnings about X factor assumptions
% looking ahead to Daddi et al. discussions.
%100711 NJE minor fixes
%110411 NJE minor edits
%120511 NJE edits after editor/RG comments and other inputs
%121211 NJE minor edits and add ref to Garcia-Burillo, 
%121211 NJE added note that one can expect change in X(CO) around 100
%121611 NJE sorted out He issue and noted our standard X(CO)
%121911 NJE minor edits
%011712 NJE edits after comments
%020312 NJE minor edits
%021312 NJE minor edits after RCK comments
%031212 NJE minor edits
%050512 NJE updated refs
%050912 NJE updated refs
%051012 NJE updated refs
%051412 NJE minor edits used ion
%060112 NJE corrected equation from Draine
%061912 NJE added refs to HI surveys using old coefficient

\subsection{Gas Tracers}\label{gastracers}

One troublesome aspect of extragalactic studies has been the determination
of the amount and properties of the gas. We discuss first the atomic 
phase, then the molecular phase, and finally tracers of dense gas.

\subsubsection{The Cold Atomic Phase}\label{atomic}

The cold, neutral atomic phase is traced by the hyperfine transition
of hydrogen, occuring at 21 cm in the rest frame. This transition
reaches optical depth of unity at 
\begin{equation}
N(\hi) = 4.57\ee{20} \cmc (T_{spin}/100 {\rm K})(\sigma_v/1 \kms),
\label{hioptdepth}
\end{equation}
where $T_{spin}$ is the excitation temperature, usually
equal to the kinetic temperature, \tk, and 
$\sigma_v$ is the velocity dispersion 
(eq. 8.11 in \citealt{2011piim.book.....D}).
% Draine, new book on Physics of ISM and IGM
For dust in the
diffuse ISM of the Milky Way, this column density corresponds to 
$\av = 0.24$ mag, and optical depth effects can be quite important. 
The warm, neutral phase is difficult to study, as lines become broad and weak,
but it has been detected \citep{1988gera.book...95K}.
%(Kulkarni and Heiles 1988, in book by Kellerman and Verschuur). 
A detailed analysis of both CNM and WNM can be found in 
\citet{2003ApJ...586.1067H}.
%Heiles and Troland, Millenium Arecibo 21 cm survey

With $I(\hi) = \int{S_{\nu}dv}$ (Jy \kms),
\begin{equation}
M(\hi)/\msun = 2.343\ee5 (1+z)^{-1} (D_L/{\rm Mpc})^2 I(\hi),
\label{himasseq}
\end{equation}
not including helium (eq. 8.21 in \citealt{2011piim.book.....D}, 
% Draine, new book on Physics of ISM and IGM
but note errata regarding $(1+z)$ factor, which was in the numerator in the
2011 edition). A coefficient of 2.356\ee5 is in common use among
\hi\ observers 
(e.g., \citealt{2008AJ....136.2563W, 2010MNRAS.403..683C, 2005AJ....130.2598G,
2004MNRAS.350.1195M}),
% Walter et al. (THINGS) B. Catinella (GASS), Giovanelli (ALFA), Meyer (HIPASS)
usually traced to \citet{1975gaun.book..309R}. 
% Roberts in Galaxies and the Universe
The difference 
is almost certainly caused by a newer value for the Einstein A value
in \citet{2011piim.book.....D}.

\subsubsection{Molecular Gas}\label{molecular}

The most abundant molecule is \hh, but its spectrum is
not a good tracer of the mass in molecular clouds. The primary reason
is not, as often stated, the fact that it lacks a dipole moment, but
instead the low mass of the molecule. The effect of low mass is
that the rotational transitions require quite high temperatures to
excite, making the bulk of the \hh\ in typical clouds invisible.
Continuum optical depth at the \mir\ wavelengths of these lines is
also an impediment.
The observed rotational \hh\ lines originate in surfaces of clouds, probing
1\% to 30\% of the gas in a survey of galaxies (e.g.,
\citealt{2007ApJ...669..959R}).
%Roussel et al. 2007 SINGS survey)

Carbon monoxide (CO) is the most commonly used
tracer of molecular gas because its lines are
the strongest and therefore easiest to observe. These advantages
are accompanied by various drawbacks, which \mw\ observations may
help to illuminate. Essentially, CO emission traces column density of
molecular gas only over a very limited dynamic range.
At the low end, CO requires protection from 
photodissociation and some minimum density to excite it, so it does not
trace all the molecular (in the sense that hydrogen is in \hh)
gas, especially if metallicity is low. 
At the high end, emission from CO saturates at modest column densities.
A more complete review by A. Bolatto of the use of CO to trace
molecular gas is scheduled for a future ARAA, so we briefly highlight
the issues here.

CO becomes optically thick at very low total column densities.
Using CO intensity to estimate the column density of a cloud is akin to
using the presence of a brick wall to estimate the depth of the building
behind it. The isotopologues of CO (\coo\ and \cooo)
provide lower optical depths, but at the cost of weaker lines.
To use these, one needs to know the isotope ratios; these are
fairly well known in the Milky Way \citep{1994ARA&A..32..191W},
% Wilson and Rood review
but they may differ substantially in other galaxies.
In addition, the commonly observed lower transitions of CO are
very easily thermalized ($\tex = \tk$) making them insensitive to
the presence of dense gas. 

The common practice in studies of \mw\ clouds is to use a combination of
CO and \coo\ to estimate optical depth and hence CO column density ($N$(CO)). 
One can then correlate $N$(CO) or its isotopes against extinction to 
determine a conversion factor 
(\citealt{1978ApJS...37..407D, 1982ApJ...262..590F}).
%Dickman 1978, Frerking et al. 1982
A recent, careful study of CO and \coo\ toward the Taurus molecular cloud 
shows that $N$(CO) traces \av\ up to 10 mag in some regions
but only up to 4 mag in other regions, and other issues cause problems below
$\av = 3$ mag \citep{2010ApJ...721..686P}.
%Pineda et al. 2010)
Freeze-out of CO and conversion to CO$_2$ ice causes further
complications in cold regions of high column density
(\citealt{2003ApJ...583..789L, 2008ApJ...678.1005P, 2007ApJ...655..332W}),
%Lee et al. study of prestellar cores) %Pontoppidan 2008 %Whittet et al. 2007
but other issues likely dominate the 
interpretational uncertainties for most extragalactic CO observations.

Since studies of other galaxies generally have only CO observations
available, the common practice has been to relate \hh\ column density
($N$(\hh)) to the integrated intensity of CO [$\ico = \int{T dv}$ (K \kms)]
via the infamous ``X-factor":
$N$(\hh)$ = \xco \ico$.  For example, \citet{1998ApJ...498..541K}
%Kennicutt 98, ApJ
used $\xco = 2.8\ee{20}$ cm$^{-2}$ (K km s$^{-1}$)$^{-1}$.

\citet{2010ApJ...721..686P} 
% Pineda et al. 2010
find $\xco = 2.0\ee{20}$ cm$^{-2}$ 
(K km s$^{-1}$)$^{-1}$ 
in the parts of the cloud where both CO
and \coo\ are detected along individual sight lines. 
CO does emit beyond the area where \coo\ is detectable and even,
at a low level, beyond the usual detection threshold for CO.
Stacking analysis of positions beyond the usual boundary of the Taurus
cloud suggest a factor of two additional mass in this transition region
\citep{2008ApJ...680..428G}.
%Goldsmith et al. 2008
Such low intensity, but large area, emission
could contribute substantially to observations of
distant regions in our galaxy and in other galaxies.
For the outer parts of clouds, where only CO is detected, and only by stacking 
analysis, \xco\ is six times higher \citep{2010ApJ...721..686P}.
Even though about a quarter of the total mass is in this outer region,
they argue that the best single value to use is $\xco = 2.3\ee{20}$ 
cm$^{-2}$ (K km s$^{-1}$)$^{-1}$, similar to values commonly adopted
in extragalactic studies.
When possible, we will use 
$\xco = 2.3\ee{20}$ cm$^{-2}$ (K km s$^{-1}$)$^{-1}$ 
for this paper.
In contrast, \citet{2010A&A...518A..45L}
% Liszt et al. 2010
argue that \xco\ stays the same even in diffuse gas.
If they are right, much of the integrated 
CO emission from some parts of galaxies could arise in diffuse, unbound gas.

Comparing to extinction measures in two nearby clouds extending up 
to $\av = 40$ mag, \citet{2010ApJ...723.1019H}
%Heiderman et al.
found considerable variations from region to region, with \ico\ totally
insensitive to increased extinction in many regions. Nonetheless, when
averaged over the whole cloud, the standard \xco\ would cause errors
in mass estimates of $\pm 50$\%, systematically overestimating the mass
for $\sigmamol < 150$ \msunpc\ and underestimating for
$\sigmamol > 200$ \msunpc.
Simulations of turbulent gas with molecule formation also suggest that
\ico\ can be a very poor tracer of \av\ \citep{2010MNRAS.404....2G}.
%Glover et al. 2010
Overall, the picture at this point is that measuring the extent of the
brick wall will not tell you anything about particularly deep or dense
extensions of the building, but if the buildings are
similar, the extent of the wall is a rough guide to the total mass of the
building (cf. \citealt{2011MNRAS.415.3253S}). 
% Shetty et al. 2011b on X factor in MW clouds

Following that thought, the luminosity of CO \jj10\ is often used as a measure
of cloud mass, with the linewidth reflecting the virial theorem in a crude
sense. In most extragalactic studies, individual clouds are not resolved,
and the luminosity of CO is essentially a cloud counting technique
\citep{1986ApJ...309..326D}.
% Dickman et al. (1986). 
Again, this idea works only if the objects being counted have rather uniform
properties.  In this approach, $M = \aco \lco$. For the standard value
for \xco\ discussed above
($\xco = 2.3\ee{20}$ cm$^{-2}$ (K km s$^{-1}$)$^{-1}$)
$\aco = 3.6$ \msun (K \kms pc$^{2})^{-1}$
without helium, and
$\aco = 4.6$ \msun (K \kms pc$^{2})^{-1}$ 
with correction for helium.

The effects of cloud temperature, density, metallicity, etc.
on mass estimation from CO were discussed by
\citet{1988ApJ...325..389M},
% Maloney and Black 1988
who concluded that large variations in the conversion factor are likely.
Ignoring metallicity effects, they would predict 
$\xco \propto n^{0.5} \tk^{-1}$, where $n$ is the mean density in the cloud.
\citet{2011MNRAS.415.3253S}
% Shetty et al 2011b on X in MW clouds
find a weaker dependence on \tk, $\xco \propto \tk^{-0.5}$.
These scalings rests on some arguments that may not apply in other galaxies.
Further modeling with different metallicities 
(\citealt{2011MNRAS.412.1686S, 2012ApJ...747..124F})
%Shetty et al. 2011, Feldmann et al. 2011)
has begun to provide
some perspective on the range of behaviors likely in other galaxies.

There is observational evidence for changes in \aco\ in centers of galaxies.
In the \mw\ central region, \citet{1998ApJ...493..730O}
%Oka et al. 1998 (2-1) 
suggest a value for \xco\ a factor of 10 lower than in the \mw\ disk.
The value of \aco\ appears to decrease by factors of 2-5 in nearby 
galaxy nuclei (e.g., \citealt{2004AJ....127.2069M, 2008ApJ...675..281M}),
% Meier and Turner 2004  Meier, Turner, and Hurt 2008
and by a factor of about 5-6 in local ULIRGs 
(\citealt{1998ApJ...507..615D,1993ApJ...414L..13D}),
% Downes et al. 1993  Downes and Solomon,
and probably by a similar factor in high-redshift, molecule-rich galaxies
\citep{2005ARA&A..43..677S}.
% Solomon and Vanden Bout, ARAA

\begin{figure}%1	% Example of Figure pull
\center
\includegraphics[scale=0.8, angle=90]{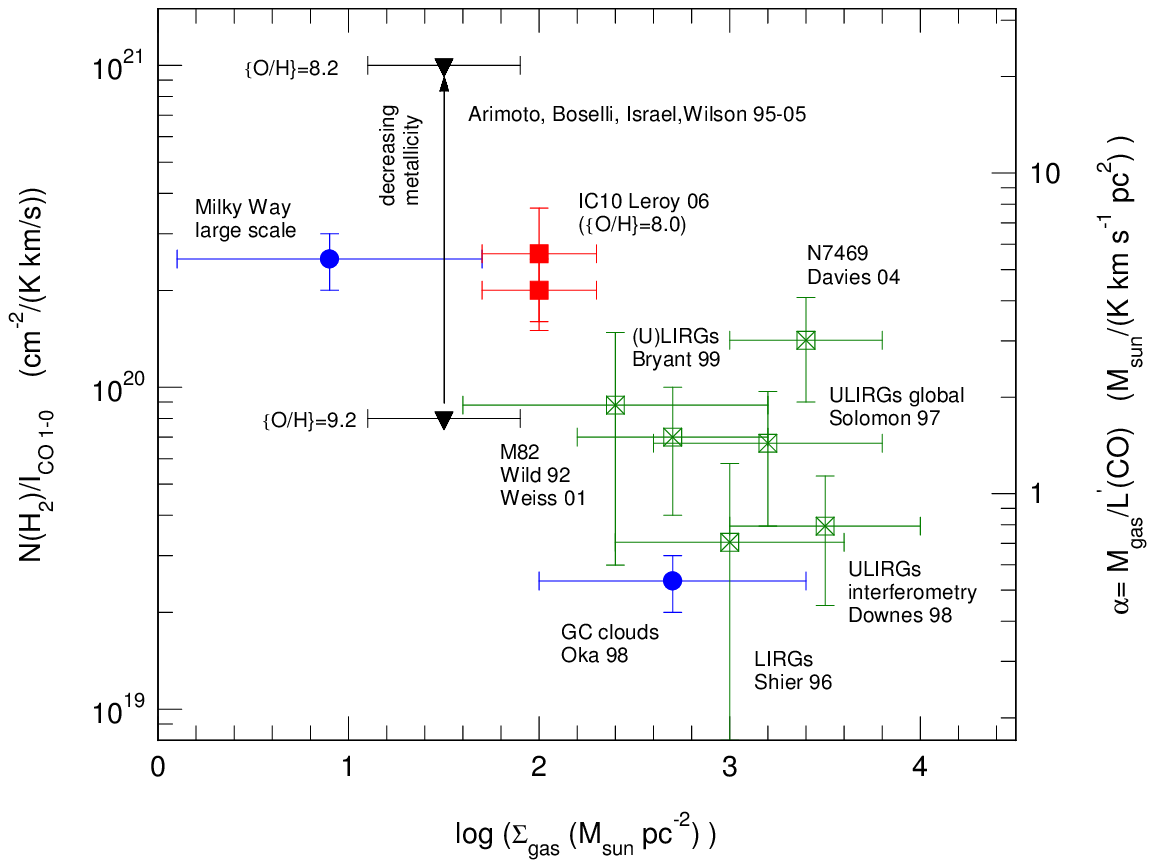}
\caption{
Compilation of the conversion factor (\xco) from the CO \jj10
integrated intensity [$I_{\rm CO}$ (K km s$^{-1}$)] or 
luminosity [$L^\prime_{\rm CO}$ (K km s$^{-1}$ pc$^2$)] to \hh\ column density 
(left vertical scale) and total (\hh\ and He) gas mass (right vertical scale), 
derived in various Galactic and extragalactic targets. 
Blue circles denote measurements in the disk and center of the Milky Way, 
based on various virial, extinction, and isotopomeric analyses.
%(Solomon et al. 1987; Strong & Mattox 1996; Dame et al. 2001; Oka et al. 1998).
Crossed green squares denote measurements in starbursts and (U)LIRGs, mainly 
based on dynamical constraints. 
%(Wild et al. 1992; Davies et al. 2004; Shier et al. 1994; Hinz & Rieke 2006; 
%Weiss et al. 2001; Solomon et al. 1997; Downes & Solomon 1998; 
%Bryant & Scoville 1999). 
Filled triangles denote conversion factors as a function of decreasing 
metallicity (vertical arrow) from  (bottom) to 8.2 (top), derived 
mainly from global (large scale) dust mass measurements in nearby galaxies 
and dwarfs by several groups. 
%(Arimoto et al. 1996; Israel 2000, 2005; Boselli et al. 2002; 
%see also Wilson 1995). 
In contrast, red filled squares mark X-factor measurements 
toward individual clouds, over the same range in metallicity.
%(Rosolowsky et al. 2003; Leroy et al. 2006).
Taken from \citet{2008ApJ...680..246T}; references to the original work
are given there.
%Tacconi et al. 2008.
Reproduced by permission of the AAS.
}
\label{tacconifig10}
\end{figure}

A compilation of measurements indicates that 
\aco\ decreases from the usually assumed value 
with increasing surface density of gas once $\sigmamol > 100 $ \msunpc\
(Fig. \ref{tacconifig10}, \citealt{2008ApJ...680..246T}).
% Tacconi et al. 2008, in appendix
This \sigmamol\ corresponds to $\av \sim 10$ mag, about the point where
CO fails to trace column density in Galactic clouds, and the point
where the beam on another galaxy might be filled with CO-emitting gas. 
Variation in \aco\ has a direct bearing on interpretation of
starbursts (\S \ref{sflobs}), and it is highly unlikely that \aco\ 
is simply bimodal over the full range of star forming environments.
Indeed, \citet{2012MNRAS.421.3127N}
% Narayanan et al. 2012 on general form of X(CO)
use a grid of model galaxies to infer a smooth function of \ico:
$\aco = {\rm min}[6.3, 10.7 \times \ico^{-0.32}] Z^{\prime -0.65}$,
where $Z^\prime$ is the metallicity divided by the solar value.

Other methods have been used to trace gas indirectly, including gamma-ray
emission and dust emission. Gamma-rays from decay of pions produced from
high-energy cosmic rays interacting with baryons in the ISM directly
trace all the matter if one knows the cosmic ray flux 
(e.g., \citealt{ 1989ARA&A..27..469B}).
% Bloemen 1989, ARAA 29, 469). 
Recent analysis of gamma-ray data from {\it Fermi}
have inferred a low value for \xco\ of $(0.87\pm0.05) \ee{20}$ 
cm$^{-2}$ (K km s$^{-1}$)$^{-1}$ in the solar neighborhood 
\citep{2010ApJ...710..133A}.
% Abdo et al. 2010, ApJ710, 133.
This value is a factor of three lower than other estimates and
early results from dust emission, using {\it Planck}, which found
$\xco = (2.54\pm 0.13)\ee{20}$ cm$^{-2}$ (K km s$^{-1}$)$^{-1}$
for the solar neighborhood. 
\citep{2011A&A...536A..19P}.
% Planck Collaboration all sky T and tau from Planck, constraints on dark gas 

\citet{2010ApJ...710..133A}
%Abdo et al. 2010  
also inferred the existence of ``dark gas", which is not
traced by either \hi\ or CO, surrounding the CO-emitting region in
the nearby Gould Belt clouds, with mass
of about half of that traced by CO.
\citet{2011A&A...536A..19P}
have also indicated the existence of
``dark gas" with mass 28\% of the atomic gas and 118\% of the CO-emitting
gas in the solar neighborhood. 
The dark gas appears around $\av = 0.4$ mag,
and is presumably the CO-poor, but H$_2$-containing, outer parts 
of clouds (e.g., \citealt{2010ApJ...716.1191W}),
%Wolfire et al. 2010 
which  can be seen in [C{II}] emission at 158 \micron\
and in fluorescent \hh\ emission at 2 \micron\
(e.g., \citealt{1998ApJ...498..735P}).
%Pak et al. 1998
It is dark only if one restricts attention to \hi\ and CO.
At low metallicities, the layer of C$^+$ grows, and a given CO
luminosity implies a larger overall cloud (see Fig. 14 in 
\citealt{1998ApJ...498..735P}).
However, 
\citet{2011A&A...536A..19P}
%Planck Collaboration astro-ph 1101.2029
also suggest that the dark gas may include some gas that is primarily
atomic, but not traced by \hi\ emission, owing to optical depth effects.

Dust emission is now being used as a tracer of gas for other galaxies.
Maps of \hi\ gas are used to break the degeneracy between the gas
to dust ratio
$\delta_{gdr}$ and $\alpha^\prime_{\rm CO}$ in the following equation:
\begin{equation}
\delta_{gdr} \Sigma_{dust} = \alpha^\prime_{\rm CO}\ico + \Sigma_{HI}.
\label{leroyeqn}
\end{equation}
\citet{2011ApJ...737...12L}
%Leroy et al. 
minimize the variation in $\delta_{gdr}$ to find the
best fit to $\alpha^\prime_{\rm CO}$ in several nearby galaxies. They determine
$\Sigma_{dust}$ from {\it Spitzer} 
images of 160 \micron\ emission, with the dust
temperature (or equivalently the external radiation field) constrained
by shorter wavelength \fir\ emission. This method should measure everything
with substantial dust that is {\it not} \hi, so includes the so-called
``dark molecular gas."
They find that 
$\alpha^\prime_{\rm CO} = 3 -9$ \msun\ pc$^{-2}$ (K km s$^{-1}$)$^{-1}$,
similar to values based on \mw\ studies, for galaxies with metallicities
down to $12 + $[O/H]$ \sim 8.4$, below which $\alpha^\prime_{\rm CO}$ increases
sharply to values like 70. Their method does not address the
column density within clouds, but only the average surface density on 
scales larger than clouds.
For galaxies not too different from the \mw, the standard 
$\alpha^\prime_{\rm CO}$
will roughly give molecular gas masses, but in galaxies with lower 
metallicity, the effects can be large. An extreme is the SMC, where
$\alpha^\prime_{\rm CO} \sim 220$ \msun\ pc$^{-2}$ (K km s$^{-1}$)$^{-1}$
\citep{2011ApJ...741...12B}.
% Bolatto et al.
Maps of dust emission at longer wavelengths should decrease the sensitivity
to dust temperature.

\subsubsection{Dense Gas Tracers}\label{densegas}

Lines other than CO \jj10\ generally trace warmer (e.g., higher J CO lines)
and/or denser (e.g., HCN, CS, ...) gas. 
%Wu et al. (2010)
\citet{2010ApJS..188..313W} showed that the line luminosities of commonly
used tracers (CS \jj21, \jj54, and \jj76; HCN \jj10, \jj32) correlate
well with the virial mass of dense gas; indeed most follow relations
that are close to linear. Since these lines are optically thick, the linear
correlation is somewhat surprising, but it presumably has an explanation
similar to the cloud-counting argument offered above for CO. 
Collectively, we refer to these
lines as ``dense gas tracers", but the effective densities increase with
$J$ and vary among species 
(see \citealt{1999ARA&A..37..311E, 2011ApJS..195....1R} 
% Evans 1999 ARAA and Reiter 2011
for definitions and tables of effective and critical densities). 
As an example for HCN \jj10,
\begin{equation}
 {\rm log}(L'_{{\rm HCN1-0}})=1.04\times {\rm log}(M_{{\rm Vir}}
(R_{{\rm HCN1-0}}))-1.35 
\label{hcnmasseq}
\end{equation}
from a robust estimation fit to data on dense clumps in the \mw\
\citep{2010ApJS..188..313W}.
%Wu et al. 2010
The line luminosities of other dense gas tracers also show strong
correlations with both virial mass \citep{2010ApJS..188..313W}
and the mass derived from dust continuum observations
\citep{2011ApJS..195....1R}.
%Reiter et al.

These tracers of dense gas are of course subject to the same caveats
about sensitivity to physical conditions, such as metallicity,
volume density, turbulence, etc., as discussed above for CO.
In fact, the other molecules are more sensitive to photodissociation
than is CO, so they will be even more dependent on metallicity
(e.g., \citealt{2011MNRAS.415.1977R}).
%Rosolowsky, Pineda, and Gao 2011
Dense gas tracers should become much more widely used in the ALMA era, but 
caution is needed to avoid the kind of over-simplification that has tarnished
the reputation of CO. Observations of multiple transitions and rare
isotopes, along with realistic models, will help 
(e.g., \citealt{2012A&A...539A...8G}).

\subsubsection{Summary}\label{gastracerssummary}

What do extragalactic observations of molecular tracers actually measure?
None are actually tracing the surface density of a smooth medium,
with the possible exception of the most extreme starbursts.
The luminosity of CO is a measure of the number of emitting structures times
the mean luminosity per structure.  
Dense gas tracers work in a similar way, but select gas above
a higher density threshold than does CO.
For CO \jj10, the structure is the
CO emitting part of a cloud, which shrinks as metallicity decreases.
For dense gas tracers, the structure is the part of a cloud dense enough
to produce significant emission, something like a clump if conditions
are like the \mw, but even more sensitive to metallicity.  Using a 
conversion factor, one gets a ``mass" in those structures. 

Estimates
of the conversion factors for CO  vary by factors of 3 for the solar 
neighborhood and at least an order of magnitude at low metallicity and by
at least half an order of magnitude for our own Galactic Center and
for extreme systems like ULIRGs.
Dividing by the area of the galaxy or the beam, 
one gets ``\sigmamol",
which really should be thought of as an area filling factor of 
the structures being probed times some crude estimate of the mass
per structure. That estimate depends on conditions in the structures,
such as density, temperature, and abundance and on the external radiation
field and the metallicity. 
Once the \lco\ translates to surface densities above that of individual
clouds ($\sigmamol > \sim 100$ \msunpc),
the interpretation may change as the area filling factor can now
exceed unity.  Large ranges of velocity in other galaxies 
(if not exactly face-on) allow \ico\ to still count clouds above this threshold,
but the clouds may no longer be identical.
The full range of inferred molecular surface densities in other galaxies 
inferred from CO, assuming a constant conversion factor, 
is a factor of \eten3\ (\S \ref{sflobs}); 
given that CO traces \sigmamol\
only over a factor of 10 at best in well-studied clouds in the \mw\
(\citealt{2010ApJ...721..686P,2010ApJ...723.1019H}),
this seems a dangerous extrapolation.

% gastracers
%massfunctions.tex
% moved from names.tex  060611 NJE
% 070911 NJE
% 080411 NJE moved some things here from milkyway.tex on cloud mass function
% 082311 NJE added some stuff on most massive clusters, IGIMF, ...
% 083111 NJE added on mass functions of clouds in nearby galaxies
% 092211 NJE minor cleanup
% 100611 NJE smoothed out after moving some to definitions
% 100611 NJE moved this to end of gas diagnostics
% 100711 NJE some clean up
% 110411 NJE minor edits
% 121211 NJE very minor edits
% 121611 NJE changed title and a few edits
% 011712 NJE edits after comments major edits re IGIMF, etc.
% 050512 NJE updated references
% 050912 NJE updated references

\subsection{Mass Functions of Stars, Clusters, 
and Gas Structures}\label{massfunctions}

The mass functions of the structures in molecular clouds (\S \ref{names})
are of considerable interest in relation
to the origin of the mass functions of clusters (clumps) and stars (cores). 
\citet{1955ApJ...121..161S}
% Salpeter 1955
fit the initial mass function (IMF) of stars to a power law in log$M_*$
with exponent $\gamma_* = 1.35$, or $\alpha_* = 2.35$ (see \S \ref{definitions}
for definitions).
More recent determinations over a wider range of masses, 
including brown dwarfs, indicate a clear flattening at lower masses, 
either as broken power laws \citep{1993MNRAS.262..545K}
%Kroupa (may not be best Kroupa ref, look for later one) 
or a log-normal distribution 
(\citealt{1979ApJS...41..513M, 2003PASP..115..763C}).
% Miller and Scalo 1979, Chabrier 2003 
A detailed study of the IMF in the nearest massive young cluster, Orion,
with stars from 0.1 to 50 \msun\ with a mean age of 2 Myr,
shows a peak between 0.1 and 0.3 \msun, depending on evolutionary
models, and a deficit of brown dwarfs relative to the field IMF
\citep{2012ApJ...748...14D}.
%Da Rio 2012 
Steeper IMFs in massive elliptical galaxies have been suggested by
\citet{2011ApJ...735L..13V}.
% Van Dokkum and Conroy
\citet{2007ARA&A..45..481Z}
%Zinnecker and Yorke 2007
summarize the evidence for a real (not statistical) 
cut-off in the IMF around 150 \msun\ 
for star formation in the \mw\ and Magellanic Clouds. 
Variations in the IMF with environment are plausible, but convincing
evidence for variation remains elusive \citep{2010ARA&A..48..339B}.
%Bastian et al. ARAA 2010

The masses of embedded young clusters \citep{2003ARA&A..41...57L},
% Lada and Lada ARAA
OB associations \citep{1997ApJ...476..144M},
% McKee and Williams 1997
and massive open clusters (\citealt{1997ApJ...480..235E,1999ApJ...527L..81Z}),
% Elmegreen and Efremov 1997, Zhang and Fall 1999
follow a similar relation with $\gamma_{cluster} \sim 1$.
Studies of clusters in nearby galaxies have also found 
$\gamma_{cluster} = 1 \pm 0.1$, with a possible  upper mass cutoff or
turn-over  around $1-2\ee6$ \msun\ \citep{2006A&A...450..129G},
% Gieles et al. 2006
perhaps dependent on the galaxy (but see \citealt{2011ApJ...727...88C}).
% Chandar et al. 2011
In contrast, the most massive known open clusters in the \mw\
appear to have a mass of $(6-8)\ee4$ \msun\ 
(\citealt{2010ARA&A..48..431P,2011MNRAS.411.1386D,2005A&A...430..481H}).
% Portegies Zwart ARAA 2010, Davis et al. 2011, Homeier and Alves 2005

The mass functions of clusters and stars are presumably related to
the mass functions of their progenitors, clouds or clumps, and cores.
For comparison to structures in clouds, we will use the distributions
versus mass, rather than logarithmic mass, so our points of comparison will 
be $\alpha_* = 2.3$ and $\alpha_{cluster} = 2$. 
From surveys of emission from CO \jj10\ in the outer galaxy, where confusion
is less problematic, \citet{2001ApJ...551..852H}
%Heyer et al. (2001)
found a size distribution of clouds, $dN/dr \propto r^{-3.2\pm0.1}$,
 with no sign of a break from the power-law form
from 3 to 60 pc. The mass function, using the definition in equation
\ref{massfunceq2}, was fitted by
$\alpha_{cloud} = 1.8\pm 0.03$ over the range 700 to 1\ee6 \msun.
Complementary surveys of the inner Galaxy found $\alpha_{cloud} = 1.5$ up to
$M_{max} = 10^{6.5}$ \msun\ \citep{2005PASP..117.1403R}.
%Rosolowsky et al. 2005
The upper cut-off around 3-6\ee6 \msun\ appears to be real, despite
issues of blending of clouds
(\citealt{1997ApJ...476..166W, 2005PASP..117.1403R}).
%Williams and McKee 1977, Rosolowsky et al.  2005 
Using \coo\ \jj10, which selects somewhat denser parts of
clouds, \citet{2010ApJ...723..492R}
%Roman-Duvall et al. (2010)
found $\alpha_{\coo} = 1.75\pm 0.23$ for $M_{cloud} > \eten{5}$ \msun.
Clumps within clouds, identified by mapping \coo\ or \cooo, have a similar
value for $\alpha_{clump} = 1.4-1.8$
\citep{1998A&A...329..249K}.
% Kramer, C. 1998 
All results agree that most clouds are small, but, unlike stars or clusters,
most mass is in the largest clouds ($\alpha_{cloud} < 2$).

Studies of molecular cloud properties in other galaxies have
been recently reviewed 
(\citealt{2010ARA&A..48..547F, 2007prpl.conf...81B}).
%Fukui 2010, Blitz et al. in PPV
Mass functions of clouds in nearby galaxies appear to show
some differences, though systematic effects introduce substantial
uncertainty (\citealt{2011ApJS..197...16W, 2008ApJ...675..330S}).
% Wong et al. 2011 on magma, Sheth dubious about differences
Evidence for a higher value of $\alpha_{cloud}$ has been found for the
LMC (\citealt{2001PASJ...53L..41F, 2011ApJS..197...16W})
% Fukui et al. 2001 (get 1.9) and Wong et al. (get > 2.2)
and M33 (\citealt{2003ApJS..149..343E,2005PASP..117.1403R}).
% Engargiola 2003 and Rosolowsky 2005 compares several gals.
Most intriguingly, the mass function appears to be steeper in the
interarm regions of M51 than in the spiral arms (E. Schinnerer
and A. Hughes, personal communication), possibly caused by the
aggregation of clouds into larger structures within arms and 
disaggregation as they leave the arms \citep{2009ApJ...700L.132K}.
% Koda et al. 2009 on M51

Much work has been done recently on the mass function of cores,
and Herschel surveys of nearby clouds
will illuminate this topic (e.g., \citealt{2010A&A...518L.102A}).
% Andre et al. 2010, Herschel 
At this point, it seems that the mass function
of cores is clearly steeper than that of clouds, much closer to that of stars
(\citealt{1998A&A...336..150M, 2008ApJ...684.1240E, 2010ApJ...710.1247S}).
%ref Motte et al. 1998, Enoch et al. 2008, Sadavoy et al. 2010), 
A turnover in the mass function appears at a mass about 3-4 times
that seen in the stellar IMF (\citealt{2007A&A...462L..17A, 2008ApJ...684.1240E}).
%Alves et al. 2007, Enoch et al. 2008 
The similarity of $\alpha_{core}$ to $\alpha_*$ supports the idea that the
cores are the immediate precursors of stars and that the mass function of stars
is set by the mass function of cores. In this picture, the offset of the turnover
implies that about 0.2 to 0.3 of the core mass winds up in the star,
consistent with simulations that include envelope clearing by outflows
(e.g., \citealt{2010ApJ...710..470D}).
Various objections and caveats to this appealing picture have been registered
(e.g., \citealt{2008ApJ...679..552S, 2010ApJ...719..561R, 2007MNRAS.379...57C,
2008A&A...482..855H}).
% Swift and Wiliams, 2008, Reid et al. 2010, Clark et al. 2007, 
%Hatchell et al. 2008 

Note that no such tempting similarity exists between the mass function of
clusters and the mass function of clumps defined by \coo\ maps, questioning
whether that observational definition properly fits the theoretical concept
of a clump as a cluster-forming entity. 
Structures marked by stronger
emission in \coo\ or \cooo\  \citep{2009ApJS..182..131R}
% Rathborne et al. 2009 on cloud and clump catalog
are not always clearly bound gravitationally.
Ground and space-based imaging of submillimeter emission by dust
(e.g., \citealt{2010A&A...518L.103M})
%Menshchikov et al. Aquila and Polaris
offer promise in this area, but information on
velocity structure will also be needed.
The data so far suggest that the structure is primarily
filamentary, more like that of clouds than that of cores.
CS, HCN and other tracers of much denser gas (\S \ref{gastracers}),
along with dust continuum emission (\S \ref{dusttracers})
have identified what might be called ``dense clumps", which do appear
to have a mass function similar to that of clusters 
(\citealt{2003ApJS..149..375S, 2006A&A...447..221B, 2005ApJ...625..891R}), 
% Shirley et al., Beltran et al., M. Reid and Wilson  2005

The comparison of mass functions supports the idea \citep{2007ARA&A..45..565M}
% McKee and Ostriker 2007 ARAA
that cores are
the progenitors of stars and dense clumps are the progenitors of clusters,
with clouds less directly related. Upper limits to the mass function of
clumps would then predict upper limits to the mass function of clusters,
unless nearby clumps result in merging of clusters.
Because most clouds and many clumps  are more massive than the most 
massive stars, it is less obvious that an upper limit to stellar
masses results from a limit on cloud or clump masses.
If clump masses are limited {\it and} if the mass of
the most massive star to form depends causally (not just statistically)
on the mass of the clump or cloud, the integrated galactic IMF (IGIMF)
can be steepened (\citealt{2003ApJ...598.1076K,2006MNRAS.365.1333W}).
%Kroupa and Weidner 2003, W and K 2006
We discuss the second requirement in \S \ref{highmass} and consequences
of this controversial idea in \S \ref{halpha} and \S \ref{localtests}.

% massfunctions

\section{SFR DIAGNOSTICS: THE IMPACT OF MULTI-WAVELENGTH OBSERVATIONS}\label{sfrdiagnostics}

%\input resolvedstars.tex

%sfr_diagnostics.tex  (Section 3)
%updated 06/09/11 RCK
%06/10/11 NJE edit to include YSO counting subsection and fix numbering
% 07/05/11 NJE added subsection labels
% 18/08/11 RCK added X-ray section
% 100911 RCK major revision
% 101411 RCK trimmed references
% 101711 RCK added bibtex citations (ugh!)
% 101911 NJE fixed latex problems
% 102111 NJE changed Orion example to use numbers from Chomiuk
% 102411 NJE inserted fig refs
% 110211 RCK updates
% 110411 NJE minor edits
% 110611 RCK minor edits
% 112411 RCK edits following Ewine's comments
% 121111 RCK ditto
% 121211 NJE minor edits based on comments
% 121611 NJE minor edits
% 121911 NJE minor edits
% 012912 RCK edits after comments
% 020112 NJE edits, some substantial
% 020312 NJE small clean up
% 020612 RCK bigger cleanup, answering NJE queries (removing when agreed)
% 020712 NJE small clean up, remove RCK, residual bf
% 031212 NJE minor edits
% 050912 NJE updated refs
% 051412 NJE minor edits used ion for H

%\section{STAR FORMATION RATE DIAGNOSTICS} 

The influx of new observations over the past decade
has led to major improvements in the calibration and validation
of diagnostic methods for measuring SFRs in galaxies.  Whereas 
measuring uncertainties of factors of two or larger in SFRs were commonplace
ten years ago, new diagnostics based on multi-wavelength data 
are reducing these internal uncertainties by up to an order
of magnitude in many instances.  These techniques have also
reduced the impact of many systematic errors, in particular
uncertainties due to dust attenuation, though others such as
the IMF (\S \ref{massfunctions}) remain important limiting factors.

A detailed discussion of SFR diagnostics and their calibrations
was given in K98, and here we highlight progress made since that
review was published.
In \S \ref{compendium}, we compile updated calibrations 
for the most commonly used indices, based on updated evolutionary
synthesis models and IMF compared to K98.
%a more realistic IMF than the extended 
%[NJE: why extended? The key point for me is that the power
%law was truncated at some arbitrary mass.]
%\citet{1955ApJ...121..161S} function
%Salpeter (1955) 
The new challenges which come
with spatially-resolved measurements of galaxies are discussed
in \S \ref{challenge}.

%resolvedstars.tex
%060611
%081811  RCK  added exgal text
%100911  RCK  general trim and cleanup
%101411  RCK  trimmed some references
%110211  RCK  minor edits
%110411 NJE minor edits
%121211 NJE minor edits
%121611 NJE fixed typo
%011812 NJE edits after comments, clarified lifetime
%031212 NJE minor edit
%050912 NJE updated refs and minor edits

\subsection{Star Counting and CMD Analysis}\label{resolvedstars}

The most direct way to measure star formation rates is to count the
number of identifiable stars of a certain age. Ideally, one would have
reliable masses and ages for each star and the mean star formation rate
would be given by

\begin{equation}
\mean{\sfr} = {\sum_{M_*=M_l}^{M_u} N(M_*, t_*) M_* /t_*}
\label{starcounteq}
\end{equation}
where $N(M_*,t_*)$ is the number of stars in a mass bin characterized by
mass $M_*$ and a lifetime bin (the time since formation)
characterized by $t_*$, and the star formation
rate would be averaged over the longest value of $t_*$ in the sum. In practice,
complete information is not available, and one needs to limit the
allowed lifetimes to find the star formation rate over a certain period.
In particular, for nearby clouds in the \mw,  nearly complete lists of
young stellar objects (YSOs) with infrared excesses are available. 
If one assumes 
that the YSOs sample the IMF in a typical way, one can derive the mean mass
of stars ($\mean{M_*}$), and the equation becomes
\begin{equation}
\mean{\sfr} = {N(YSOs) \mean{M_*} /t_{excess}}.
\label{ysocounteq}
\end{equation}
With currently favored IMFs, $\mean{M_*} = 0.5$ \msun.
The main source of uncertainty is in $t_{excess}$, the duration of an
infrared excess (\S \ref{lowmass}).

For young clusters one can determine mean ages from fitting isochrones
to a color-magnitude diagram, and measure \sfr\ assuming coeval
formation as long as some stars have not yet reached the main sequence.
Some clusters have measurements of stars down to very low masses
(e.g., the Orion Nebula Cluster, \citealt{1997AJ....113.1733H}),
but most need corrections for unseen low-mass stars.
For older clusters one can derive the total mass, but not (directly) the
duration of star formation, which has to be assumed.

In principle similar techniques can be applied on a galaxy-wide
basis.  Beyond the Magellanic Clouds, current instruments cannot
resolve individual YSOs in most regions, and as a result most studies
of SFRs in galaxies are based either on measurements of massive O-type
and/or Wolf-Rayet stars (see K98) 
or on measurements of the entire color-magnitude
diagram (CMD).  
Considerable progress has been made recently in using spatially-resolved
mapping of CMDs to reconstruct spatially
and age-resolved maps of star formation in nearby galaxies.
Many of these use the distribution of blue helium-burning stars
in the CMD (e.g., \citealt{2002AJ....123..813D}), but more recent
% Dohm-Palmer et al. 2002)
analyses fit the entire upper CMDs to synthetic stellar
populations to derive estimates of spatially resolved stellar
age distributions with formal uncertainties (e.g., 
\citealt{2002MNRAS.332...91D, 2008ApJ...689..160W}).  This 
%Dolphin 2002; Weisz et al. 2008)
technique does not have sufficient age resolution to determine
accurate SFRs for ages less than $\sim$10 Myr, but when
applied to high quality datasets such as those which can be
obtained with HST, they can provide sufficient age resolution
to constrain the temporal behaviors and changes in the spatial
distributions of formation over the past 100 Myr or longer 
(e.g., \citealt{2008ApJ...689..160W, 2011ApJ...734L..22W}).
%Weisz et al. 2008; Williams et al. 2011).

\subsection{Ultraviolet Continuum Measurements: The Impact of GALEX}\label{uvcont}

The near-ultraviolet emission of galaxies longward
of the Lyman-continuum break directly traces the photospheric
emission of young stars and hence is one of the most direct tracers
of the recent SFR.  For a conventional 
IMF, the peak contribution to the integrated UV luminosity of a young
star cluster arises from stars with masses of several solar masses.
Consequently this emission traces stars formed over the past 10--200 Myr,
with shorter timescales at the shortest wavelengths 
(see Table \ref{sfrequations}).

For extragalactic studies, this subject has been revolutionised by
the launch of the Galaxy Evolution Explorer (GALEX) mission 
\citep{2005ApJ...619L...1M}.  It imaged approximately two thirds of the sky in
%Martin et al. 2005)
far-ultraviolet (FUV; 155\,nm) and near-ultraviolet (NUV; 230\,nm)
channels to limiting source fluxes $m_{AB} \sim 20.5$, and obtained 
deeper full-orbit or multi-orbit imaging ($m_{AB} \ge 23$) for 
selected galaxies and fields such as those in the Sloan Digital 
Sky Survey (SDSS).  The spatial resolution
(4.5\arcsec\ to 6\arcsec\ FWHM) of the imager makes it an especially powerful
instrument for integrated measurements of distant galaxies and 
resolved mapping of the nearest external galaxies.  
Although most scientific applications of GALEX data to
date have arisen from its imaging surveys, a series of spectroscopic surveys
and pointed observations of varying depths were
carried out as well \citep{2005ApJ...619L...1M}.
%(Martin et al. 2005).

The main impacts of GALEX for this subject are summarized
in \S5.  Broadly speaking its largest impacts were in providing
integrated UV fluxes (and hence SFR estimates) for hundreds of
thousands of galaxies, and in exploiting the dark sky from space
at these wavelengths to reveal star formation at low surface
brightnesses and intensities across a wide range of galactic environments.

Other spaceborne instruments have also provided important datasets
in this wavelength region, including the XMM Optical Monitor 
\citep{2001A&A...365L..36M}
%(Mason et al. 2001) 
and the Swift UV/Optical Telescope \citep{2005SSRv..120...95R}.
%(Roming et al. 2005).
Although these instruments were
primarily designed for follow-up of X-ray and gamma-ray observations,
they also have been used to image nearby and distant galaxies,
with the advantage of higher spatial resolution ($\sim$1$^{\prime\prime}$ FWHM).
Several important studies have also been published over the past
decade from observations made with the Ultraviolet Imaging 
Telescope on the ASTRO missions \citep{1997PASP..109..584S, 2001ApJS..132..129M}.
%(Stecher et al. 1997, Marcum et al. 2001).
The Hubble Space Telescope continues to be a steady source
of observations (mainly in the NUV) for targeted regions in
nearby galaxies.

The primary disadvantage of the ultraviolet is its 
severe sensitivity to interstellar dust
attenuation.  The availability of new data from GALEX and other
facilities has stimulated a fresh look at this problem.  If the
intrinsic color of the emitting stellar population is known
{\it a priori}, the FUV--NUV color or the UV spectral slope (usually
denoted $\beta$) can be used to estimate the dust
attenuation, and numerous calibrations have been published (e.g., 
\citealt{1994ApJ...429..582C, 2004MNRAS.349..769K, 2005ApJ...619L..55S, 
2007ApJS..173..392J, 2007ApJS..173..267S, 2007ApJS..173..256T, 
2011ApJ...741..124H}).
%Calzetti et al. 1994, 
%Kong et al. 2004, Seibert et al. 2005, Johnson et al. 2007,
%Salim et al. 2007, Treyer et al. 2007, Hao et al. 2011).
The accuracy of these prescriptions rests
heavily on the presumed (but uncertain) intrinsic
colors, the shape of the dust extinction curve, and the
complicated effects of geometry and scattering when averaging
over a large physical region (e.g., \citealt{2001ApJ...551..269G}).
%Gordon et al. 2001).

The abundance of high quality far-infrared observations of
nearby galaxies has made it possible to test how well the
UV colors correlate with independent estimates of the dust
attenuation from the IR/UV flux ratios.  Earlier observations 
of bright starburst galaxies suggest a tight relation between
the logarithmic IR/UV ratio (often termed the ``IRX") and
UV color (e.g., \citealt{1999ApJ...521...64M}), and indeed this
%Meurer et al. 1999 
``IRX--$\beta$ relation" provided the primary means for 
calibrating the color versus attenuation relation.  Subsequent
observations of a wider range of galaxies, however, has
revealed a much larger scatter in the relation
(e.g., \citealt{2012A&A...539A.145B}).
% Boquien et al. 2012
When galaxies of all types are considered,
the scatter in actual FUV attenuations for a fixed FUV--NUV
color can easily be two orders of magnitude. Even
when the comparison is restricted to galaxies with 
intrinsically high SFRs (as might be observed at high redshift),
the uncertainties can easily reach an order of magnitude.
As a result, more weight is being applied to
alternative estimates of attenuation corrections and SFRs
based on combinations of UV with IR measurements (\S \ref{composite}).

\subsection{Emission-Line Tracers}\label{halpha}

The remaining SFR indicators to be discussed here
rely on measuring starlight that has been reprocessed
by interstellar gas or dust, or 
on tracers related to the death of massive stars.  
The most widely applied of these are the optical and
near-infrared emission lines from ionized gas surrounding 
massive young stars.  For a conventional 
IMF, these lines trace stars with masses greater than $\sim$15\,\msun,
with the peak contribution from stars in the range 30--40\,\msun.
As such, the lines (and likewise the free-free radio continuum)
represent a nearly instantaneous measure of the SFR,
tracing stars with lifetimes of $\sim$3--10 Myr 
(Table \ref{sfrequations}).  

The application of \halpha\ and other emission line SFR
tracers has expanded dramatically in the last decade, through
very large spectroscopic surveys of samples of local and distant
galaxies, narrow-band emission-line imaging surveys, and 
large imaging surveys of nearby galaxies designed to study
spatial variations in the SFR.  Advances in near-infrared 
instrumentation have also led to the first systematic surveys
in the Paschen and Brackett series lines, as well as for
\halpha\ observed at high redshifts.  Results from several
of these surveys are discussed and referenced in \S5.  

The \halpha\ emission line remains the indicator of choice
for observations of local and distant galaxies alike, but for
moderate redshifts, the bluer visible lines have been applied,
in particular the [\ion{O}{2}] forbidden line doublet at
372.7\,nm.  This feature is
subject to severe systematic uncertainties from dust attenuation
and excitation variations in galaxies.  K98 published a single
SFR calibration for the line, but subsequent analyses 
(e.g., \citealt{2001ApJ...551..825J, 2004AJ....127.2002K, 2006ApJ...642..775M})
%Jansen et al. 2001, Kewley et al. 2004, Moustakas et al. 2006)
have shown that the systematic effects must be removed or
at least calibrated to first order for reliable measurements.
Even then, the uncertainties in the [\ion{O}{2}]-based SFRs are much
larger than for \halpha.

Over the past decade, increasing attention has been given
to measurements of the redshifted Ly$\alpha$
line ($\lambda_{rest}$ = 121.6\,nm) as a tracer of star-forming
galaxies, especially at high redshifts where it provides unique
sensitivity to both low-mass star-forming galaxies and intergalactic
gas clouds or ``blobs" 
(e.g., \citealt{2009ApJ...696.1164O, 2010ApJ...723..869O}).
% Ouchi et al. 2009, 2010
The strength of the line (8.7 times stronger than \halpha\ for Case B
recombination) makes it an attractive tracer in principle, but 
in realistic ISM environments the line is subject to strong
quenching from the combination of resonant trapping and eventual
absorption by dust,
usually quantified in terms of a  Ly$\alpha$ escape fraction. 
As a result Ly$\alpha$ surveys to date have been mainly used
for identifying large samples of distant star-forming galaxies.
Applying the line as a quantitative SFR tracer requires an accurate
measurement of the escape fraction.  Several ongoing studies are
quantifying this parameter by comparing  Ly$\alpha$ fluxes of
galaxies with independent SFR tracers such as \halpha\ or the 
UV continuum (e.g., \citealt{2009A&A...506L...1A, 2009ApJ...704L..98S,  
2011ApJ...736...31B, 2011ApJ...730....8H}).
% Atek et al. 2009, Scarlota et al. 2009, Blanc et al. 2011, Hayes et al. 2011) 
These show that the escape fraction varies wildly between galaxies
with a range of more than two orders of magnitude (order 0.01 to 1),
but also increases systematically with redshift.  It is possible
that Ly$\alpha$ will prove to be a powerful SFR tracer for the
highest-redshift objects, but in view of the large scatter and
systematic uncertainties associated with its use we have chosen
not to include a SFR calibration of the line in this review.

Other workers have investigated the efficacy of the infrared fine-structure
cooling lines, which arise in \hii\ regions or PDRs,
as quantitative SFR tracers.  
\citet{2007ApJ...658..314H}
%Ho \& Keto (2007)
compiled data from ISO and Spitzer on the [\ion{Ne}{2}]12.8\,$\mu$m
and [\ion{Ne}{3}]15.6\,$\mu$m lines, and found that the sum of 
the line fluxes correlates well with hydrogen recombination line
fluxes, with a scatter of $\sim$0.3 dex.  In a similar vein,
\citet{2002A&A...385..454B} and \citet{2006A&A...453...77R} 
%Boselli et al. (2002) and Rodriguez-Fernandez et al. (2006)
have investigated the applicability of the [\ion{C}{2}]158\,$\mu$m line
as a SFR measure.  They found good general correspondance between
the [\ion{C}{2}] luminosity and other measures of the SFR such as 
ionized gas and dust emission, but the scatter in the relationships
is at least a factor of ten.  Even larger variations in 
$L_{{\rm [CII]}}/L_{{\rm IR}}$
were found in a more diverse sample of star-forming galaxies by
\citet{1997ApJ...491L..27M}.
%Malhotra et al. (1997).  
The availability of a rich new set of [\cii] observations from Herschel,
combined with the detection of redshifted 
[\cii] emission in submillimeter galaxies (SMGs) from ground-based
instruments, has sparked a resurgence of interest in this application.

The largest systematic errors affecting \halpha-based SFRs
are dust attenuation and sensitivity to the population of
the upper IMF in regions with low absolute SFRs.  
For regions with modest attenuations the ratios of 
Balmer recombination lines (Balmer decrement) can be used
to correct for dust, and this method has been applied in a
number of large spectrophotometric surveys of nearby galaxies
(e.g., \citealt{2002AJ....124.3135K, 2004MNRAS.351.1151B, 2006ApJ...642..775M}).
% Kewley et al. 2002, Brinchmann et al. 2004, Moustakas \& Kennicutt 2006
The Balmer decrements offer only approximate corrections for 
attenuation because of variations on scales smaller than the resolution.
These variations may lead to an underestimate of the extinction because
lines of sight with low extinction are more heavily weighted within
the beam.  
This problem can be addressed partly by
adopting a reddening law which compensates in part for these 
departures from a pure foreground scheme geometry 
(e.g., \citealt{2000ApJ...539..718C}).
% Charlot \& Fall 2000
The attenuation of the emission lines is found to be systematically
higher than that of the continuum starlight at the same wavelengths
(e.g., \citealt{1994ApJ...429..582C}),
% Calzetti et al. 1994
which presumably reflects the higher concentrations of dust
in the young star-forming regions.

As is the case with UV-based SFRs, the availability of far-infrared
maps and luminosities for nearby galaxies has also made it possible
to calibrate {\it multi-wavelength methods} for applying dust attenuation
corrections to these measurements (\S3.7).  They reveal that the
Balmer decrement provides
reasonably accurate attenuation corrections in normal galaxies,
where attenuations are modest (typically 0--1 mag at \halpha),
and care is taken to correct the emission-line fluxes
for underlying stellar absorption.
The Balmer decrement method for estimating dust attenuation breaks 
down badly, however in circumnuclear starbursts or other
dusty galaxies  (e.g., \citealt{2006ApJ...642..775M}).
% Moustakas et al. 2006
With the advent of large-format integral-field
spectrographs there is promise of applying Balmer decrement
measurements on a spatially-resolved basis in galaxies
(e.g., \citealt{2010fsgc.confE.153B, 2012A&A...538A...8S}). 
% Blanc et al. 2010, Sanchez et al. 2012

The accuracy of SFRs derived from emission lines will also
degrade in regions where the SFR is so low that one enters
the regime of small number statistics in the population of
massive ionizing stars.  If the IMF itself were completely
blind to the SFR, we would expect such effects to 
become apparent below \halpha\ luminosities of order 
$10^{38}$\,ergs\,s$^{-1}$, or SFRs of order 0.001 \msun\,yr$^{-1}$
(e.g., \citealt{2003A&A...407..177C}). 
%Cervino et al. 2003  
At the very least, this effect
will produce a much larger scatter in ionizing flux per
unit SFR in this regime.  In low-SFR regions, this sampling
noise can be exacerbated by the short lifetimes of the
ionizing stars, producing large temporal fluctuations in
\halpha\ emission even for a fixed longer-term SFR.
As a result, other tracers (e.g., FUV emission) tend to provide 
more accurate and sensitive measurements of the SFR at
low star formation levels.  

%RCK revised paragraph below to clean up and link to IMF section
%RCK also updated the Weisz et al. reference
Can these effects cause systematic
errors in the SFRs?  \citet{2007ApJ...671.1550P, 2009MNRAS.395..394P} 
%Pflamm-Altenburg et al. (2008, 2009)
have investigated the effect of an IGIMF (\S \ref{massfunctions}) on
SFR tracers, and shown that the systematic depletion of massive
stars in low-SFR environments could cause \halpha\ to substantially
underestimate the actual SFR.  Interestingly a systematic deficit of \halpha\
emission in dwarf galaxies with low SFRs and in low 
SFR density regions is observed 
(\citealt{2000MNRAS.312..442S, 2001ApJ...548..681B, 2009ApJ...706..599L, 
2009ApJ...695..765M}).
%(Sullivan et al. 2000, Bell \& Kennicutt 2001, Lee et al. 2009, Meurer et al. 2009
Recent work suggests however that 
the systematic dependence of the \halpha/UV ratio may be 
produced instead by temporal variations in SFRs, without having
to resort to modifying the IMF itself 
\citep{2011ApJ...741L..26F, 2012ApJ...744...44W}.
%Fumagalli et al. 2011), Weisz et al. 2012).

\subsection{Infrared Emission: The Impact of Spitzer and Herschel}\label{infrared}

Interstellar dust absorbs approximately half the starlight
in the universe and re-emits it in the infrared, so measurements
in the IR are essential for deriving a complete inventory of 
star formation.  This section focusses on the
transformational results which have come from the Spitzer
Space Telescope \citep{2004ApJS..154....1W} 
%(Werner et al. 2004) 
and the Herschel Space Observatory \citep{2010A&A...518L...1P}.
%(Pilbratt et al. 2010).  
A review in this journal of extragalactic science from
Spitzer can be found in \citet{2008ARA&A..46..201S}.
%Soifer, Helou, \& Werner (2008).

Three other space missions are beginning to influence this subject:
the {\it AKARI} mission
\citep{2007PASJ...59S.369M},
%Murakami et al. 2007
the {\it Wide-field Infrared Survey Explorer} (WISE) mission
\citep{2010AJ....140.1868W},
%Wright et al 2010
and the {\it Planck} mission
\citep{2011A&A...536A...1P}.
% Planck collab 2011c)
These observatories conducted all-sky surveys, with {\it AKARI} imaging
in the $2.4 - 160$ \micron\ region, WISE in the range of $3.4 - 22$
\micron, and {\it Planck} at $350 - 850$ \micron\ (with several bands
extending to longer wavelengths). Much of the science from these missions
is just beginning to emerge, but they will provide very important results
in this subject in the coming decade.

While early applications of dust-based SFR measurements effectively
(and necessarily) assumed a one-component dust model, 
subsequent observations show 
that the dust emission is comprised of
distinct components, each with different spatial distributions and couplings
to the young stars.  At wavelengths of $\sim$5--20\,$\mu$m,
the emission is dominated by molecular bands arising from 
polycyclic aromatic hydrocarbons (PAHs).  
Longward of $\lambda \sim 20$ \micron, the emission is
dominated by thermal continuum emission from the main dust grain population.
Emission from small grains transiently heated by intense radiation fields 
(usually in or near star-forming regions) is important out to about
60 \micron, whereas at longer wavelengths, emission from
larger grains with steady state temperatures dominates
\citep{2003ARA&A..41..241D}.
%(Draine 2003).   

\begin{figure}    
\center
\includegraphics[scale=0.6]{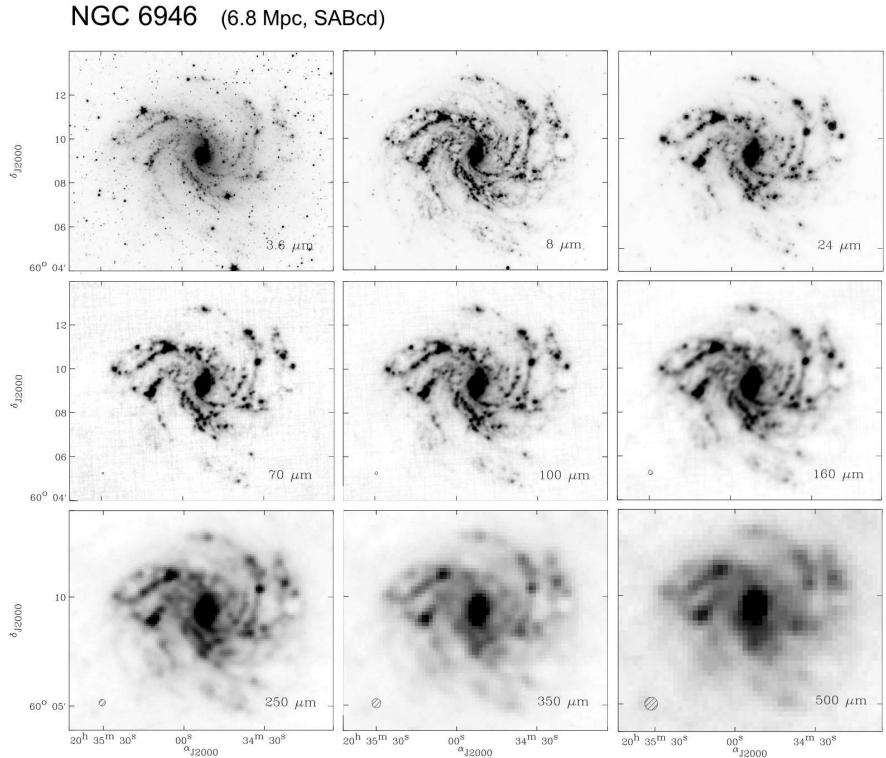}
\caption{
%\footnotesize
A montage of infrared images of NGC\,6946 from
{\it Spitzer} (SINGS) and {\it Herschel} (KINGFISH).  
{\it Top panels}:  {\it Spitzer} IRAC
images at 3.6\,\micron\ and 8.0\,\micron, and MIPS
image at 24\,\micron.  The emission at these wavelengths
is dominated by stars, small PAH dust grains, and
small dust grains heated by intense radiation fields, respectively.
{\it Middle panels}:  {\it Herschel} PACS images at 70\,\micron,
100\,\micron, and 160\,\micron, processed with the Scanamorphos
map making package.  Note the excellent
spatial resolution despite the longer wavelengths, and
the progressive increase in contributions from diffuse
dust emission (``cirrus") with increasing wavelength.
{\it Bottom panels}:  {\it Herschel} SPIRE images at 250\,\micron,
350\,\micron, and 500\,\micron. These bands trace increasingly
cooler components of the main thermal dust emission, with
possible additional contributions from ``submillimeter
excess" emission at the longest SPIRE wavelengths. FWHM beam sizes
for the respective {\it Herschel} bands are shown in the lower
left corner of each panel.  
%Figure taken from \citet{2011PASP..123.1347K}.
This figure
originally appeared in the Publications of the Astronomical Society of the
Pacific \citep{2011PASP..123.1347K}. Copyright 2011,
Astronomical Society of the Pacific; reproduced with permission of the
Editors.
%Kennicutt et al. 2011
}

\label{ngc6946}
\end{figure}

The distribution of these different emission components is illustrated in
Figure \ref{ngc6946}, which shows Spitzer and Herschel images of the nearby 
star-forming galaxy, NGC\,6946, at wavelengths ranging from 3.6--500\,$\mu$m.
At 24\,$\mu$m, the emission peaks around the youngest star-forming
regions and \hii\ regions, with a more diffuse component extending between
these regions.  As one progresses to longer wavelengths, the prominence
of the diffuse component increases.  Recent measurements with 
Herschel show that this is mainly a physical change, and not an
artifact of lower spatial resolution at longer wavelengths 
\citep{2011AJ....142..111B}.
% Boquien et al. 2011
This  diffuse emission is analogous to the
``infrared cirrus" emission observed in our own Galaxy.  Interestingly 
PAH emission appears to correlate the most strongly with the 
longer-wavelength component of the thermal dust emission 
(e.g., \citealt{2008MNRAS.389..629B}), 
%(e.g., Bendo et al. 2008), 
though it often also appears as resolved
shells around the young star-forming regions 
\citep{2004ApJS..154..253H}.
%(e.g., Helou et al. 2004).

These variations in the morphologies in the different dust
emission components translate into considerable variations in
the dust SEDs within and between galaxies (e.g., 
\citealt{2002ApJ...576..159D, 2005ApJ...633..857D, 2007ApJ...656..770S}),
%Dale \& Helou 2002, Dale et al. 2005, Smith et al. 2007) 
and as a consequence
the conversion of infrared luminosities into SFRs must change
for different IR wavelengths.  Most early applications of the 
dust emission as a SFR tracer were based on the integrated total-infrared
(TIR) emission.  
%For this paper we adopt the integral over the wavelength
%range 3--1100\,$\mu$m \citep{2002ApJ...576..159D}.
%(Dale \& Helou 2002).  
This parameter has the
physical advantage of effectively representing the bolometric
luminosity of a completely dust-enshrouded stellar population.
The TIR-based SFR calibration derived in K98, applicable in the
limits of complete dust obscuration and dust heating fully dominated
by young stars, is still in widespread use today.
For most galaxies however this complete wavelength coverage
will not be available, so many workers have calibrated monochromatic
SFR indices, usually tuned to one of the Spitzer or Herschel bands,
including 24\,$\mu$m (e.g., \citealt{2005ApJ...632L..79W, 2006ApJ...650..835A, 
2007ApJ...666..870C, 2007ApJ...667L.141R, 2009ApJ...692..556R},
%Wu et al. 2005, Alonso-Herrero et al. 2006,
%Calzetti et al. 2007, Relano et al. 2007, Rieke et al. 2009, 
and references therein), and 70 and 160\,$\mu$m 
\citep{2010ApJ...719L.158C}.
%(Calzetti et al. 2010).
The latter paper contains an excellent discussion comparing the
various calibrations in the literature at the time.

As with all of the SFR indicators, the dust emission is subject
to important systematic effects.
% which need to be considered when it is applied.  
Just as the UV and visible tracers miss radiation
that has been attenuated by dust, the infrared emission misses
the starlight that 
is {\it not} absorbed by dust (e.g., \citealt{2001A&A...366...83H}).
%(e.g., Hirashita et al. 2001).  
As discussed earlier, dust attenuates only about half of the
integrated starlight of galaxies on average, so the infrared emission
will systematically underestimate the SFR if the missing fraction of
star formation is not incorporated into the calibrations.  This 
``missing" unattenuated component varies from essentially zero
in dusty starburst galaxies to nearly 100\%\ in dust-poor dwarf
galaxies and metal-poor regions of more massive galaxies.
Another major systematic error
works in the opposite direction; in most galaxies, evolved stars
(e.g., ages above $\sim$100--200 Myr) contribute significantly to 
the dust heating, which tends to cause the IR luminosity to
overestimate the SFR.   
The fraction of dust heating from young stars varies by a large
factor among galaxies; in extreme circumnuclear starburst galaxies or individual
star-forming regions, nearly all of the dust heating arises from
young stars, whereas in evolved galaxies with low
specific SFRs, the fraction can be as low as $\sim$10\%\ (e.g.,
\citealt{1992ApJ...396L..69S, 1996ApJ...460..696W, 2008MNRAS.386.1157C}).
%Savuage \& Thuan 1992, Walterbos \& Greenawalt 1996, Cortese et al. 2008).  
In practical terms this means that
the conversion factor from dust luminosity to SFR--- even in the
limit of complete dust obscuration--- is not fixed, but rather
changes as a function of the stellar population mix in galaxies.
The difference in conversion factor
between starbursts and quiescient galaxies with constant SFR,
for example, is about a factor of 1.3--2 
(e.g., \citealt{1992ApJ...396L..69S, 2009ApJ...703.1672K, 2011ApJ...741..124H}).
%Sauvage \& Thuan 1992, Kennicutt et al. 2009, Hao et al. 2011). 

The calibration of the mid-IR PAH emission as a quantitative
SFR tracer deserves special mention.  This index is of particular
interest for studies of galaxies at high redshift, because 
the observed-frame 24\,$\mu$m fluxes of galaxies at $z = 1-3$ 
tend to be dominated by redshifted PAH emission.  A number of
studies have shown that the PAH luminosity scales relatively
well with the SFR in metal-rich luminous star-forming galaxies
(e.g., \citealt{2001A&A...372..427R, 2004ApJ...613..986P,
2004A&A...419..501F, 2005ApJ...632L..79W,
2007ApJ...667..149F, 2007ApJ...666..870C}),  
%(e.g., Roussel et al. 2001, Peeters et al. 2004, Forster-Schreiber et al. 2004,
%Wu et al. 2005, Farrah et al. 2007, Calzetti et al. 2007),
but the PAH bands weaken dramatically below metal abundances of
approximately 1/4 to 1/3 $Z_\odot$ (e.g., \citealt{2000NewAR..44..249M, 
2005ApJ...628L..29E, 2007ApJ...666..870C, 2007ApJ...656..770S}),
%Madden 2000, Engelbracht et al. 2005, Calzetti et al. 2007, 
% Smith et al. 2007), 
rendering them problematic as quantitative SFR tracers in this regime.

The best way to overcome these systematic biases is to combine
the IR measurements with UV or visible-wavelength SFR tracers,
to measure the unattenuated starlight directly and to constrain
the dust-heating stellar population (\S3.7).  However, in
cases where only IR observations are available one can attempt
to incorporate corrections for these effects into the SFR calibrations
themselves, and this approach has been taken by most
authors (e.g., \citealt{2010ApJ...714.1256C} 
%Calzetti et al. 2010 
and references therein).
Fortunately, for most galaxies with moderate to high specific
SFRs, the effects of partial dust attenuation and cirrus dust
heating by evolved stars appear to roughly compensate for each
other.  For example,  \citet{2002AJ....124.3135K} 
%Kewley et al. (2002) 
showed that for a sample
of spiral galaxies with integrated emission-line spectra, the 
TIR-based calibration of K98 for dusty starburst galaxies was
in good agreement with SFRs based on attenuation-corrected 
\halpha\ line fluxes.  However, one must avoid applying
these IR-based recipes in environments where they are bound to fail,
for example, in low-metallicity and other largely dust-free
galaxies or in galaxies with low specific SFRs and a strong
radiation field from more evolved stars.

\subsection{Radio Continuum Emission}\label{radio}

The centimeter-wavelength radio continuum emission of galaxies
consists of a relatively flat-spectrum, free-free component,
which scales with the ionizing luminosity (subject to a weak
electron temperature dependence) and a steeper spectrum synchrotron
component, which overwhelmingly dominates the integrated radio
emission at  $\nu \leq 5$ GHz.  The
free-free component can be separated with multi-frequency radio
measurements or high-frequency data (e.g., 
\citealt{1983AJ.....88.1736I, 1997A&A...322...19N, 2011ApJ...737...67M}), 
%Israel \& van der Hulst 1983, Niklas et al. 1997, Murphy et al. 2011
to provide a photoionization-based measure of the SFR, without
the complications of dust attenuation which are encountered
with the Balmer lines.  

At lower frequencies the integrated emission is dominated
by the synchrotron emission from
charged particles produced by supernovae. 
%but there is no {\it a priori} theory for deriving a 
A SFR calibration has not been derived from first principles,
but observations have repeatedly
confirmed a tight correlation between this non-thermal emission
and the far-infrared emission of galaxies, which favors its
application as a SFR tracer (e.g., 
\citealt{1985ApJ...298L...7H, 1992ARA&A..30..575C}).
%Helou et al. 1985, Condon 1992).  
Moreover, improvements to receiver
technology with the Expanded Very Large Array (EVLA) and other
instruments have made the radio continuum a primary means of
identifying star-forming galaxies at high redshift and estimating
their SFRs.  As a result it is appropriate to include it in this
discussion of SFR tracers.  

Current calibrations of the radio continuum versus SFR relation
are bootstrapped from the far-infrared calibrations, using
the tight radio--IR correlation.  
%Due to the 
The steep synchrotron spectrum makes this calibration strongly wavelength
dependent, and most are referenced to 1.4\,GHz (e.g., 
\citealt{2001ApJ...554..803Y, 1992ARA&A..30..575C, 2003ApJ...586..794B}).
%Condon 1992, Yun et al. 2001, Bell et al. 2003).
The calibration adopted here in \S3.8 is derived in similar
fashion, but adapted to the Kroupa IMFs.

As described above, the IR--based SFR calibrations break
down severely in faint galaxies with low dust contents,
yet the radio--IR correlation remains tight and nearly
linear over the entire luminosity range.  How can this 
arise?  The likely explanation can be found in \citet{2003ApJ...586..794B}
%Bell (2003,
and references therein, where it is shown that the decrease
in dust opacity in low-mass galaxies is accompanied by a
decline in synchrotron emission relative to other tracers of
the SFR.  This is seen most directly as a decline in the ratio
of non-thermal radio emission (still dominant in the 1--5 GHz region)
to the free-free thermal radio emission.
Since the thermal radio emission is directly coupled to the
stellar ionization rate and SFR, the relative decline in 
synchrotron luminosity must reflect a physical decline per
unit SFR.  If correct then the continuity of the radio--IR
relation to low luminosities is the result of a ``cosmic
conspiracy" \citep{2003ApJ...586..794B}, 
%(Bell 2003), 
and one should beware of applying
the method in galaxies fainter than $\sim$0.1\,L$^*$.

\subsection{X-ray Emission}\label{othersfr}

Over the past decade, the integrated hard X-ray emission
of galaxies has been increasingly applied as a SFR tracer.
The component of X-ray emission that does not arise from
AGN accretion disks is dominated by massive X-ray binaries,
supernovae and supernova remnants, and massive stars,
all associated with young stellar populations and recent
star formation.  Furthermore the observed 2--10 keV fluxes
of galaxies are observed to be strongly correlated with
their infrared and non-thermal radio continuum fluxes
(e.g., \citealt{2002AJ....124.2351B, 2003A&A...399...39R, 2011MNRAS.417.2239S}),
%(e.g., Bauer et al. 2002; Ranalli et al. 2003; 
%Symeonidis et al. 2011), 
strengthening the link to the SFR.

Since there is no way to calibrate the relation between
X-ray luminosity and SFR from first principles, the calibration
is usually bootstrapped from the infrared or radio.  
%Ranalli et al. (2003) 
\citet{2003A&A...399...39R} derived such a calibration for
integrated 2--10 keV X-ray luminosities, referenced
to the K98 calibrations and IMF, and this relation is still
widely applied today.  \citet{2004A&A...419..849P} 
%Persic et al. (2004) 
derived an alternate calibration in terms of the hard X-ray binary
luminosity alone, which is useful for nearby resolved 
galaxies.  Calibrations of the SFR in terms of X-ray
luminosity and stellar mass have been published by
\citet{2004ApJ...602..231C} and \citet{2010ApJ...724..559L}.
%Colbert et al. (2004) and Lehmer et al. (2010).  
%We refer
%the reader to these papers for a detailed explanation
%of the calibrations.  
For simplicity, we have listed in
Table \ref{sfrequations} 
the widely-applied relation of \citet{2003A&A...399...39R}, 
%Ranalli et al. (2003),
but adjusted to the Kroupa IMF used for the other calibrations.

\subsection{Composite Multi-Wavelength Tracers}\label{composite}

Large multi-wavelength surveys of galaxies
allow tests and calibrations of SFR indices that 
combine information from more than one tracer and exploit
the complementary strengths of different wavelengths.
Currently this capability is mainly limited to nearby
galaxies, but with the expansion of far-infrared to millimeter
surveys of high-redshift galaxies, opportunities to apply
multi-wavelength diagnostics to distant galaxies should
expand in the coming decade (e.g., \citealt{2011ApJ...726L...7O, 
2012ApJ...744..154R}).
%Overzier et al. 2011, Reddy et al. 2012

The most widely explored of these methods have combined
UV (usually FUV) observations with infrared measurements
to construct dust-corrected SFRs, using an approximate 
energy-balancing approach.  In its simplest form one can
use a linear combination of UV and IR luminosities to correct the UV fluxes for 
dust attenuation:

\begin{equation}
L_{\rm UV}({\rm corr}) = L_{\rm UV}({\rm observed}) + \eta\,L_{\rm IR} 
\end{equation}

\noindent
where the luminosities are usually calculated from flux densities using
the definition $L = \nu\,L_\nu$, and the
coefficient $\eta$ is dependent on the bandpasses
chosen for the UV and IR measurements.  The most common
form of this correction uses GALEX FUV (155\,nm) and 
total-infrared luminosities, hence:

\begin{equation}
L_{\rm FUV}({\rm corr}) = L_{\rm FUV}({\rm observed}) + \eta\,L_{\rm TIR}
\end{equation}

\noindent
Other prescriptions
sometimes adopt a higher order polynomial dependence on 
$L_{\rm{IR}}$ (e.g., \citealt{2005ApJ...619L..51B}), 
%Buat et al. 2005), 
but for brevity we only discuss the linear combinations here.
In most cases, $\eta < 1$,  
%because the $L = \nu\,L_\nu$
because only part of the dust-heating radiation is contained in the FUV band,
and in many galaxies there is significant dust heating and TIR emission
arising from stars other than the UV-emitting population (IR cirrus).
The coefficient $\eta$ can be calibrated theoretically using evolutionary
synthesis models, or empirically, using independent measurements
of dust-corrected SFRs 
(e.g., \citealt{2010ApJ...719.1191T, 2011ApJ...741..124H}).
%Treyer et al. 2010, Hao et al. 2011

Early applications of
this method were largely restricted to luminous starburst
galaxies and star-forming regions, for which UV and IR
data could be obtained prior to the advent of GALEX,
Spitzer, and Herschel (e.g., \citealt{1999A&A...352..371B,
1999ApJ...521...64M, 2000ApJ...533..236G}),
%Buat et al. 1999, Meurer et al. 1999, Gordon et al. 2000
yielding values of $\eta \sim 0.6$.
Subsequent analyses extending to normal star-forming galaxies
(e.g., \citealt{2003A&A...410...83H, 2004MNRAS.349..769K, 2005MNRAS.360.1413B,
2005ApJ...619L..51B, 2010ApJ...719.1191T, 2011ApJ...741..124H}) 
%(e.g., Hirashita et al. 2003, 
%Kong et al. 2004, Burgarella et al. 2005, Buat et al. 2005,
%2011, Treyer et al. 2010, Hao et al. 2011) 
typically produce values of $\eta$ which are somewhat
lower (e.g., 0.46 for \citealt{2011ApJ...741..124H}), 
%Hao et al. 2011), 
which almost 
certainly reflects the larger contribution to dust
heating from starlight longward of the FUV.  Considering
the wide differences in dust heating populations between
starburst and normal galaxies, however, this difference
of $\sim$30\%\ is hardly crippling, especially when it
reduces dramatically the much larger systematic errors
from UV or IR-based SFRs alone.

\begin{figure}
\center
\includegraphics[scale=0.8]{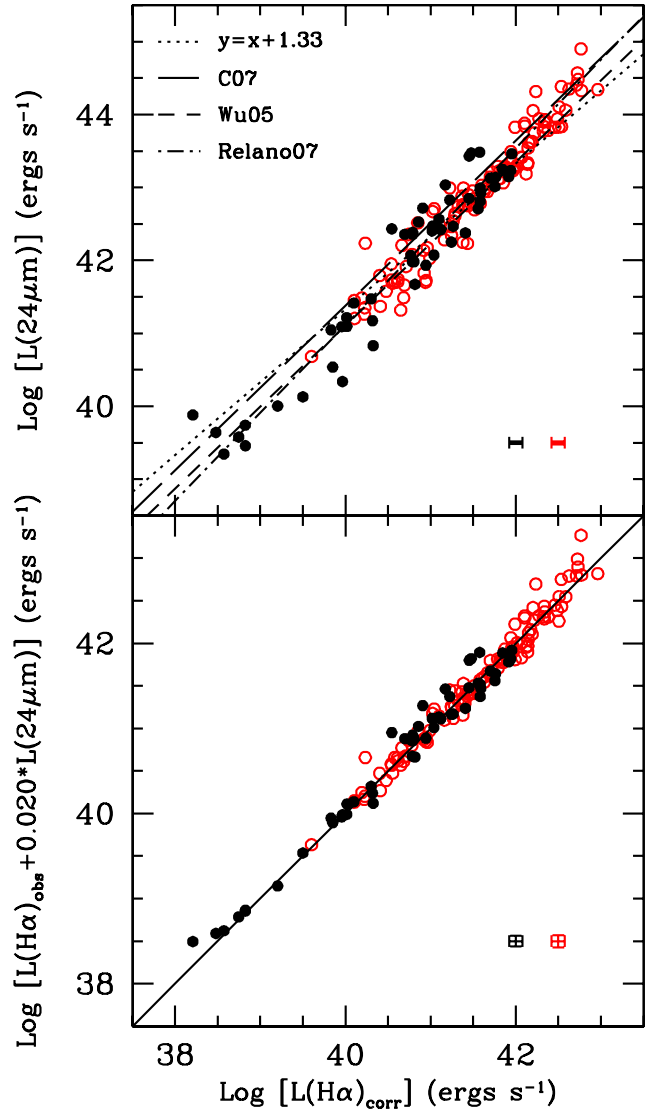}
\caption{
Top: relation between observed 24 \micron\ IR luminosity and dust-corrected
\halpha\ luminosity for nearby galaxies.  The dust corrections were
derived from the absorption-corrected H$\alpha$/H$\beta$ ratios in optical 
spectra. The dotted line shows a linear relation for comparison, while 
the other lines show published fits to other samples of galaxies.  
Bottom:  linear combination of (uncorrected)
\halpha\ and 24 \micron\ luminosities compared to the same Balmer-corrected
\halpha\ luminosities.  Note the tightness and linearity of the relation 
over nearly the entire luminosity range. Taken from \citet{2009ApJ...703.1672K};
reproduced by permission of the AAS.
}

\label{sfr_dustcorr}
\end{figure}

A similar energy-balancing approach has been applied to
derive dust-corrected emission-line luminosities of
galaxies \citep{2007ApJ...666..870C, 2008ApJ...686..155Z,
2009ApJ...703.1672K, 2010ApJ...719.1191T}.
%(Calzetti et al. 2007, 
%Zhu et al. 2008, Kennicutt et al. 2009, Treyer et al. 2010).
Here the equivalent parameter to $\eta$ above is calibrated
using independent estimates of the \halpha\ attenuation,
usually the recombination line decrement as measured from
the Pa$\alpha$/\halpha\ or \halpha/H$\beta$ ratio.
\citet{2009ApJ...703.1672K} 
%Kennicutt et al. (2009) 
have shown that this approach
can be more broadly applied to estimate attenuation corrections
for other emission lines (e.g., [\ion{O}{2}]), and it can be
calibrated using TIR fluxes, single-band IR fluxes, or
even 1.4\,GHz radio continuum fluxes.  An illustration 
of the effectiveness of the method is shown in Figure 
\ref{sfr_dustcorr},
taken from \citet{2009ApJ...703.1672K}.
%Kennicutt et al. (2009).  
The top panel shows
the consistency between an IR-based tracer (24\,$\mu$m luminosity)
used alone and the dust-corrected \halpha\ luminosity (via the
integrated Balmer decrement).  There is a general correlation
but considerable scatter and a pronounced non-linearity,
which reflects a general correlation of average dust attenuation
with the SFR itself.  The lower panel
plots the best fitting linear combination
of \halpha\ and IR (in this case 24\,$\mu$m) luminosities against
the Balmer-decrement-corrected \halpha\ luminosities.  The scatter
between the tracers now is reduced severalfold, and the non-linearity
in the comparison with IR luminosities alone is essentially removed.

These composite SFR indicators are not without systematic
uncertainties of their own.  Questions remain about the systematic
reliability of the independent attenuation calibrations (in
particular with the effects of dust geometry on the Balmer decrements),
and most of the recipes still retain a dependence on stellar
population age, via the infrared term; for example the best
fitting values of $\eta$ for \hii\ regions and young starbursts
differ from those of normal star forming galaxies by a factor
of $\sim$1.5, consistent with expectations from evolutionary 
synthesis models, as discussed in \S3.8 \citep{2009ApJ...703.1672K}.
%(Kennicutt et al. 2009).  
Nevertheless they
represent a major improvement over single-wavelength tracers, and
provide a valuable testing ground for calibrating and
exploring the uncertainties in the monochromatic indicators
(e.g., \citealt{2010ApJ...714.1256C}).
%Calzetti et al. 2010).

Table \ref{multitab}
lists examples of dust attenuation corrections 
using combinations of FUV and \halpha\ fluxes with various
infrared and radio tracers, taken from \citet{2011ApJ...741..124H}
and \citet{2009ApJ...703.1672K}, respectively.  These can
be applied in combination with the monochromatic SFR zeropoint calibrations
listed in Table \ref{sfrequations} (\S \ref{compendium})
 to derive dust-corrected SFR measurements.

\subsection{An Updated Compendium of Integrated SFR Calibrations}\label{compendium}

K98 presented calibrations for SFRs derived from UV continuum, 
TIR, \halpha\ emission-line, and [\ion{O}{2}] emission-line luminosities,
which have come into common usage in the field.  The considerable
expansion of the subject to other wavelengths and SFR diagnostics
since that time motivates a revisit of these calibrations.

Nearly all of these calibrations are based on evolutionary
synthesis models, in which the emergent SEDs are derived for synthetic
stellar populations with a prescribed age mix, chemical composition,
and IMF.  The K98 calibrations employed a mix of models from the
literature, and assumed a single power-law IMF \citep{1955ApJ...121..161S},
%Salpeter 1955
with mass limits of 0.1 and 100\,M$_\odot$.
This IMF gave satisfactory SFR calibrations relative to that
of more realistic IMFs for \halpha, but for other wavelengths, the
relative calibrations using different tracers are sensitive to
the precise form of the IMF.  Today most workers calibrate SFR
tracers using a modern IMF with a turnover below $\sim$1\,M$_\odot$,
for example the IMF of \citet{2003ApJ...598.1076K},
%Kroupa and Weidner 2003 
with a Salpeter slope ($\alpha_* = -2.35$) from 1--100\,M$_\odot$ 
and $\alpha_* = -1.3$ for 0.1--1\,M$_\odot$.  The calibrations
presented here use this IMF, but the IMF fit from \citet{2003PASP..115..763C}
% Chabrier 2003
yields nearly identical results (e.g., \citealt{2011AJ....142..197C}).
%Chomiuk and Povich
The past decade has also seen
major improvements in the stellar evolution and atmospheric
models which are used to generate the synthetic SEDs.  The  
results cited here use the Starburst99 models of \citet{1999ApJS..123....3L},
%Leitherer et al. 1999
which are regularly updated in the on-line version of the package.

Table \ref{sfrequations} presents in compact form calibrations for a suite
of SFR tracers in the form:

\begin{equation}
%$\log$\,SFR (M$_\odot$\,yr$^{-1}$) = $\log L_x - C_x$
\log \sfr (\msunyr) = \log L_x - \log C_x
\end{equation}

\noindent
The table lists for each tracer the units of luminosity ($L_x$),
the logarithmic SFR calibration constant $C_x$, and the
primary reference(s) for the calibration.  For methods presented
in K98, we also list the scaling constant between the new
(Kroupa IMF, SB99 model) SFRs and those from K98.  
As in K98,
the recipes for infrared dust emission are in the limit
of complete dust attenuation and continuous star formation
over a period of 0--100 Myr, which is appropriate for a
typical dust-obscured starburst galaxy.  {\it Virtually
all of the SFR calibrations presented in Table \ref{sfrequations} are taken
from the literature, and we strongly encourage users of these 
calibrations to cite the primary sources, whether or not they 
choose to cite this paper as well.}

The second column of Table \ref{sfrequations} lists the approximate age
sensitivity of the different star formation tracers.  
These were estimated using the Starburst99 models in
the approximation of constant star formation.  The
second number lists the mean stellar age producing the
relevant emission, while the third column lists the
age below which 90\%\ of the relevant emission is 
produced.  For the dust emission in the infrared
these ages can only be estimated, because they depend
on the detailed star formation history over periods of
up to 100 Myr and longer, and they are also convolved with
the level of dust attenuation as a function of stellar
age.  The numbers are given for the assumed starburst
timescales above; for normal galaxies the 90th percentile
age can be 500 Myr or longer.  

All of the results were calculated for solar
metal abundances, and readers should beware that all of
the calibrations are sensitive to metallicity.  These
have been estimated by several authors using evolutionary
synthesis models (e.g., \citealt{2002MNRAS.337.1309S, 2010A&A...523A..64R}),
%Smith et al. 2002, Raiter et al. 2010
but the precise dependences are sensitive to the details
of the input stellar models, in particular how effects of
stellar rotation are modelled.  The models cited above
(which do not adopt a dependence of rotation on metallicity)
show that a decrease in metal abundance by a factor of ten
increases the FUV luminosity of a fixed mass and IMF
population by $\sim 0.07\pm0.03$ dex (for the IMFs assumed here),
while the ionizing luminosity is more sensitive, increasing
by $\sim 0.4\pm0.1$ dex for a tenfold decrease 
in $Z/Z_\odot$.  The change in infrared luminosities for
completely obscured regions should roughly track that for
the FUV luminosity, but in most low-metallicity environments,
the dust opacity will be severely reduced, producing a
sharp fall in infrared emission for a given SFR.

\subsection{The Challenge of Spatially-Resolved Star Formation Rates in Galaxies}\label{challenge}

Nearly all of the diagnostic methods described up to now have
been designed for measuring integrated SFRs of galaxies or for
regions such as starbursts, containing thousands (or more) of massive
stars.  The resulting SFR prescriptions implicitly assume that 
local variations in stellar age mix, IMF population, and gas/dust
geometry largely average out when the integrated emission of a 
galaxy is measured.  
%The general consistency observed between
%SFRs measured with independent tracers lends support to these assumptions.

With the advent of high-resolution maps of galaxies in
the UV, IR, emission lines, and radio  (and integral-field
spectroscopic maps in the visible and near-infrared), one would like of course
to extend these diagnostic methods to create fully-sampled spatially-resolved
``SFR maps" of galaxies.  This extrapolation to much smaller
regions within galaxies (or to galaxies with extremely low SFRs), however, 
is not at all straightforward. At smaller linear scales,
nearly all of the statistical approximations cited above
begin to break down.  

First, when the SFR in the region studied drops
below  $\sim$0.001-0.01 \msun\,yr$^{-1}$ (depending on the SFR
tracer used), incomplete sampling of the stellar IMF will lead 
to large fluctuations in the tracer luminosity for a fixed SFR.
These begin to become problematic for luminosities of
order $10^{38} - 10^{39}$ ergs\,s$^{-1}$ for emission-line, UV, or
IR tracers, and they are especially severe for the ionized gas tracers,
which are most sensitive to the uppermost parts of the stellar IMF.  
For actively star forming normal disk galaxies,
this onset of stochasticity typically occurs on spatial
scales of order 0.1--1 kpc, but the region can be considerably larger
in galaxies, or parts of galaxies, with lower SFRs.
Note that this breakdown on local scales occurs regardless of
whether the IMF itself varies systematically.

As a prime example of this stochasticity, consider what 
astronomers in M51 would observe if they examined the solar
neighborhood in the Milky Way from a distance of 10 Mpc.
In a pixel of 100 pc radius centered on the Sun they would 
observe no molecular gas and no star
formation. If they degraded their resolution to a radius of 300 pc,
they would pick up all the Gould Belt clouds, but no localized \halpha\
emission. 
%Assuming they had a super-ALMA, they would measure
%$\sigmagas = 0.18$ \msunpc, but they would miss $\sigmasfr = 2.6\ee{-3}$
%\sfrkpc.
Within a slightly larger radius of 500 pc, their beam would include Orion,
with its O stars and \hii\ region. (For this example, we consider only
the stars and emission from
the Orion Nebula Cluster (ONC) stars and emission, 
for which we have good numbers, not the full
Orion clouds and OB associations.)
This roughly doubles the number of YSOs.
Our M51 observers would measure $\sigmamol = 0.11$ \msunpc, but if they applied
the relation between SFR and \halpha\ luminosity from K98,
they would derive $\sigmasfr$ a factor of 10 lower
than the actual SFR in Orion, based on the actual stellar content of the ONC
and a relatively long timescale of 3 Myr \citep{2011AJ....142..197C},
% Chomiuk and Povich
and they would miss the total SFR within 500 pc by a factor of 20 because
the other clouds produce no O stars.
This severe underestimate for Orion itself
from \halpha\ emission results because the Orion cluster
is too small to fully populate the IMF; its earliest spectral type
is O7 V, and so the star cluster produces relatively little ionizing
luminosity relative to its total mass and SFR.
% RCK see what you think of above attempt
% NJE needed to add back the point about the other clouds
% RCK reworded sentence below for clarity
If the distant observers used the total infrared emission of Orion
instead, they would still underestimate its  
SFR by a factor of 8 \citep{2012ApJ...745..190L}
% Lada et al 2012
(a factor of 10 with the newer conversions in Table \ref{sfrequations}).
%RCK  Moved point below to here, where it fits better
 
%RCK Major rejigging of paragraph, hopefully a lot clearer
% NJE Looks fine
Secondly, when the spatial resolution of the SFR measurements
encompass single young clusters, such as Orion, the assumption
of continuous star formation which is embedded in the global
SFR recipes breaks down severely.  In such regions that emission
at all wavelengths will be dominated by a very young population
with ages of
typically a few Myr, which will be shorter than the averaging times assumed
for all but the emission-line tracers.  These changes affect
both the relative luminosities of star formation tracers in 
different bands and of course the interpretation of the 
``star formation rate" itself.  Technically speaking the luminosities
of young star-forming clusters only provide information on the
masses of the regions studied, and converting these to SFRs 
requires independent information on the ages and/or age spreads
of the stars in the region.

%Secondly, on small scales the starlight in
%a star-forming region is likely to be dominated by a 
%single-age population, so the continuous SFR approximation used
%in the calibrations in Table 1 breaks down, and the emission instead
%is a strong function both of the total mass of stars formed as
%well as the age of the region.  This can also lead to dramatic
%differences in the SFRs estimated from different tracers if
%the steady-state calibrations are applied.  Indeed the notion
%of a ``star formation rate" itself for a single-age population
%is an ambiguous concept at best.

% changed to 100-200 since MCs are rarely larger than 100 pc NJE
%Third, on small scales $< 1$ kpc, the scatter is likely to increase
%because few star forming region will be averaged. 

% RCK I commented out above because the same point is made under "first"
% RCK Again rejigged text for clarity and better referencing
% NJE looks good
A third measuring bias will set in if the resolution of the
measurements becomes smaller than the Str\"omgren diameters of
H\,II regions or the corresponding dust emission nebulae, which
tend to match or exceed those of the H\,II regions 
(e.g., \citealt{2008ApJ...681.1341W, 2007ApJ...668..182P}).
% Watson et al. 2008, Prescott et al. 2007
This scale varies from $<$100\,pc in the Milky Way to 
200--500\,pc in actively star-forming galaxies such as NGC\,6946
(Figure \ref{ngc6946}).  On these small scales indirect tracers of the SFR
(e.g., \halpha, IR) tend to trace the surfaces and bubbles of 
clouds rather than the young stars.  
%, producing a
%stars themselves breaks down.  Emission lines such as \halpha\
%do not map stars at all, but rather the distribution of ionized
%gas in \hii\ regions surrounding the stars.  Likewise  
%infrared dust emission traces the structure of dust clouds surrounding
%young stars.  ``SFR maps" at these wavelengths will trace 
%ionization fronts and zones,
%PDRs, and interstellar clouds but not necessarily the
%young stars themselves.  Correlating {\bf the emission of gas and dust
%on such small linear scales} 
%with the distribution of cold atomic and molecular gas
%often will yield 
%producing tight ``Schmidt laws" which carry no physical information
%on the star formation itself, but instead compare the 
%surface densities of different ISM components against each other.
%In other cases, where the ionizing photons have destroyed the
%molecular gas, the nominal SFR tracers may anti-correlate with
%the molecular gas tracers.
%RCK  Added this point back, but much shorter and hopefully clearer
By the same token, much of the \halpha\ and dust emission in galaxies,
typically 30--60\%, is emitted by diffuse ionized gas and dust
located hundreds of parsecs or more from any young stars 
(e.g., \citealt{2007ApJ...661..801O, 2007ApJ...655..863D}).
Such emission produces a false positive signal of star formation,
and care is needed to account for its effects when mapping the 
SFR within galaxies.

Taken together, these factors complicate, but do not prevent, the
construction of 2D maps of SFRs, so long as the methodology is 
adapted to the astrophysical application.  
For example, radial profiles of SFR distributions
in galaxies may still be reliably derived using the integrated SFR
calibrations, providing that large enough annuli are used to 
assure that the IMF is fully populated in aggregate, and the
SFR is regarded to be averaged over time scales of order 100 Myr.
Likewise a 2D distribution of star formation over the last 
5 Myr or so can be derived from a short-lived tracer such
as \halpha\ emission, if the spatial coverage is limited to
very young regions, and structure on scales smaller than individual
\hii\ regions is ignored.  Maps of integrated \halpha\ and
IR emission can be used to study the population of star-forming
regions and the star formation law, when restricted to young regions
where the physical association of gas and stars is secure
(\S \ref{localtests}).  An alternative approach is to apply
direct stellar photospheric tracers, such as the ultraviolet continuum
emission on large scales (bearing in mind the variable 
10--300 Myr time scales traced by the UV), or resolved stellar tracers such as 
YSOs or deep visible-wavelength color-magnitude
diagrams, to map the distributions of young stars directly
(\S \ref{resolvedstars}).  
Likewise it should be possible to use pixel-resolved SEDs of
galaxies in the UV--visible to derive dust-corrected UV maps,
and possibly apply local corrections for age in the maps.
With such a multiplicity of approaches, we anticipate major 
progress in this area, which will be invaluable for 
understanding in detail the patterns of
star formation in galaxies and connecting to detailed studies within
the MW.

% written by Rob as one file
% calls resolvedstars.tex
% includes subsections with labels uvcont, halpha, infrared, radio, othersfr
% composite, resolved, compendium, challenge, 

\section{THE LOCAL PERSPECTIVE: FROM THE INSIDE LOOKING OUT}\label{local}

%\input localintro.tex
% localintro
%lowmass.tex
%updated 032011 NJE
%070111 NJE
%070911 NJE
%090611 NJE drastic cuts to focus on what we need
%092211 NJE small edits
%100711 NJE added tdep or SFR/Mass for local clouds
%110411 NJE minor edits, and condensed some
%121211 NJE minor edit
%121611 NJE added desc of Flat SED, since we use it later
%020312 NJE very minor edits
%021412 NJE added binaries
%031212 NJE minor edits

\subsection{Outline of low-mass star formation}\label{lowmass}

The formation of low-mass stars can be studied in greatest detail because
it occurs in relatively nearby clouds and sometimes in isolation from other
forming stars. The established paradigm provides a point of comparison for
more massive, distant, and clustered star formation. 

Physically, an individual star forming event, which may produce
a single star or a small number multiple system,
proceeds from a prestellar core, 
which is a gravitationally bound starless core (see \S \ref{names}, 
\citealt{2007prpl.conf...17D}
%Di Francesco et al. 2007 Protostars and Planets
for definitions),
a dense region, usually within a larger molecular cloud. 
Prestellar cores are centrally condensed,
and can be modeled as Bonnor-Ebert spheres
(\citealt{1999MNRAS.305..143W, 2001ApJ...557..193E, 2005MNRAS.360.1506K}),
%Ward-Thompson et al. 1999, Evans et al. 2001, Kirk et al. 2005
%Ward-Thompson et al. 1999, Evans et al. 2001, Kirk et al. 2005
which have nearly power-law ($n(r) \propto r^{-2}$)
envelopes around a core with nearly constant density.
 
Collapse leads to the formation of a first hydrostatic core in a small region
where the dust continuum emission becomes optically thick; this short-lived
core is still molecular and grows in mass by continued infall from the outer
layers (\citealt{1969MNRAS.145..271L, 1995ApJ...439L..55B,
2007PASJ...59..589O, 2005fost.book.....S}).
%Larson, Boss and Yorke, Omukai, Stahler and Palla book)
When the temperature reaches 2000 K, the molecules in the first
hydrostatic core collisionally dissociate, 
leading to a further collapse to the true
protostar, still surrounded by the bulk of the core, often referred to as the
envelope. Rotation leads to flattening and a centrifugally supported disk
\citep{1984ApJ...286..529T}.
% Terebey, Shu and Cassen.
Magnetic fields plus rotation lead to winds and jets 
(\citealt{2007prpl.conf..261S, 2007prpl.conf..277P}),
%Shang et al. PPV, Pudritz et al. PPV
which drive molecular outflows from young stars of essentially all masses
(\citealt{2007prpl.conf..245A, 2004A&A...426..503W}).
%Arce et al. 2007, Wu et al. 2004

Stage 0 sources have protostars, disks, jets, and envelopes
with more mass in the envelope than in the star plus disk 
\citep{1993ApJ...406..122A}. 
%Andre et al. 1993
Stage I sources are similar to Stage 0, but with less mass in the envelope 
than in star plus disk. Stage II sources lack an envelope, but have a disk, 
while Stage III sources have little or no disk left.
There are several alternative
quantitative definitions of the qualitative terms used here
(\citealt{2006ApJS..167..256R, 2008A&A...486..245C}).
%Robitaille et al. 2006
%Crapsi et al 

Stages I through III are usually associated with SED Classes I through III
defined by SED slopes rising, falling slowly, and falling more rapidly
from 2 to 25 \micron\ \citep{1984ApJ...287..610L}. 
%Lada and Wilking 1984
A class intermediate between I and II, with a slope near zero
(Flat SED) was added by \citet{1994ApJ...434..614G}.
%Greene et al. 94.
In fact, orientation effects in the earlier stages
can confuse the connection between class and stage; without detailed studies
of each source, one cannot observationally determine the stage without 
ambiguity.  For this reason, most of the subsequent discussion will use 
classes despite their
questionable correspondence to physical configurations.
\citet{2009ApJS..181..321E}
%Evans et al. 2009
give a historical account of the development of the class and 
stage nomenclature.

For our present purposes, the main point is to establish time scales for the
changing observational signatures, so the classes are useful.
Using the boundaries between classes from \citet{1994ApJ...434..614G},
%Greene et al. 94.
\citet{ 2009ApJS..181..321E}
%Evans et al. 2003) 
found that the combined Class 0/I phases last $\sim 0.5$ Myr,
with a similar duration for Flat SEDs,
assuming a continuous flow through the classes for at least
the  Class II duration of $2\pm 1$  Myr 
(e.g., \citealt{2009AIPC.1158....3M}),
% Mamajek, 2009, AIPC
which is essentially the age when about half the stars in a dated cluster 
lack infrared excesses; hence all durations can be thought of as half-lives.

To summarize, we would expect a single star-forming core to be dominated 
by emission at \fir\ and \smm\ wavelengths for about 0.5 Myr, 
\nir\ and \mir\ radiation for about 1.5 Myr and \nir\ to visible light 
thereafter. However, the duration of
emission at longer wavelengths can be substantially lengthened by material in
surrounding clumps or clouds not directly associated with the forming star.

The star formation rates and efficiencies can be calculated for 
a set of 20 nearby clouds ($\mean{d} = 275$ pc) 
with uniform data from {\it Spitzer}.
The star formation rate was calculated from equation \ref{ysocounteq}
%$\sfr = N(YSOs)\times \mean{M_*}/t$, where $N(YSOs)$ is the number of objects
%with infrared excess, \mean{M_*} is the mean stellar mass, and $t$ is the 
%duration of the infrared excess.  Taking $t = 2\pm1$ Myr and 
%$\mean{M_*} = 0.5$ \msun, 
and the assumptions in \S \ref{resolvedstars} by
\citet{2010ApJ...723.1019H}, 
%Heiderman et al. (2010) 
who found a wide range of values for \sfr, with a mean value for the 20 
clouds of $\sfr = 39\pm18$ \msunmyr. 
The star formation efficiency ($\epsilon = \mstar/(\mstar + \mgas)$)
can only be calculated over the last 2 Myr because surveys are incomplete
at larger ages;
averaging over all 20 clouds, only 2.6\% of the cloud mass has turned
into YSOs in that period and $\mean{\sigmasfr} = 1.2$ \msunkpc\ 
\citep{2010ApJ...723.1019H}.
%(Heiderman et al. 2010). 
For individual clouds, $\epsilon$ ranges from 2\% to 8\%
(\citealt{2009ApJS..181..321E, 2011ApJS..194...43P}). 
% Evans et al and D. Peterson et al. on Cor Aust
The final efficiency will depend on how
long the cloud continues to form stars before being disrupted; cloud lifetimes
are poorly constrained (\citealt{2007ARA&A..45..565M, 2007ARA&A..45..339B}),
% McKee-Ostriker review and pp. 345-6 of Bergin and Tafalla review 
and they may differ for clouds where massive stars form.

For comparison to extragalactic usage, the mean $\tdep = 1/\epsilon^{\prime}$
(\S \ref{definitions}) is about 82 Myr for seven local clouds, 
longer than most estimates of cloud lifetime, and much longer than 
either \mean{\tff} (1.4 Myr) or \mean{\tcross} (5.5 Myr) 
\citep{2009ApJS..181..321E}. The \mean{\tdep} for local clouds is, on the
other hand, about 10\% of the \tdep\ for the \mw\ molecular gas as a whole
(\S \ref{milkyway}).

% lowmass
%highmass.tex
%updated 042211 NJE
%070911 NJE moved info on most massive stars here from sftheory
%072611 NJE added stuff on clumps, Kauffmann and Pillai, Dunham, ...
% may eventually move to synthobs section
% 090611 NJE shortened a bit and cleaned up
% 090911 NJE added ref to Faundez et al. and Bronfman
% 092211 NJE minor edits
% 102111 NJE added ref to Heidermanfig2 (Perseus) in context of Bressert
% 110411 NJE minor edits
% 121211 NJE minor edits
% 121611 NJE minor edits
% 011812 NJE edits after comments
% 020312 NJE very minor edits
% 021312 NJE minor edit re RCK comments
% 021712 NJE added ref to update on massive star binarity
% 031212 NJE minor edits
% 050512 NJE updated refs

\subsection{Formation of Clusters and High-mass Stars}\label{highmass}

The study of young clusters provides some distinct opportunities. By averaging
over many stars, a characteristic age can be assigned, with perhaps more
reliability than can be achieved for a single star. 
As noted (\S \ref{lowmass}), the set point
for ages of all SED classes is determined by the fraction of \mir\ excesses
in clusters. This approach does implicitly assume that the cluster forms
``coevally", by which one really means that the spread in times of formation
is small compared to the age of the cluster. Because objects ranging from
prestellar cores to Class III objects often coexist \citep{2007ApJS..171..447R},
%e.g., Rebull et al,
this assumption is of dubious reliability for clusters with ages less than
about 5 Myr, exactly the ones used to set the timescales for earlier classes.
Note also that assuming coeval formation in this sense directly contradicts
the assumption of continuous flow through the classes. We live with these
contradictions.

Most nearby clusters do not sample very far up the IMF. The nearest young
cluster that has formed O stars is 415-430 pc away in Orion
(\citealt{2007A&A...474..515M, 2007PASJ...59..897H}),
%Menten et al., Hirota et al.
and most lie at
much larger distances, making study of the entire IMF difficult.

Theoretically, the birthplace of a cluster is a clump 
\citep{2000prpl.conf...97W}.
The clumps identified by \coo\ maps in nearby clouds are dubious candidates
for the reasons given in \S \ref{massfunctions}.
The structures identified by \citet{2009A&A...508L..35K} in the power-law
% Kainulainen et al. 2009
tail of the probability distribution function are better candidates, 
but most are still unbound. 
Objects found in some surveys of dust continuum emission
(\S \ref{dusttracers}) appear to be still better candidates. For example,
analysis (\citealt{2011ApJS..195...14S, 2011ApJ...741..110D})
% Schlingman and Dunham et al
of the sources found in millimeter-wave continuum emission
surveys of the \mw\ (\S \ref{dusttracers}) indicates typical volume densities
of a few\ee{3} \cmv, compared to about 200 \cmv\ for \coo\ structures.
Mean surface densities are about 180 \msunpc, higher than those of
gas in the the power-law tail ($\sigmagas \sim 40-80$ \msunpc).
These structures have a wide range of sizes, with a median of 0.75 to
1 pc.

Studies of regions with signposts of massive star formation,
using tracers requiring higher densities 
(e.g., \citealt{1992ApJS...78..505P, 2002ApJ...566..945B})
%Plume et al. 1993, Beuther, ...) 
or millimeter continuum emission from dust
(e.g., \citealt{2002ApJ...566..945B, 2002ApJS..143..469M})
% Beuther, Mueller
have identified slightly smaller structures ($r \sim 0.5$ pc) that are 
much denser, indeed denser on average than cores in nearby clouds. 
In addition to having a mass distribution consistent with that of clusters
(\S \ref{massfunctions}),
many have \fir\ luminosities consistent with the 
formation of clusters of stars with masses ranging up to those of O stars.
The properties of the \citet{1992ApJS...78..505P}
%Plume et al.  1992
sample have been studied in a series of papers 
(\citealt{1997ApJ...476..730P, 2002ApJS..143..469M, 2003ApJS..149..375S}),
%Plume et al. 1997, Mueller et al., Shirley et al. 2003) 
with the most recent summary in \citet{2010ApJS..188..313W}.
%Wu et al. 2010.
The properties of the clumps depend on the tracer used. Since the HCN \jj10
line is used in many extragalactic studies, we will give the properties
as measured in that line. The mean and median FWHM are 1.13 and 0.71 pc;
the mean and median infrared luminosities are  4.7\ee5 and 1.06\ee5 \lsun;
the mean and median virial mass are 5300 and 2700 \msun; 
the mean and median surface densities are 0.29 and 0.28 gm cm$^{-2}$; and
the mean and median of the average volume densities ($n$) 
are 3.2\ee4 and 1.6\ee4 \cmv.
As lines tracing higher densities are used, the sizes and masses decrease,
while the 
surface densities and volume densities increase, as expected for centrally
condensed regions. 
Very similar results were obtained from studies of millimeter continuum 
emission from dust \citep{2004A&A...426...97F}
% Faundez et al. 2004
toward a sample of southern hemisphere sources surveyed in CS \jj21\
\citep{1996A&AS..115...81B}.
% Bronfman et al.

\begin{figure}%4	% Example of Figure pull
%\epsfscale1200		% Figure enlarged to 120%
\center
\includegraphics[scale=0.6]{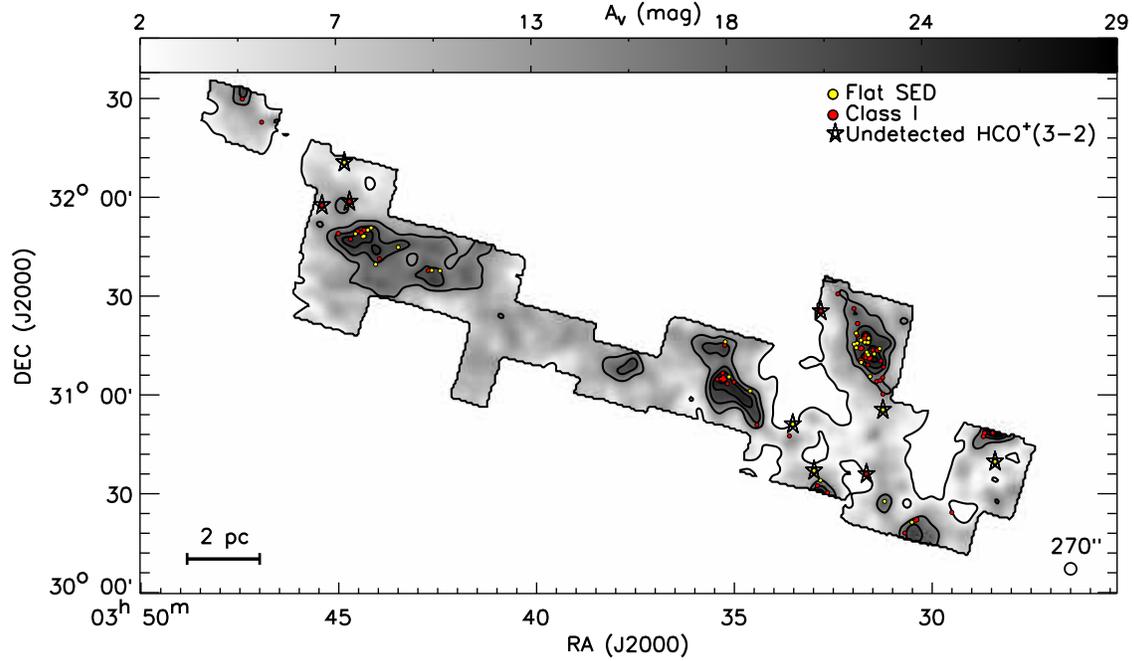}
\caption{
Example of the strong concentration of star formation in regions of high
extinction, or mass surface density in the Perseus molecular cloud. 
The gray-scale with black contours
is the extinction map ranging from 2 to 29 mag in intervals of 4.5 mag.
The yellow filled circles are Flat SED sources and the red filled circles
are Class I sources. Sources with an open star were not detected in
\hcop\ \jj32\ emission and are either older sources that may have
moved from their birthplace or background galaxies. Essentially all
truly young objects lie within contours of $\av \geq 8$ mag.
Taken from \citet{2010ApJ...723.1019H}; reproduced by permission of the AAS.
}
\label{Heidermanfig2}
\end{figure}

Embedded clusters provide important
testing grounds for theories of star formation. Using criteria of
35 members with a stellar density of 1 \msun pc$^{-3}$ for a cluster, 
\citet{2003ARA&A..41...57L}
%Lada and Lada 2003
argue that most stars form in clusters, and 90\% of those are in rich clusters
with more than 100 stars, but that almost all clusters ($> 93$\%) dissipate 
as the gas is removed, a process they call ``infant mortality." 
With more complete surveys enabled
by Spitzer, the distributions of numbers of stars in clusters and stellar
densities are being clarified. \citet{2007prpl.conf..361A}
%Allen et al. 2007 (PPV) 
found that about 60\% of young stars within 1 kpc of the Sun
are in clusters with more than 100 members, but
this number is heavily dominated by the Orion Nebula cluster.
Drawing on the samples from Spitzer surveys of nearly all clouds within
0.5 kpc, \citet{2010MNRAS.409L..54B}
% Bressert et al. (2010) 
found a continuous distribution of surface
densities and no evidence for a bimodal distribution, with distinct
``clustered" and ``distributed" modes. They found that the fraction of stars
that form in clusters ranged from 0.4 to 0.9, depending on which definition
of ``clustered" was used.  Even regions of low-mass star formation, often
described as distributed star formation, are quite clustered and the youngest
objects (Class I and Flat SED sources, see \S \ref{lowmass}) are very strongly
concentrated to regions of high extinction, especially after the samples
have been culled of interlopers (Fig. \ref{Heidermanfig2}, \S \ref{localtests}).
Similarly, studies of three clouds found that 75\% of prestellar cores
lay above thresholds in \av\ of 8, 15, and 20 mag, while most of the
cloud mass was at much lower extinction levels \citep{2007ApJ...666..982E}.
% Enoch et al. 2007 summary of three clouds with bolocam

A Spitzer study \citep{2009ApJS..184...18G}
%Gutermuth et al. 2009 
of 2548 YSOs in 39 nearby ($d < 1.7$ kpc), previously known 
(primarily from the compilation by \citealt{2003AJ....126.1916P})
%Porras et al. 2003
young clusters, but excluding Orion and NGC2264, found the following 
median properties: 26 members, core radius of 0.39 pc, stellar 
surface density of 60 pc$^{-2}$, and embedded in a clump 
with $\ak = 0.8$ mag, which corresponds to $\av = 7.1$ mag.
Translating to mass surface density using $\mean{\mstar} = 0.5$ \msun, the 
median stellar mass surface density would be 30 \msunpc\ and the gas surface
density would be 107 \msunpc. The distributions are often elongated, with
a median aspect ratio of 1.82. The median spacing between YSOs, averaged over
all 39 clusters, is $0.072 \pm 0.006$ pc, comparable to the size of 
individual cores,
and a plausible scale for Jeans fragmentation.
The distributions are all skewed toward low values, with a tail to higher values.

While various definitions of ``clustered" have been used, one physically
meaningful measure is a surface density of $\sim 200$ YSOs pc$^{-2}$ 
\citep{2005ApJ...632..397G},
%Gutermuth et al. 2005) 
below which individual cores are
likely to evolve in relative isolation (i.e., the timescale for infall
is less than the timescale for core collisions). With their sample, 
\citet{2010MNRAS.409L..54B}
%Bressert et al. 2010
found that only 26\% were likely to interact faster than they collapse.
However, that statistic did not include the Orion cluster, which exceeds
that criterion.

The fact that the Orion cluster dominates the local star formation warns
us that our local sample may be unrepresentative of the Galaxy as a whole. 
Leaving
aside globular clusters, there are young clusters that are much more massive
than Orion. \citet{2010ARA&A..48..431P}
% Portegies Zwart et al. (2010) 
have cataloged massive ($\mstar \ge \eten4$ \msun ), young (age $ \le 100$ Myr) 
clusters (12) and associations (13), but 
none of these are more distant than the Galactic Center, so they are clearly
undercounted. The mean half-light radius of the 12 clusters is $1.7\pm 1.3$ pc
compared to $11.2 \pm 6.4$ pc for the associations. For the clusters,
$\mean{\rm{log} \mstar(\msun)} = 4.2 \pm 0.3$.  
Three of the massive dense clumps from the 
\citet{2010ApJS..188..313W}
%Wu et al.  (2010) 
study have masses above \eten4 \msun\ and are plausible precursors
of this class of clusters.
Still more massive clusters can be found in other galaxies, and
a possible precursor ($M_{\rm cloud} > 1\ee5$ \msun\ within a 2.8 pc
radius) has recently been identified near the center of the \mw\
\citep{2012ApJ...746..117L}.
%Longmore et al. 2012 on G0.253+0.016. 

The topic of clusters is connected to the topic of massive stars because
70\% of O stars reside in young clusters or associations 
\citep{1987ApJS...64..545G}.
%(Gies 1987) 
Furthermore, most of the field population can be identified as runaways 
\citep{2005A&A...437..247D},
%de Wit et al. 2005)
with no more than 4\% with no evidence of having formed in a cluster.
While there may be exceptions (see \citealt{2007ARA&A..45..481Z}
%Zinnecker and Yorke 2007 
for discussion),
the vast majority of massive stars form in clusters.
The most massive star with a dynamical mass (NGC 3603-A1) weighs in
at $116\pm31$ \msun\ and is a 3.77 day binary
with a companion at $89\pm16$ \msun\ \citep{2008MNRAS.389L..38S}.
%Schnurr et al. 2008
Still higher initial masses (105-170 \msun) for the stars in NGC3603
and even higher in R136 (165-320 \msun) have been suggested
\citep{2010MNRAS.408..731C}.
%Crowther et al. 2010
Many of the most massive stars exist in tight (orbital periods of
a few days) binaries \citep{2007ARA&A..45..481Z}.
%Zinnecker and Yorke 2007
For a recent update on massive binary properties, see
\citet{2011IAUS..272..474S}.
% Sana and C. Evans in IAU

There is some evidence, summarized by \citet{2007ARA&A..45..481Z},
% Zinnecker and Yorke 2007 
that massive stars form only in the most massive molecular clouds, 
with $max(\mstar) \propto \mcloud^{0.43}$ 
suggested by \citet{1982MNRAS.200..159L}.
%Larson 1982 
Roughly speaking, it takes $M_{cloud} = \eten5$ \msun\ to make a 50 \msun\ 
star.  Recognizing that clumps are the birthplaces of clusters and that
efficiencies are not unity could make the formation of massive stars
even less likely. The question (\S \ref{massfunctions}) 
is whether the absence of massive stars in clumps or clusters of 
modest total mass is purely a sampling effect 
(\citealt{2011ApJ...741L..26F, 2010ApJ...719L.158C}
% Fumagalli et al. in updated to pub ref in njebib.bib
% Calzetti et al. 2010
and references therein)
or a causal relation \citep{2006MNRAS.365.1333W}.
If causal, differences in the upper mass limit to clouds or clumps 
(\S \ref{massfunctions}) in a galaxy could limit the formation of the most
massive stars.
\citet{1982MNRAS.200..159L}
%Larson 1982
concluded that his correlation could be due to sampling. As a concrete example,
is the formation of a 50 \msun\ star as likely in an ensemble of 100
clouds, each with \eten3 \msun, as it is in a single \eten5 \msun\ cloud?
Unbiased surveys of the \mw\ for clumps and massive stars (\S \ref{milkyway})
could allow a fresh look at this question, with due regard for the difficulty
of distinguishing ``very rarely" from ``never."

% highmass
%sftheory.tex
%updated 042611 NJE
%070111
%070511
%070711 added ref to Brogan
%070911 NJE moved obs info on most massive stars to highmass
%090611 NJE some clean up, did not shorten, added some from localintro
%090811 NJE added criterion for formation of massive stars
%092011 NJE added ref to Myers, Krumholz new papers, few other minor edits
%092211 NJE minor edits
%110411 NJE minor edits
%121211 NJE minor edits
%121611 NJE minor edits
%011812 NJE edits after comments, quite a few, but mostly minor
%012412 NJE added Walsh ref re motions
%020312 NJE very minor edits
%031212 NJE minor edits
%051412 NJE minor edits

\subsection{Theoretical Aspects}\label{sftheory}

The fundamental problem presented to modern theorists of star formation 
has been to explain the low efficiency of star formation on the scale
of molecular clouds.  Early
studies of molecular clouds concluded that they were gravitationally bound
and should be collapsing at free fall (e.g., \citealt{1974ApJ...189..441G}).
%Goldreich and Kwan
\citet{1974ARA&A..12..279Z}
%Zuckerman and Palmer 
pointed out that such a picture would produce stars at 30 times 
the accepted average recent rate of star formation in the
Milky Way (\S \ref{milkyway}) if stars formed 
with high efficiency. Furthermore, \citet{1974ApJ...192L.149Z}
%Zuckerman and Evans (1974) 
found no observational evidence for large scale collapse
and suggested that turbulence, perhaps aided by magnetic fields, 
prevented overall collapse. This suggestion led eventually to a picture
of magnetically subcritical clouds that formed stars only via a redistribution
of magnetic flux, commonly referred to as ambipolar diffusion 
(\citealt{1987ARA&A..25...23S, 1991psfe.conf..449M}),
%(Shu et al.  1987; Mouschovias 1991 in first Crete meeting book), 
which resulted in cloud lifetimes about 10 times the free-fall time. 
If, in addition, only 10\% of the cloud became supercritical, a factor
of 100 decrease in star formation rate could be achieved.

Studies of the Zeeman effect in OH have now provided enough measurements
of the line-of-sight strength of the magnetic field to test that picture.
While there are still controversies, the data indicate that most clouds
(or, more precisely, the parts of clouds with Zeeman measurements)
are supercritical or close to critical, but not strongly subcritical
\citep{2010ApJ...725..466C}.
% Crutcher et al. major work
Pictures of static clouds supported by magnetic fields are currently
out of fashion \citep{2007ARA&A..45..565M}, but magnetic fields
are almost certain to play a role in some way (for a current
review, see Crutcher 2012, this volume).
Simulations of turbulence indicate a fairly rapid decay, even
when magnetic fields are included \citep{1998ApJ...508L..99S}.
%  Stone-Ostriker 
As a result, there is growing support for a more dynamical picture in which
clouds evolve on a crossing time \citep{2000ApJ...530..277E}.
% Elmegreen 2000 
However, this picture must still deal with the Zuckerman-Palmer problem. 

There are two main 
approaches to solving this problem at the level of clouds, 
and they are essentially extensions
of the two original ideas, magnetic fields and turbulence, into larger
scales. One approach, exemplified by \citet{2011MNRAS.414.2511V}
%Vazquez-Semadini et al. (2011) 
argues that clouds are formed in colliding flows of the warm, 
neutral medium. They simulate the outcome of these flows with 
magnetic fields, but without feedback from star formation.
Much of the mass is magnetically subcritical and star formation 
happens only in the supercritical parts of the cloud.
A continued flow of material balances the mass lost to star formation so
that the star formation efficiency approaches a steady value, in rough
agreement with the observations. 

A second picture, exemplified by \citet{2011MNRAS.413.2935D},
%Dobbs et al. 2011
is that most clouds and most parts of clouds are not gravitationally bound
but are transient objects. Their simulations include feedback from
star formation, but not magnetic fields. During cloud collisions, material is
redistributed, clouds may be shredded, and feedback removes gas. Except
for a few very massive clouds, most clouds lose their identity on the timescales
of a few Myr. In this picture, the low efficiency simply reflects the fraction
of molecular gas that is in bound structures.

It is not straightforward to determine observationally if clouds are bound,
especially when they have complex and filamentary boundaries. 
\citet{2001ApJ...551..852H}
%Heyer et al.  (2001) 
argued that most clouds in the outer galaxy with $M > \eten4$ \msun\ 
were bound, but clumps and clouds with $M < \eten3$ \msun\ were often not bound.
\citet{2009ApJ...699.1092H}
%Heyer et al. (2009) 
address the issues associated with determining accurate
masses for molecular clouds.
In a study on the inner Galaxy, \citet{2010ApJ...723..492R}
%Roman-Duval et al. (2010) 
concluded that 70\%
of molecular clouds (both in mass and number) were bound.

If molecular clouds are bound and last longer than a few crossing times,
feedback or turbulence resulting from feedback is invoked. For low-mass
stars, outflows provide the primary feedback 
(e.g., \citealt{2006ApJ...640L.187L})
% Li and Nakamura 2006
while high-mass stars add radiation pressure and expanding \hii\ regions
(e.g., \citealt{2011ApJ...729..133M}).
% Murray 2011
In addition, the efficiency is not much higher in most clumps, so 
the problem persists to scales smaller than that of clouds.

Another longstanding problem of star formation theory has been to explain
the IMF (\S \ref{massfunctions}). 
This topic has been covered by many reviews, so we emphasize only
two aspects. The basic issue is that the typical conditions in star forming
regions suggest a characteristic mass, either the Jeans mass or the 
Bonnor-Ebert mass, around 1 \msun\ (e.g., \citealt{2007prpl.conf..149B}).
Recently, \citet{2011ApJ...743..110K} 
%Krumholz 2011 on IMF
has derived a very general expression for a characteristic mass of
0.15 \msun, with only a very weak dependence on pressure.
However, we see stars down to the
hydrogen-burning limit and a continuous distribution of brown dwarfs
below that, extending down even to masses lower than those seen in 
extrasolar planets \citep{2006ApJ...644..364A}.
%Allers et al. 2006
An extraordinarily high density would
be required to make such a low mass region unstable. 

On the other end,
making a star with mass over 100 times the characteristic mass
is challenging.
\citet{2007prpl.conf..165B} 
%Beuther et al. in PPV
provide a nice review of both observations
and theory of massive star formation.
While dense clumps with mass much greater than 100 \msun\ are seen, they
are likely to fragment into smaller cores. 
Fragmentation can solve the problem of
forming low mass objects, but it makes it hard to form massive objects.
Simulations of unstable large clumps in fact tend to fragment so strongly 
as the mean density increases that they overproduce brown dwarfs, but
no massive stars (e.g., \citealt{1998ApJ...501L.205K, 
2006ApJS..163..122M}).
%Klessen et al. 1998 and other refs, ending with Martel et al. 2006).  
This effect is caused by the assumed isothermality of the gas, because
the Jeans mass is proportional to $(\tk^3/n)^{0.5}$. Simulations
including radiative feedback, acting on a global scale, have shown 
that fragmentation can be suppressed and massive stars formed 
(e.g., \citealt{2009MNRAS.397..232B, 2010ApJ...710.1343U,
2010ApJ...713.1120K}).
%(e.g., Bate et al. 2009,  Urban et al. 2010, Krumholz et al. 2010).
\citet{2008Natur.451.1082K}
% Krumholz et al on 1 gm per cmsq criterion
have argued that a threshold clump surface density of 1 gm cm$^{-2}$ is needed
to suppress fragmentation, allowing the formation of massive stars.

The radiative feedback, acting locally, can in principle also limit the mass
of the massive stars through radiation pressure. This can be a serious
issue for the formation of massive stars in spherical geometries, but
more realistic, aspherical simulations show that the radiation is
channeled out along the rotation axis, allowing continued accretion
through a disk 
(\citealt{2002ApJ...569..846Y, 2009Sci...323..754K, 2011ApJ...732...20K}).
% Yorke and Sonnhalter  2002  Krumholz et al.  2009, Kuiper et al. 2011

The picture discussed so far is basically a scaled-up version of
the formation of low mass stars by accretion of material from a
single core (sometimes called Core Accretion).
An alternative picture, called Competitive Accretion, 
developed by \citet{1998ApJ...501L.205K}, \citet{2003MNRAS.339..577B},
and \citet{2003MNRAS.343..413B},
%Klessen 98, Bate, Bonnell and Bromm 2003, Bonnell, et al. 2003
builds massive stars from the initial low-mass fragments.
The pros and cons of these two models are discussed in a joint
paper by the leading protagonists \citep{2009sfa..book..288K}.
%Krumholz and Bonnell in book
They agree that the primary distinction between the two is the
original location of the matter that winds up in the star:
in Core Accretion models, the star gains the bulk of its mass
from the local dense core, continuing the connection of the core
mass function to the initial stellar mass function to massive stars;
in Competitive Accretion models, the more massive stars collect
most of their mass from the larger clump by out-competing other,
initially low mass, fragments. A hybrid picture, in which
a star forms initially from its parent core, but then continues
to accrete from the surrounding clump, has been advanced 
(\citealt{2009ApJ...706.1341M, 2011ApJ...743...98M}),
% Myers 2009, 2011
with some observational support
(e.g., \citealt{2011ApJ...726...97L}).
% Longmore et al. 2011a on G8.68

Observational tests of these ideas are difficult. The Core
Accretion model requires
that clumps contain a core mass function extending to massive cores.
Some observations of massive dense clumps have found substructure 
on the scale of cores
(\citealt{2007A&A...466.1065B,2009ApJ...707....1B}),
% Beuther et al. , Brogan et al., both SMA 
but the most massive cores identified in these works are 50-75 \msun.
Discussions in those references illustrate the difficulties in doing this
with current capabilities.
ALMA will make this kind of study much more viable, 
but interpretation will always be tricky.
The  Competitive Accretion  model relies on a continuous flow of
material to the densest parts of the clump to feed the growing oligarchs.
Evidence for overall inward flow in clumps is difficult; 
evidence for it is found in some surveys
(\citealt{2003ApJ...592L..79W, 2005A&A...442..949F, 2011ApJ...740...40R}),
%Wu et al. Fuller et al. Reiter et al.
but not in others \citep{2006MNRAS.367..553P}.
% Purcell et al. 2006
Since special conditions are required to produce an inflow signature,
the detection rates may underestimate the fraction with inflow.
The dense clumps are generally found to be centrally condensed
(\citealt{2002ApJ...566..945B, 2002ApJS..143..469M, 2000ApJ...537..283V}), 
%van der Tak, et al. 2000,  Beuther et al, Mueller et al.
which can also suppress over-fragmentation.

Other possible tests include the coherence of outflows in clusters,
as these should be distorted by sufficiently rapid motions.
The kinematics of stars in forming clusters,
which should show more velocity dispersion in the competitive accretion
models, provide another test. 
Astrometric studies are beginning to be able to constrain these motions
\citep{2010ApJ...716L..90R},
% Rochau et al. 2010) 
along with cleaner separation of cluster members from field objects.
Relative motions of cores within clumps appear to be very low
($< 0.1$ /kms), challenging the Competitive Accretion model
\citep{2004ApJ...614..194W}.
%Walsh et al. on shifts of N2H+ (cores) relative to 13CO (clumps)

Theoretical studies include evolutionary tracks of
pre-main-sequence stars (e.g., \citealt{2000ARA&A..38..337C}
%Chabrier and Baraffe 2000
for low mass stars and substellar objects).
These evolutionary calculations are critical for determining the
ages of stars and clusters.
For more massive stars, it is essential to include accretion in the
evolutionary calculations \citep{1992ApJ...392..667P}
% Palla and Stahler 1992
and high accretion rates strongly affect the star's evolution
(\citealt{2007ARA&A..45..481Z, 2009ApJ...691..823H}).
%Zinnecker and Yorke, Hosokawa and Omukai
Assumptions about accretion and initial conditions
may also have substantial consequences for the usual methods of determining
ages of young stars (\citealt{2009ApJ...702L..27B, 2010A&A...521A..44B,
2011ApJ...738..140H}).
% Baraffe et al. and Baraffe and Chabrier on Lithium.
%Hosokawa et al. 2011

% sftheory
% localsum.tex
% 121211 Initialized to provide summary of Local SF section re Ewine suggestion

\subsection{Summary Points from the Local Perspective}\label{localsum}

The main lessons to retain from local studies of star formation as
we move to the scale of galaxies are summarized here.

\begin{enumerate}

\item Star formation is not distributed smoothly over molecular clouds
but is instead highly concentrated into regions of high extinction or
mass surface density, plausibly associated with the theoretical idea
of a cluster-forming clump. This is particularly apparent when prestellar
cores or the youngest protostars are considered 
(e.g., Figure \ref{Heidermanfig2}). In contrast, most of the mass in
nearby molecular clouds is in regions of lower extinction.

\item By counting YSOs in nearby clouds, one can obtain reasonably
accurate measures of star formation rate and efficiency without
the uncertainties of extrapolation from the high-mass tail of the IMF. 
The main source of uncertainty is the ages of YSOs.

\item There is no obvious bimodality between ``distributed" and ``clustered"
star formation, but dense clusters with massive stars tend to form in 
regions of higher mean density and turbulence, which are centrally 
condensed. Plausible precursors of quite massive (up to \eten4 or 
perhaps \eten5 \msun) clusters can be found in the \mw.

\item Despite much theoretical progress, the challenge of explaining the
low efficiency of star formation, even in regions forming only low-mass
stars, remains. Similarly, an understanding of the full IMF, from brown
dwarfs to the most massive stars, remains elusive.

\end{enumerate}

% localsum

\section{THE GALACTIC PERSPECTIVE: FROM THE OUTSIDE LOOKING IN}\label{outside}

%milkyway.tex
%updated 042611 NJE
% 061011 NJE
%070511 NJE
%070811 NJE
%070911 NJE
%072211 NJE added info on filling factors of phases
%072611 NJE some more refs on SFR and discussion of filling factor
%080411 NJE moved size and mass func to massfunctions.tex
% 082811 NJE added info on galactic distribution
% 082911 NJE added discussion of outer MW SF and CMZ
% 083011 NJE major reorganization and tightening
% 083111 NJE added a few points about radial, axial distributions
% 0908911 NJE added note on W43 at very end
% 090911 NJE added note on Casassus re SF in outer galaxy
%0915011 NJE updated from Leo's email
% 092211 NJE minor edits
% 100711 NJE minor cuts
% 102111 NJE added Chomiuk ref and cut a lot of previous stuff
% 102111 NJE added ref to Milkyway model
% 110411 NJE minor edits
% 121211 NJE added referencs, updated info on CMZ
% 121611 NJE minor edits
%     inserted placeholder for discussion of Figure of MW radial
% 121911 NJE minor edits
% 011812 NJE edits after comments, added depletion time for MW and disc
% re Eve's points
% 012012 NJE further edits incorporating fig with radial distributions
% 012312 NJE further edits, include info from Longmore
% 012612 NJE further edits
% 020312 NJE edited to give values within 13.5 kpc
% 021312 NJE edited after RCKs comments; main issues total mass and He corr.
% 021412 NJE fixed ref to Kalberla re HI, and slight edits
% 021512 NJE fixed atomic mass based on Kalberla data, my integ.
% 021712 NJE few small edits
% 031212 NJE minor edits
% 050512 NJE updated references
% 050912 NJE minor edits
% 051412 NJE minor edits, use ion

\subsection{The Milky Way as a Star-forming Galaxy}\label{milkyway}

As our nearest example, the Milky Way (hereafter denoted as \mw)
has obvious advantages in studies of
star formation in galaxies. However, living inside the \mw\ presents serious
problems of distance determination and selection effects, compared to 
studies of other galaxies.  We can ``look in from the outside"
only with the aid of models. After reviewing surveys briefly,
we consider properties of gas and star formation within the \mw,
bearing in mind the issues raised in \S \ref{gastracers} and \S
\ref{sfrdiagnostics}, first as a whole, and then the radial distribution,
and finally some notes on non-axisymmetric structure.

\subsubsection{Surveys}

Recent and ongoing surveys of the Galactic Plane
at multiple wavelengths are revitalizing the study of the
\mw\ as a galaxy.
Surveys of \hi\ from both northern \citep{2003AJ....125.3145T}
%Taylor et al. Canadian GPS
and southern \citep{2005ApJS..158..178M} 
%McClure-Griffiths et al. Southern Galactic Plane Survey
hemispheres, along with a finer resolution survey of parts of the
Galactic Plane \citep{2006AJ....132.1158S}
% Stil et al. VLA Galactic Plane Survey
have given a much clearer picture of the atomic gas in the \mw\
(for a review of \hi\ surveys, see \citealt{2009ARA&A..47...27K}).
Based on these surveys, properties of the cool atomic clouds have been
analyzed by \citet{2003ApJ...585..801D}.
% Dickey et al. 2003

Numerous surveys of the \mw\ have been obtained in CO \jj10\
(e.g., \citealt{2001ApJ...547..792D, 1988ApJ...324..248B, 1988ApJ...327..139C}).
%Dame et al. 2001, Bronfman et al. 1988, Clemens et al. 1988
A survey of the inner part of the \mw\ in \coo\ 
\citep{2006ApJS..163..145J}
%Jackson et al. GRS
has helped with
some of problems caused by optical depth in the main isotopologue.

Surveys for tracers of star formation have been made in radio
continuum (free-free) emission \citep{1970A&AS....1..319A}
% Altenhoff, W. et al. 1970
and recombination lines 
(\citealt{2011ApJS..194...32A, 1989ApJS...71..469L}), 
% Anderson et al., Lockman et al.1989
in water masers 
(\citealt{1988A&AS...76..445C, 2011MNRAS.416.1764W}),
% Cesaroni et al. 1988 north
%Walsh et al. 2011 (HOPS) south
and methanol masers (\citealt{2005A&A...432..737P, 2009MNRAS.392..783G}).
% Pestalozzi et al. 2005 compendium of various surveys
%Green et al. on method and early results
In addition to being signposts of (mostly massive) star formation,
the masers provide targets for astrometric studies using VLBI.

Complementary surveys of clouds have been done using \mir\ extinction
(e.g., \citealt{1996A&A...315L.165P, 1998ApJ...494L.199E, 2009A&A...505..405P}),
% Perault et al. 1996, Egan et al. 1998, Peretto and Fuller 2009
which can identify Infrared Dark Clouds (IRDCs) against the Galactic
background emission. The MIPSGAL survey \citep{2009PASP..121...76C}
%Carey et al. description of the survey
should also provide a catalog of very opaque objects.
At longer wavelengths, the dust is usually in emission,
and millimeter continuum emission from dust (e.g., \citealt{2011ApJS..192....4A,
2010ApJS..188..123R, 2009A&A...504..415S})
%Aguirre et al.  Rosolowsky et al., Schuller et al.
provides a different sample of objects.

Each method has its own selection effects and
efforts are ongoing to bring these into a common framework. A key ingredient
is to determine the distances, using spectral lines and methods to break the
distance ambiguity in the inner galaxy; initial work is underway
(\citealt{2011A&A...526A.151R, 2011ApJS..195...14S, 2011ApJ...741..110D,
2011ApJS..197...25F}).
% Russeil et al. 2010 (HIGAL), Schlingman et al. and Dunham et al.
% Foster et al 2011 (MALT90)
Ultimately, these studies should lead to a better definition of the
total amount and distribution of dense gas (\S \ref{definitions}).
The Herschel HIGAL survey \citep{2010A&A...518L.100M}
%Molinari et al. 2010 in Herschel early science
will add wavelengths into the \fir\ and lead to a far more
complete picture. Quantities like $\lfir/M_{cloud}$ and $\lfir/M_{dense}$, 
as used in extragalactic studies, can be calculated for large samples.

\subsubsection{The Milky Way as a Whole}\label{wholegalaxy}

\begin{figure}%4	% Example of Figure pull
\center
\includegraphics*[scale=0.6]{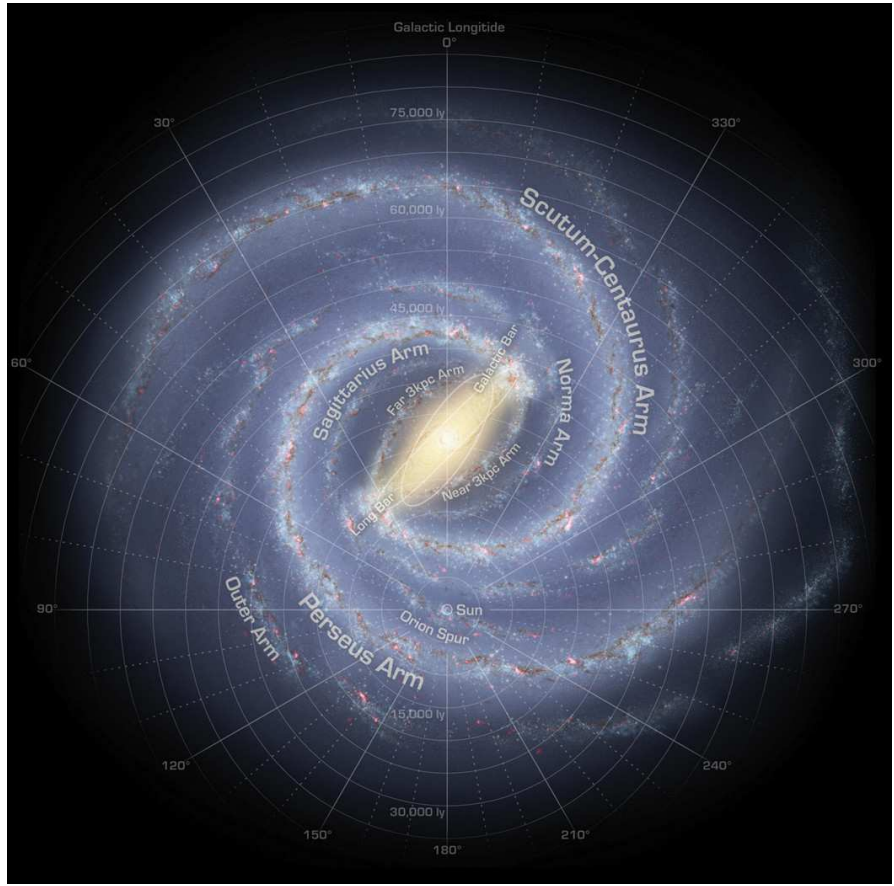}
\caption{
Sketch of approximately how the Galaxy is likely to appear viewed face-on, 
with Galactic coordinates overlaid and the locations of spiral arms 
and the Sun indicated.
This sketch was originally made by Robert Hurt of the Spitzer Science Center 
in consultation with Robert Benjamin at the University of 
Wisconsin-Whitewater. The image is based on data obtained from the 
literature at radio, infrared, and visible wavelengths. As viewed from a 
great distance our Galaxy would appear to be a grand-design two-armed 
barred spiral with several secondary arms: the main arms being the 
Scutum-Centaurus and Perseus arms and the secondary arms being Sagittarius, 
the outer arm, and the 3 kpc expanding arm.
Adapted from a figure in Churchwell et al. (2009) by R. Benjamin.
The original figure appeared in the Publications of the Astronomical 
Society of the Pacific, Copyright 2009,
Astronomical Society of the Pacific; reproduced with permission of the
Editors.
}
\label{MWwithcoord}
\end{figure}

\begin{figure}
\center
\includegraphics*[scale=0.6, angle = 90]{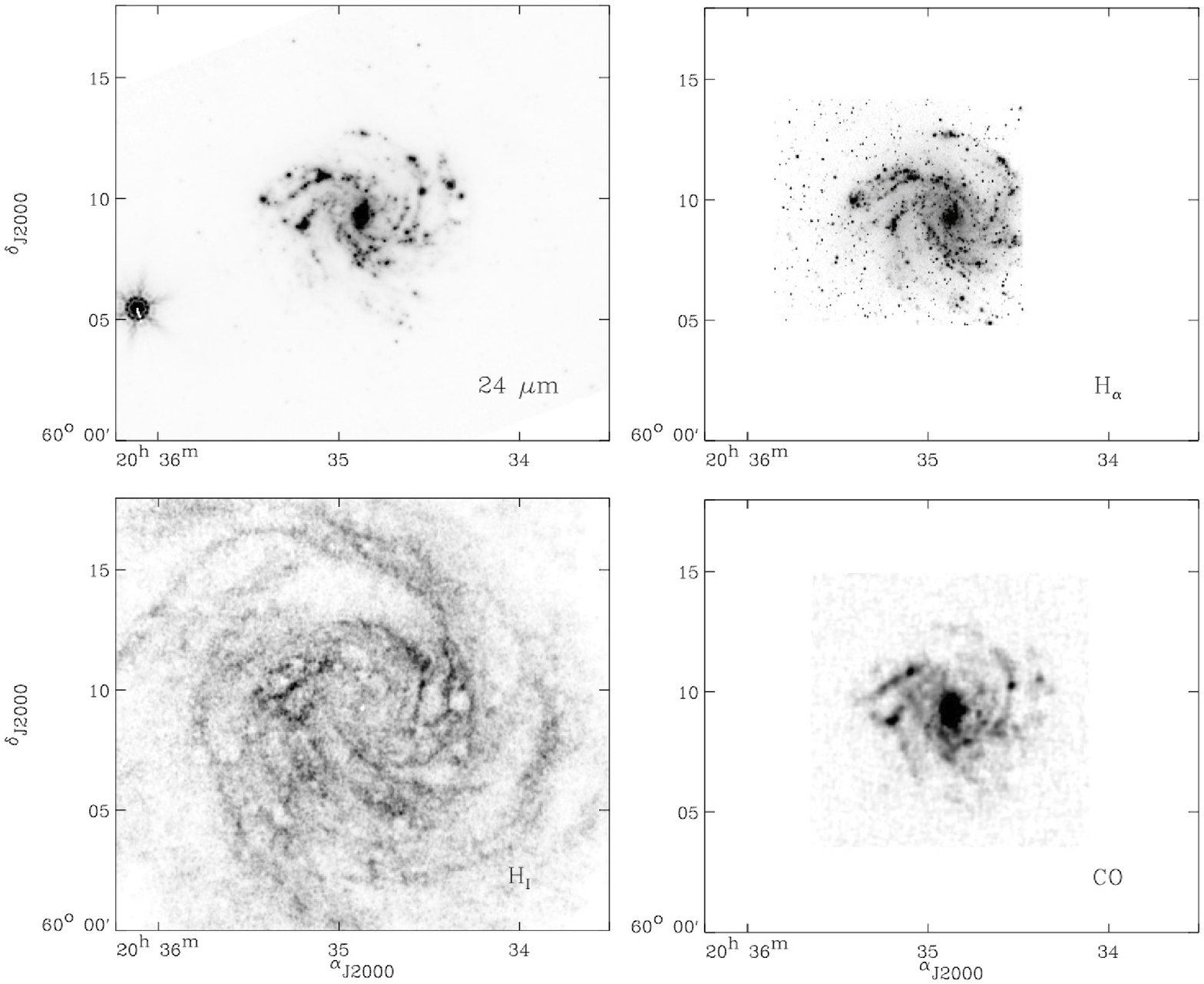}

\caption{
High-resolution maps of star formation tracers and cold
gas components in NGC 6946.  Top left:  Spitzer 24 \micron\ dust
emission from the SINGS/KINGFISH project \citep{2011PASP..123.1347K};
%Kennicutt et al. 2011
Top right: \halpha\ emission from \citet{2004A&A...426.1135K};
%Knapen et al. 2004
Bottom left: \hi\ emission from the VLA THINGS survey \citep{2008AJ....136.2563W};
%Walter et al. 2008
Bottom right: CO \jj21\ emission from the HERACLES survey 
\citep{2009AJ....137.4670L}.
%Leroy et al. 2009
Note that the \hi\ map extends over a much wider area than the CO, \halpha,
and 24\,$\mu$m observations. 
%Upper right panel \citep{2004A&A...426.1135K}
%reproduced with permission of ESO; 
%all other panesl reproduced by permission of the AAS.
}

\label{n6946optir}
\end{figure}

The Milky Way is a barred spiral galaxy 
(\citealt{1988gera.book..295B, 2001ApJ...547..792D, 2005ApJ...630L.149B}).
%Burton 88, Dame et al. 2001, Benjamin 2005
The number and position of spiral arms are still topics for debate, but
the best currently available data favor a grand-design, two-armed,
barred spiral with several secondary arms. A conception
of what the \mw\ would look like from the outside \citep{2009PASP..121..213C}
is shown in Figure \ref{MWwithcoord}.
% Churchwell 2009 GLIMPSE Summary
We can expect continuing improvements in the model of the \mw\ 
as VLBI astrometry improves distance determinations of star forming regions
across the \mw\ (\citealt{2009ApJ...700..137R, 2011AN....332..461B}).
%Reid et al. 2009 new gal model, Brunthaler et al. 2011, even newer constants.
As we will be comparing the \mw\ to NGC 6946, we also show images of
that galaxy at \halpha, 24 \micron, \hi, and CO \jj21\ in Fig. 
\ref{n6946optir}.

Based on a model including dark matter (\citealt{2009ARA&A..47...27K,
2007A&A...469..511K, 2008A&A...487..951K}),
%(Kalberla et al. 2007, Kalberla ARAA, Kalberla and Dedes 2008 for \hi\ dist),
the \mw\ mass within 60 kpc of the center is $M(tot) = 4.6\ee{11}$ \msun,
with $M(baryon) = 9.5\ee{10}$ \msun.  
The total mass of atomic gas (\hi\ plus He) is $M(atomic) = 8\ee9$ \msun,
and the warm ionized medium contains $M(WIM) = 2\ee9$ \msun. 
The mass fraction of the HIM (\S \ref{ismintro}) is negligible.
With a (perhaps high) estimate for the molecular mass 
of $M(mol) = 2.5\ee9$ \msun, 
\citet{2009ARA&A..47...27K}
% Kalberla and Kerp
derive a gas to baryon ratio of 0.13. 
As explained in \S \ref{mwradial}, for consistent comparison to the radial
distribution in NGC~6946, we will use a constant 
$\xco = 2.0\ee{20}$, correct for helium, and assume an outer radius 
of the star forming disk of 13.5 kpc to define masses, surface densities, etc.
Within $\rgal = 13.5$ kpc, $M(mol) = 1.6\ee9$ \msun\ and 
$M(atomic) = 5.0\ee9$ \msun\ with these conventions.
% multiplied previous value by 1.36 for He (021412)

The volume filling factor of the CNM is about 1\% \citep{2005ARA&A..43..337C}
% Cox ARAA
and that of molecular clouds, as traced by the \coo\ survey
\citep{2010ApJ...723..492R},
%Roman-Duval
has been estimated at about 0.5\% (M. Heyer, personal communication)
in the inner galaxy, but much lower overall.
Denser ($n \sim$ few\ee{3} \cmv) structures
found in millimeter continuum surveys, roughly corresponding to clumps, 
appear to have a surface filling factor of \eten{-4} and 
a volume filling factor of about \eten{-6} (M. K. Dunham, 
personal communication), but better estimates should be available soon.

The star formation rate of the \mw\ has been estimated from 
counting \hii\ regions, which can be seen across the \mw\ and extrapolating
to lower mass stars (\citealt{1987sbge.proc....3M, 1997ApJ...476..144M,
2010ApJ...709..424M}).
These methods average over the effective lifetime of
massive stars, about 3-10 Myr (Table \ref{sfrequations}).
Estimates of \sfr\ from a model of the total \fir\ emission of the
\mw\ (e.g., \citealt{2006A&A...459..113M})
are somewhat less sensitive to the high end of the IMF (\S
\ref{sfrdiagnostics}) and average over a longer time.
%Misiriotis 2006, based on modeling COBE data.
An alternative approach, based on counting likely YSOs in the GLIMPSE
survey of the Galactic plane \citep{2010ApJ...710L..11R}
% Robitaille
is much less biased toward the most massive stars, 
but is limited by sensitivity, issues of identification
of YSOs, and models of extinction. 
\citet{2011AJ....142..197C}
%Chomiuk and Povich
have recently reviewed all methods
of computing \sfr\ for the \mw\ and conclude that they are consistent
with $\sfr = 1.9\pm 0.4$ \msunyr. However they conclude that
resolved star counts give SFRs that are  factors of 2-3 times higher
(see \S \ref{localtests} for further discussion).

With  a radius of active star formation of 13.5 kpc (\S \ref{mwradial}), 
$\sfr = 1.9$ \msunyr\ yields $\mean{\sigmasfr} = 3.3\ee{-3}$ \msunkpc.
Taking $M(mol, \rgal<13.5) = 1.6\ee9$ \msun, 
$\tdep = 0.8$ Gyr, 10 times longer than \mean{\tdep} in local clouds
(\S \ref{lowmass}). 
%If the local clouds are typical, this difference 
%suggests that final efficiencies ($\epsilon$) in the clouds are about 6\%, 
%no more than twice the current efficiencies. One explanation would
%be that the rest of the cloud is dispersed back to the pool of unbound 
%molecular, atomic, or ionized gas, which forms star-forming
%structures only on much longer timescales. 
For all the gas in the (60 kpc) \mw,
$\tdep = 6.6$ Gyr, or 5.5 Gyr if the WIM is excluded.
If we attribute all non-gaseous baryons in the \citet{2009ARA&A..47...27K}
%Kalberla and Kerp
model (8.25\ee{10} \msun) to stars and stellar remnants, ignore
recycling, and take an age of 1\ee{10} yr, we would derive an average
star formation rate of 8.25 \msunyr, about 4 times the current rate.
% The numbers above may need to be corrected for He... NJE
%There is evidence for 2 to 3 bursts of star formation at
%0-1 Gyr, 2-5 Gyr, and 7-9 Gyr ago \citep{2000A&A...358..869R}.
%Rocha-Pinto et al. (2000).
%\citet{2000MNRAS.316..605H}
% Hernandez et al. 2000
%also found evidence for an oscillation in SFR with period 0.5 Gyr
%over the last 3 Gyr.

\subsubsection{Radial Distributions} \label{mwradial}

\begin{figure}
%\center
%\begin{center}
%\begin{tabular}[htp]{c}
%\includegraphics*[totalheight=5.0cm, viewport=20 355 540 700]{fig_Schmidt_global.eps} &
%\includegraphics*[totalheight=5.0cm, viewport=20 355 540 700]{fig_gao.eps} \\
\includegraphics[scale=0.5, angle=-90]{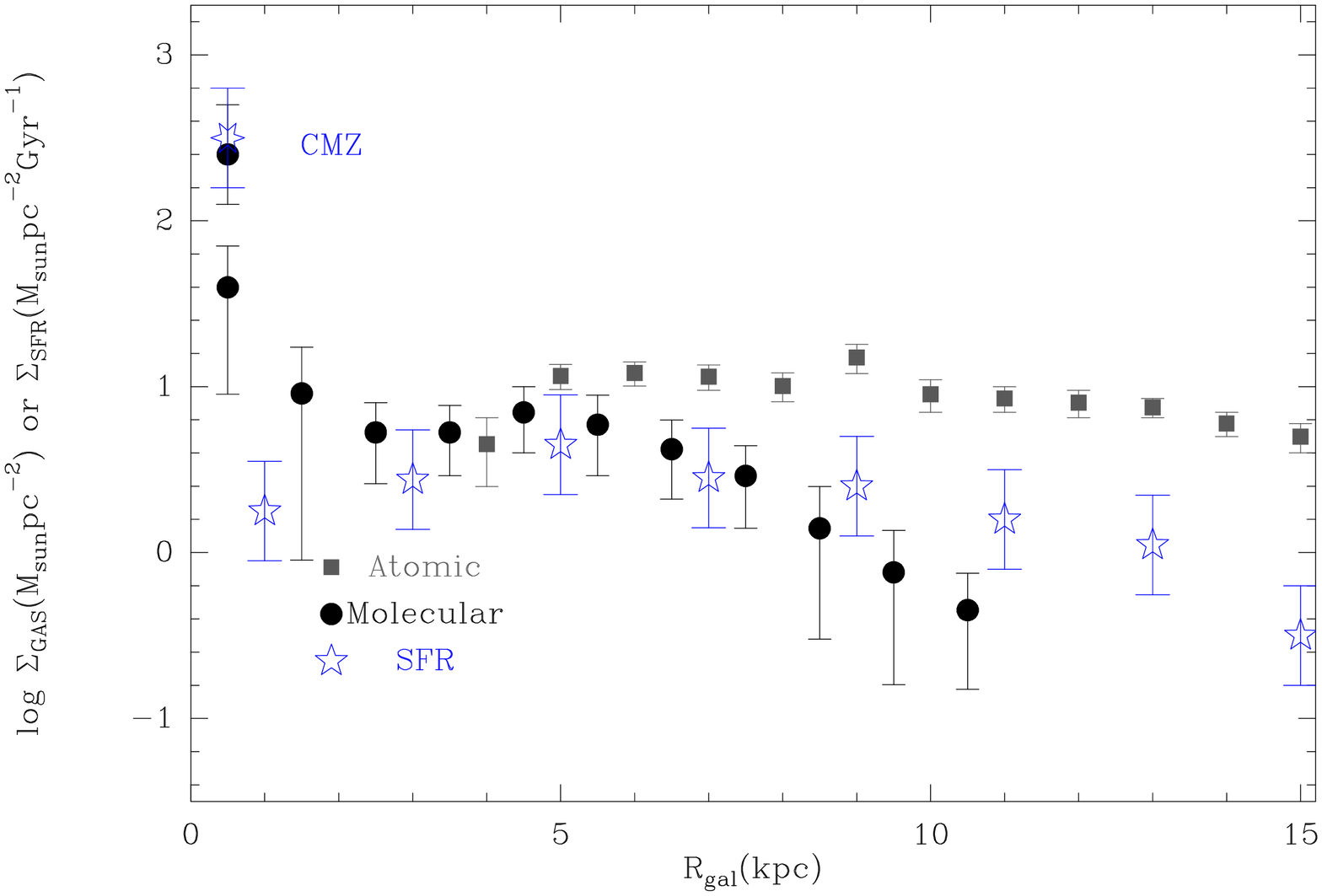} \\
\includegraphics[scale=0.5, angle=-90]{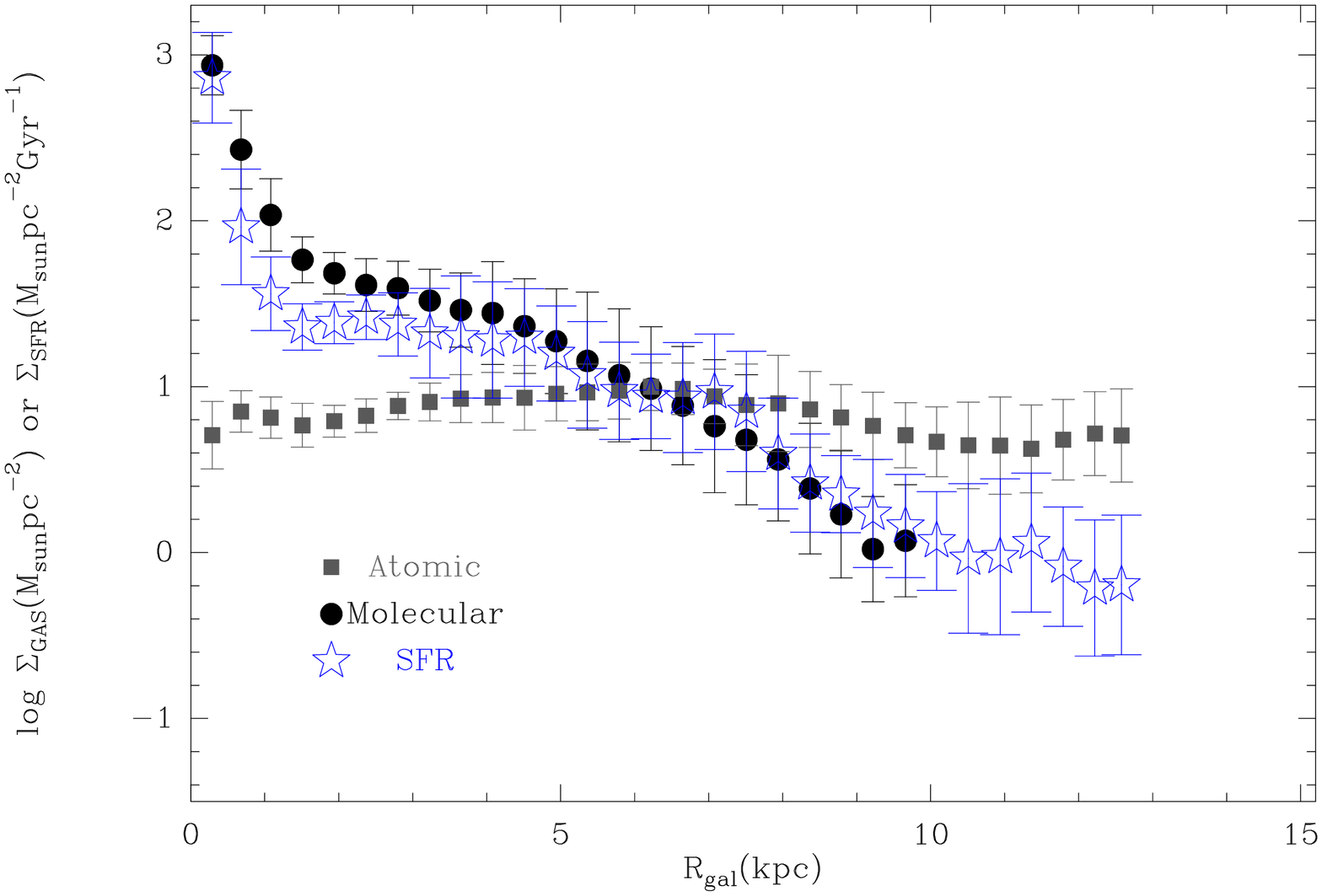} \\
%\end{tabular}
%\end{center}

\caption{ 
\footnotesize 
Top:  The radial distribution of surface densities of atomic gas, molecular
gas, and star formation rate for the \mw.
The atomic data were supplied by P. Kaberla,
\citep{2008A&A...487..951K}, 
% Kalberla and  Dedes, et al. in Astr Ap
but we corrected for helium. The discontinuity around 8.5 kpc reflects
the inability to model \hi\ near the solar circle.
The molecular data are taken from Table 1 in \citet{2006PASJ...58..847N},
% Nakanishi and Sofue 2006
but scaled up to be consistent with the bottom panel, using 
$\xco = 2.0\ee{20}$ and multiplying by 1.36 to include helium. The
molecular surface density in the innermost bin is highly uncertain
and may be overestimated by using a constant value of \xco\ (\S 
\ref{gastracers}). The \sigmasfr\ is read from Figure 8 in 
\citet{2006A&A...459..113M}, 
% Misiriotis et al. 
but it traces back, through a complex history, to \citet{1982VA.....26..159G},
% Guesten and Mezger, Vistas in Astronomy 
based on radio continuum emission from \hii\ regions.
That result was in turn scaled to the latest estimate of total
star formation rate in the \mw\ of 1.9 \msunyr\ \citep{2011AJ....142..197C}.
% Chomiuk and Povich
Separate points for \sigmamol\ and \sigmasfr\ are plotted 
for the CMZ ($\rgal < 250$ pc). 
An uncertainty of 0.3 was assigned arbitrarily to the 
values of log \sigmasfr.
Bottom: Same as the top panel, but for NGC~6946.
It is based on a figure in \citet{2011AJ....142...37S}
% Schruba et al.
but modified by A. Schruba to show radius in kpc for ready comparison
to the plot for the \mw. Reproduced by permission of the AAS.
All gas distributions for both galaxies include a correction for helium.
}

\label{MWradial}
\end{figure}

The \mw\ also offers a wide range of conditions, from the far
outer Galaxy to the vicinity of the Galactic Center, for detailed study.
The hurdle to overcome, especially in the inner \mw, is to assign
accurate distances.
The radial distributions of atomic and molecular gas, along with
the star formation rate surface density (here in units of
\msun\ Gyr$^{-1}$ pc$^{-2}$), are plotted in Figure \ref{MWradial},
along with a similar plot for NGC~6946 \citep{2011AJ....142...37S}.
% Schruba et al.
For consistency, we have used the same \xco\ and  correction for helium
(for both molecular and atomic gas)
as did \citet{2011AJ....142...37S}.
% Schruba et al.
These distributions for the \mw\ have considerable uncertainties, and should
be taken with appropriate cautions, especially in the innermost regions,
as discussed below.
\citet{2006A&A...459..113M} 
% Misiriotis et al.
presents a collection of estimates of \sigmasfr\ and a model.
They all show a steady decline outward from a peak at $\rgal \sim 5$ kpc, but
the situation in the innermost galaxy is unclear. The data on \hii\ 
regions that forms the basis of most estimates are very old, and a
fresh determination of $\sigmasfr(\rgal)$ is needed.

In comparison, NGC~6946 has a similar, rather flat, distribution of 
atomic gas, with $\Sigma$(atomic) $\sim 10$ \msunpc. 
The molecular gas distribution
is similar to that of the \mw, except for a clearer and stronger peak within
$\rgal = 2$ kpc. 
The distribution of \sigmasfr\ follows the molecular gas in both
galaxies, more clearly so in NGC 6946.

We discuss the \mw\ distributions from the outside, moving inward.
The average surface density of molecular gas drops precipitously beyond
13.5 kpc, after a local peak in the Perseus arm \citep{1998ApJS..115..241H},
% Heyer et al. 2nd quadrant map.
as does the stellar density \citep{1996A&A...313L..21R},
% Ruphy et al.
thereby defining the far outer Galaxy, and the radius used in our
whole Galaxy quantities (\S \ref{wholegalaxy}).
The atomic surface density also begins to drop around 13-17 kpc,
following an exponential with scale length 3.74 kpc 
\citep{2009ARA&A..47...27K}.
Molecular clouds in the far outer galaxy are rare, but can be found 
with large-scale surveys.
Star formation does continue in the rare molecular clouds
\citep{1989A&AS...80..149W}, and Figure \ref{MWradial}
% Wouterloot and Brand
would suggest a higher ratio of star formation to molecular
gas, but there are issues of incompleteness in the outer galaxy.
A study of individual regions \citep{2002ApJ...578..229S} 
found a value of $\lfir/M$
similar to that for the inner Galaxy. They conclude that the
star formation process within a cloud is not distinguishable from 
that in the inner Galaxy; the low global rate of star formation is
set by the inefficient conversion of atomic to molecular gas in the
far outer Galaxy. This result is consistent with the results of
\citet{2011AJ....142...37S}
%Schruba et al,
for other galaxies, as exemplified by NGC 6946.

Inside about 13 kpc, the atomic surface density is roughly constant at
10-15 \msunpc, while the average molecular 
surface density increases sharply, passing 1 \msunpc\ somewhere
near the solar neighborhood \citep{2001ApJ...547..792D},
% Dame et al. 2001 "new survey"
to a local maximum around $\rgal =$ 4-5 kpc \citep{2006PASJ...58..847N}.
% Nakanishi and Sofue
After a small decrease inside 4 kpc, the surface density rises 
sharply within $\rgal = 1$ kpc.
\citet{2006PASJ...58..847N}
%Nakanishi and Sofue
and thus Figure \ref{MWradial} use a constant \xco, which could overestimate
the molecular mass near the center (\S \ref{gastracers}).

The innermost part of the \mw, $\rgal < 250$ pc, known as the Central 
Molecular Zone (CMZ), potentially provides an opportunity for a
close look at conditions in galactic nuclei without current AGN
activity, provided that the issues of non-circular motions, 
foreground and background confusion, and possible changes in 
\xco\ \citep{1998ApJ...493..730O}
% Oka et al. 1998
can be overcome.
Despite their conclusion that \xco\ is lower,
\citet{1998ApJ...493..730O}
% Oka et al 1998 CO 2-1 paper
argue that the CMZ has a molecular mass of 2-6\ee7 \msun.
Analysis of dust emission yields a mass of 5\ee7 \msun\ 
\citep{2000ApJ...545L.121P},
% Pierce-Price 2000
which leads to $\sigmagas = 250$ \msunpc\ for the CMZ.
A TIR luminosity of 4\ee8 \lsun\ \citep{1997ApJ...480..173S}
% Sodroski et al. 1997
for the CMZ would imply $\sfr \sim 0.06$ \msunyr\ using the conversion
in Table \ref{sfrequations}, roughly consistent with other 
recent estimates \citep{2012A&A...537A.121I},
% Immer et al. 2012
based on analysis of point infrared sources, but
new data from the surveys mentioned above should allow
refinement of these numbers.
Evidence for variations  in star formation rate 
on short timescales is discussed by \citet{2009ApJ...702..178Y}.
% Yusef-Zadeh et al. 2009
Using $\sfr = 0.06$ \msunyr\ and $\rgal = 0.25$ kpc for the CMZ, 
$\sigmasfr = 300$ \msun\ Gyr$^{-1}$ pc$^{-2}$.
These values of \sigmamol\ and \sigmasfr\ for the CMZ are
plotted separately in Figure \ref{MWradial} and identified as ``CMZ." 
If correct, the CMZ of the \mw\ begins to look a bit more like 
the inner few kpc of NGC 6946.
The CMZ provides an excellent place to test scaling relations, including
those for dense gas, in detail, if the complicating issues can be understood.
Preliminary results suggest that $\sigmasfr$ is similar to that expected
from \sigmamol\ (see Fig. \ref{MWradial}), but lower than expected
from dense gas relations (S. Longmore, personal communcation).

The properties of clouds and clumps also may vary with \rgal.
\citet{2009ApJ...699.1092H}
% Heyer et al. 2009
reanalyzed cloud properties in the inner Galaxy 
based on CO and \coo\ surveys. For the
area of the cloud defined roughly by the 1 K CO detection threshold, 
he found a median $\sigmamol(cloud) = 42$ \msunpc\ for clouds, 
substantially less than
originally found by \citet{1987ApJ...319..730S}, 
% Solomon et tal.
and only a few clouds
have $\sigmamol(cloud) > 100$ \msunpc. 
Analysis based on detection thresholds 
for \coo\ yield a median $\sigmamol(\coo\ {\rm clump}) = 144$ \msunpc\ 
for ``\coo\ clumps"
(\citealt{2010ApJ...723..492R}, who refer to them, however as clouds).
The mean in 0.5 kpc bins of surface density per \coo\ clump is roughly
constant at 180 \msunpc\ from $\rgal = 3$ to 6.6 kpc, beyond
which it drops sharply toward the much lower values in the solar neighborhood.
Assuming the region sampled is representative, \citet{2010ApJ...723..492R}
have plotted the azimuthally averaged surface density of \coo\ clumps
versus Galactocentric radius; it peaks at 2.5 \msunpc\ at 4.5 kpc,
declining to $<0.5$ \msunpc\ beyond 6.5 kpc. 
The clouds do appear to
be associated with spiral arms, so azimuthal averaging should be taken
with caution.
The cloud mass function of 246 clouds in the far outer galaxy
has a power-law slope of 
$\alpha_{cloud} = -1.88$, 
similar to but slightly steeper than that found inside the solar circle
(\S \ref{massfunctions})
and a maximum mass of about \eten4 \msun\
\citep{2002ApJ...578..229S}.
% Snell et al. clouds and IR properties
The lower value for the maximum mass of a cloud seems to
result in a concomitant limit on the number of stars formed in
a cluster, but no change is inferred for the intrinsic
IMF \citep{2000A&A...358..514C}.
% Casassus et al. 2000
A piece of a spiral arm at $R_{gal} = 15$ kpc has been recently identified in 
\hi\ and CO, containing a molecular cloud with $M = 5\ee4$ \msun\
\citep{2011ApJ...734L..24D}.
% Dame and Thaddeus.

\subsubsection{Non-axisymmetric Structure}

The molecular gas surface density shows a strong local maximum
around $\rgal = 4.5$ kpc in the northern surveys (Galactic quadrants
I and II), while southern surveys (quadrants III and IV) show a 
relatively flat distribution of \sigmamol\ from about 2 to 7 kpc
\citep{2000A&A...358..521B}. 
% Bronfman 2000
This asymmetry is likely associated with
spiral structure \citep{2006PASJ...58..847N} and a long (half-length
of 4.4 kpc) bar \citep{2005ApJ...630L.149B}.
% Benjamin 2005.
The star formation rate, measured from \lfir,  
does not, however, reflect this asymmetry; in fact, it is larger
in quadrant IV than in I, suggesting a star formation efficiency
up to twice as high, despite the lower peak \sigmamol\
 \citep{2000A&A...358..521B}.
% Bronfman 2000
Interestingly, the distribution of denser gas, traced by millimeter-wave
dust continuum emission \citep{2012ApJ...747...43B}, 
%Beuther et al. from ATLASGAL
is also symmetric on kpc scales, but asymmetries appear within the
CMZ region \citep{2010ApJ...721..137B}.
%Bally et al. BGPS
Major concentrations of molecular clouds with very active star formation
appear to be associated with regions of low shear \citep{2006ApJ...641..938L},
%Luna et al. 2006
and with the junction of the bar and spiral arms
\citep{2011A&A...529A..41N}.
%Nguyen Luong et al.

% milkyway
%demographics.tex
%080711 RCK (and many before)
% 081111 made a subsection by cutting out from Rob's exgal.tex
% 101011 RCK significant edits and trimming
% 101411 RCK updated references
% 101811 RCK citations
% 101911 NJE fixed latex problems
% 102411 NJE added fig refs
% 110411 NJE minor edits
% 110611 RCK minor edits
% 121211 NJE minor edits
% 012612 NJE updated note on point for Milky Way in Figure
% 012912 RCK edits in response to comments
% 020712 NJE minor edit
% 021212 RCK removed redundant text on SSFR with definitions.tex
% 031212 RCK edited text for new demographics figure
% 031212 NJE minor edits

\subsection{Demographics of Star-Forming Galaxies Today}\label{demographics}

The recent influx of multi-wavelength data 
has expanded the richness of information available on global
star formation properties of galaxies, and transformed the
interpretive framework from one based on morphological types
to a quantitative foundation based on galaxy luminosities, masses, and other
physical properties.  

The role of galaxy mass as a fundamental determinant of 
the star formation history of a galaxy has been long recognized
(e.g., \citealt{1996A&A...312L..29G}), 
%Gavazzi \& Scodeggio 1996), 
but data from SDSS and subsequent 
surveys have reshaped our picture of the population of star-forming
galaxies (e.g., \citealt{2003MNRAS.341...54K, 2004ApJ...600..681B,
2006MNRAS.373..469B, 2004MNRAS.351.1151B}). 
%Kauffmann et al. 2003; Baldry et al. 2004, 2006; Brinchmann et al. 2004).
The integrated colors of galaxies,
which are sensitive to their star formation histories, show a strongly
bimodal dependence on stellar mass, with a relatively tight ``red
sequence" populated by galaxies with little or no current star formation,
and a somewhat broader ``blue sequence" or ``blue cloud" of actively
star-forming galaxies.  The dominant population shifts from blue to red
near a transition stellar mass of $\sim 3 \times 10^{10}$\,\msun\ 
\citep{2003MNRAS.341...54K}.
%(Kauffmann et al. 2003).  
The relative dearth of galaxies in the ``green valley"
between the red and blue sequences suggests a deeper underlying
physical bimodality in the galaxy population {\bf and} a rapid evolution of
galaxies from blue to red sequences.  

\begin{figure}
\center
\includegraphics[scale=1.2]{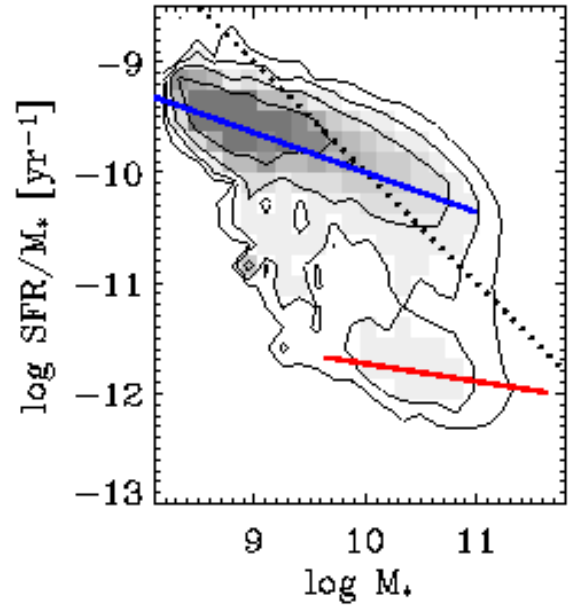}

\caption{
Relation between specific star formation rate ($\sfr/M_*$) 
and galaxy mass for galaxies
in the SDSS spectroscopic sample, and with SFRs measured from GALEX ultraviolet
luminosities.  Gray contours indicate the (1/Vmax-weighted) number distribution 
of galaxies in this bivariate plane.  The blue solid line shows the fit to the 
star-forming sequence fit, and the red line shows approximate position of 
non-star-forming red sequence on this diagram. The dotted line shows the locus
of constant $SFR\,=1\ M_{\odot }$ yr$^{-1}$.  Figure adapted from 
\citet{2007ApJS..173..315S}. Reproduced by permission of the AAS.
}

\label{galex}
\end{figure}

A similar bimodality characterizes the mass dependence of the 
SFR per unit galaxy mass (SSFR).  
The mass dependence of the SSFR has been explored by numerous investigators
(e.g., \citealt{2004MNRAS.351.1151B, 2007ApJS..173..267S, 
2007ApJS..173..315S, 2007ApJ...671L.113L}). 
%Brinchmann et al. 2004; Salim et al. 2007; Schiminovich et al. 2007;
%Lee et al. 2007
Figure \ref{galex}  shows an example from \citet{2007ApJS..173..315S}, 
%Schiminovich et al. (2007)
based in this case on dust-corrected FUV measurements of SDSS galaxies.  

%[Figure fig_galex here]

A clear separation between the blue and red sequences is evident,
and the dispersion of SSFRs within the blue sequence is surprisingly small,
suggesting that some kind of self-regulation
mechanism may be at work among the actively star-forming galaxies.  
The bimodality is not absolute; there is
a clear tail of less active but significantly star forming galaxies
between the two sequences.  These represent a combination of relatively
inactive (usually early-type) disk 
galaxies and unusually active spheroid-dominated systems (\S5.4).  The sharp
increase in the fraction of inactive galaxies above stellar masses of order
a few times 10$^{10}$\,\msun\ is also seen.   

The blue sequence in Figure \ref{galex} is not horizontal; the SSFR clearly
increases with decreasing galaxy mass.  The slope ($-0.36$ for the
data in Figure \ref{galex}) varies somewhat between
different studies, possibly reflecting the effects of different sampling
biases.  The negative slope implies that lower-mass galaxies are
forming a relatively higher fraction of their stellar mass today, and
thus must have formed relatively fewer of their stars (compared to
more massive galaxies) in the past.  The most straightforward explanation
is that the dominant star-forming galaxy population in the Universe
has gradually migrated from more massive to less massive galaxies over
cosmic time.  Direct evidence for this ``downsizing" is seen in observations
of the SSFR vs mass relation in high-redshift galaxies  
(e.g.,  \citealt{2007ApJ...660L..47N}).
%Noeske et al. 2007a, b).
 
\begin{figure}
\center
\includegraphics[scale=0.8, angle=-90]{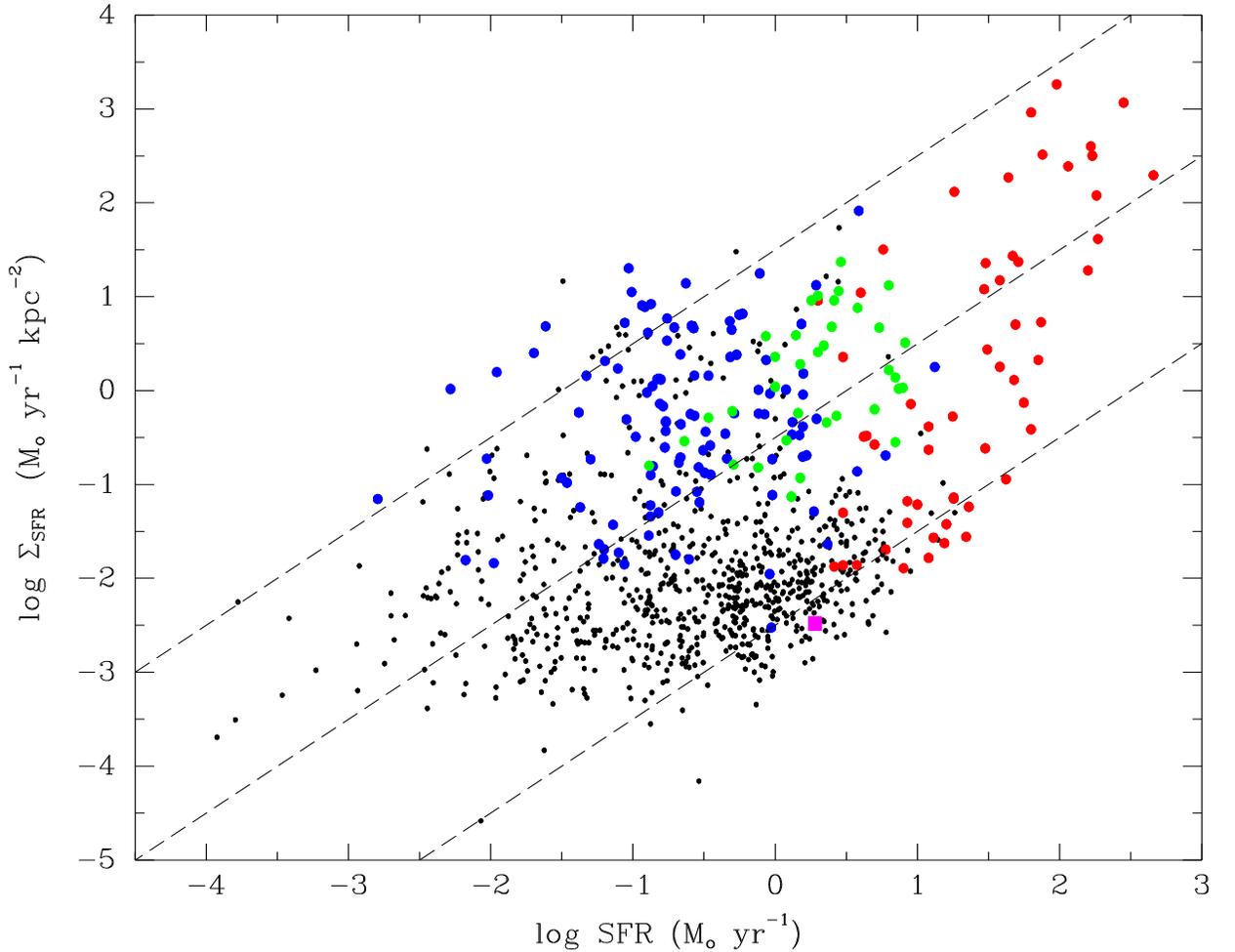}

\caption{
%\footnotesize
Distribution of integrated star formation properties of
galaxies in the local Universe.  Each point represents an individual
galaxy or starburst region, with the average SFR per unit area 
(SFR intensity) plotted as a function of the absolute SFR.  The 
SFR intensities are averaged over the area of the main star-forming
region rather than the photometric area of the disk.  Diagonal lines
show loci of constant star-formation radii, from 0.1 kpc (top) to 
10 kpc (bottom).  
Several sub-populations of galaxies are shown:
normal disk and irregular galaxies measured in \halpha\ and corrected
for dust attenuation (solid black points) from the surveys of
\citet{2003A&A...400..451G}, \citet{2004A&A...414...23J},
\citet{2005AJ....129.2597H}, and \citet{2008ApJS..178..247K}; 
luminous star-formation dominated infrared galaxies (LIRGs)
and ultraluminous infrared galaxies (ULIRGs), plotted as red points,
from \citet{2002ApJS..143...47D} and \citet{2000AJ....119..991S};
blue compact starburst galaxies measured in \halpha\ (blue points)
from \citet{2003ApJ...591..827P} and \citet{2005ApJ...627L..29G};
and circumnuclear star-forming rings
in local barred galaxies measured in Pa$\alpha$ (green points)
as compiled by \citet{2004ARA&A..42..603K}.
The \mw\ (magenta square) lies near the high end of normal spirals in total
star formation rate, but it is near the average in \sigmasfr\
(\S \ref{milkyway}).
The relative numbers of galaxies plotted does not reflect their relative 
space densities.
In particular the brightest starburst galaxies are extremely rare.
}
%
%Points plotted include normal disk and irregular galaxies 
%(solid black points), luminous and ultraluminous infrared galaxies
%(red points), blue compact starburst galaxies (blue points), and
%circumnuclear starbursts in local barred galaxies (green points).
%The position of the Milky Way
%is shown by the magenta square, using the values from \S \ref{milkyway}.  
%The relative numbers of galaxies of
%different types does not reflect the relative space densities.  In 
%particular the brightest starburst galaxies are extremely rare,
%and their numbers are grossly overestimated in this figure; the main
%purpose of the comparison is to illustrate the physical diversity of
%the present-day population of star-forming galaxies, as discussed in the text.}

\label{demographicsfig}
\end{figure}

Another instructive way to examine the statistical properties of 
star-forming galaxies is to compare absolute SFRs and 
SFRs normalized by mass or area.
Several interesting trends can be seen in
Figure \ref{demographicsfig}, which plots integrated 
measurements of the SFR per unit area as a function of the
absolute SFRs.  
The first is the extraordinary range in SFRs, more than seven orders of
magnitude, whether measured in absolute terms or normalized per unit
area or (not shown) galaxy mass.  Much of this range is contributed 
by non-equilibrium systems (starbursts).  
Normal galaxies occupy a relatively tight range of SFRs per unit area,
reminiscent of the tightness of the blue sequence when expressed in
terms of SSFRs.  The total SFRs of the quiescent star-forming galaxies
are also tightly bounded below a value of $\sim$20\,\msun\,yr$^{-1}$.
Starburst galaxies, and infrared-luminous and ultraluminous systems
in particular, comprise most of the galaxies in the upper 2--3 decades
of absolute SFRs and \sigmasfr.

\begin{figure}
\center
\includegraphics[scale=0.6]{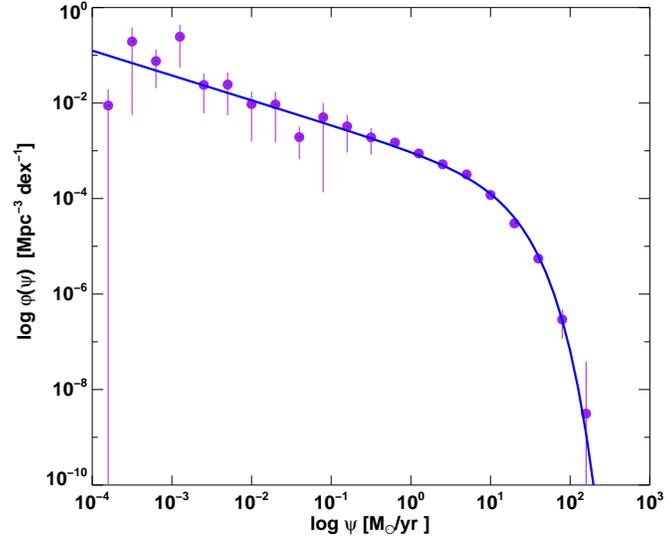}

\caption{Volume-corrected SFR distribution function for disk and
irregular galaxies in the local Universe, 
as derived from a combination of large flux-limited ultraviolet
and infrared catalogs and a multi-wavelength survey of the local 11 Mpc
volume.  Vertical lines indicate uncertainties due to finite sampling;
these are especially important for low SFRs, where the statistics 
are dominated by dwarf galaxies, which can only be observed over small volumes.
Statistics on gas-poor (elliptical, dwarf spheroidal) galaxies are not 
included in this study.  The blue line shows a maximum-likelihood 
Schechter function fit, as described in the text.  Figure is original,
but similar to one in
\citet{2011MNRAS.415.1815B}.}

\label{bothwell}
\end{figure}

The distribution of SFRs along the X-axis of Figure \ref{demographicsfig} 
(after correcting for volume completeness biases) is simply the
SFR distribution function (e.g., \citealt{1995ApJ...455L...1G,
2005ApJ...619L..59M}).
%Gallego et al. 1995, Martin et al. 2005b).  
Figure \ref{bothwell}  shows a recent determination of this distribution from
\citet{2011MNRAS.415.1815B}, 
%Bothwell et al. (2011), 
based on flux-limited UV and TIR samples
of galaxies and SFRs derived using the methods of 
\citet{2011ApJ...741..124H}, 
%Hao et al. (2011),
and corrections for AGN contamination following 
\citet{2010ApJ...723..895W, 2011ApJ...734...40W}.
%Wu et al. (2010, 2011).
The SFR function is well fitted by a \citet{1976ApJ...203..297S}
%Schechter (1976) 
exponentially
truncated power-law, with a faint-end slope $\alpha = -1.5$ and 
characteristic SFR, 
SFR$^*$ = 9\,\msun.  Although SFRs as high as $\sim$1000\,\msun\,yr$^{-1}$
are found in present-day ULIRGs, the contribution of LIRGs and ULIRGs
to the aggregate star formation today is small ($<10$\%\, also see 
\citealt{2011MNRAS.410..573G}).
%Goto et al. 2011).  
Steady-state star formation in galaxies dominates today. 
The value of SFR$^*$ increases  rapidly with $z$, and galaxies 
with $\lbol > 10^{11}$ \lsun\ (i.e., LIRGs) become dominant by redshifts $z > 1$
(e.g., \citealt{2005ApJ...632..169L}).  
%Le Floc'h et al. 2005).
This change partly reflects
an increase in merger-driven star formation at higher redshift, but
at early epochs even the steady-state star formation in massive
galaxies attained levels of order tens to hundreds of solar masses
per year, thus placing those galaxies in the LIRG and ULIRG regime.
This suggests that most of the decrease in the cosmic SFR in recent
epochs has been driven by downsizing in the level of steady-state
star formation (e.g., \citealt{2005ApJ...625...23B, 2009ApJ...697.1971J}).
% Bell et al. 2005, Jogee et al. 2009

% demographics
%starbursts.tex
%080711 RCK (and many before)
%081111 NJE made a separate file from Rob's section
%101011 RCK significant revision and trim
%101411 RCK updated references
%101811 RCK more citation insertions
%102411 NJE added figure ref
%110411 NJE minor edits
%110611 RCK minor edits
%121511 NJE minor edits
%020812 NJE minor edits, added ref to SMGs
%021212 RCK minor edits at beginning reflecting text move to definitions.tex
%031212 NJE minor edits
%050912 NJE update refs

\subsection{The High-Density Regime: Starbursts}
\label{starbursts}

As highlighted in \S \ref{definitions}, 
the starburst phenomenon encompasses a wide
range of physical scales and host galaxy properties, and no precise
physical definition of a starburst has been placed into wide use.  
%Although
%a universally adopted definition does not exist, the term is commonly
%applied to galaxies with SFRs which cannot be sustained for
%longer than a small fraction (e.g., $\le$10\%) of the Hubble time,
%i.e., with gas consumption timescales of less than $\ll$1 Gyr.
%Although many different classes of starburst galaxies have been
%defined, based largely on subjective criteria, spanning a very
%large range in absolute SFRs, nearly all share the common properties
%of high SFR surface densities ($\sigmasfr \ge 0.1$ \msunkpc,
%Fig. \ref{demographicsfig})  and (almost by definition)
%SSFRs of more than a few times $10^{-10}$\,yr$^{-1}$ (i.e.,
%birthrate parameter $b \gg 1$).  
Part of the explanation can be seen in Fig. \ref{demographicsfig};
although the most extreme starburst galaxies have properties that
are well separated from those of normal star-forming galaxies, 
there is no clear physical break in properties; starbursts instead
define the upper tails of the overall distributions in SFRs within
the general galaxy population.

Much of the attention in this area continues to focus on the
infrared-luminous and ultraluminous systems, because they probe
star formation in the most extreme high-density circumnuclear
environments seen in the local Universe.  A major breakthrough
in recent years has been the use of mid-infrared spectroscopy
to distinguish dust heated by massive stars from that heated
by a buried AGN (e.g., \citealt{1998ApJ...498..579G, 
2000A&A...359..887L, 2006ApJ...646..161D, 2007ApJ...656..148A}).
%Genzel et al. 1998, Laurent et al. 2000, Dale et al. 2006, Armus et al. 2007).
Regions heated by AGNs are distinguished
by the appearance of highly ionized atomic species, as well
as suppressed mid-infrared PAH emission relative to stellar-heated
dust.  Application of these diagnostics has made it possible 
to construct clean samples
of LIRGs and ULIRGs that are dominated by star formation.

With the Spitzer and Herschel observatories, it has been possible
to extend imaging and spectroscopy of the most luminous infrared-emitting
galaxies to intermediate redshifts 
(e.g., \citealt{2005A&A...434L...1E, 2007ApJ...658..778Y}).
%Elbaz et al. 2005, Yan et al. 2007).
As mentioned earlier the fraction of star formation in LIRGs and ULIRGs increases
sharply with redshift, but imaging and spectroscopy of these objects
reveals a marked shift in the physical characteristics of the population.
Whereas at low redshift, ULIRGs (and many LIRGs) are dominated by 
very compact circumnuclear starbursts triggered by mergers, at
higher redshifts the LIRG/ULIRG population becomes increasingly
dominated by large disk galaxies with extended star formation.
As a consequence of cosmic downsizing (i.e., ``upsizing" with increasing
redshift), by redshifts $z \sim 1$ the populations of LIRGs and ULIRGs
become increasingly dominated by the progenitors of present-day 
normal galaxies, rather than by the transient merger-driven starbursts
which dominate the present-day populations of ULIRGs, though examples
of the latter are still found, especially in the population of 
submillimeter-luminous galaxies (SMGs) 
(e.g., \citealt{2005ApJ...622..772C}).
% Chapman et al. 2005

Visible-wavelength observations of these high-redshift galaxies
with IFU instruments, both with and without adaptive optics correction,
have revealed many insights into the physical nature of this population
of starburst galaxies (e.g., 
\citealt{2008ApJ...687...59G, 2011ApJ...733..101G, 2009ApJ...706.1364F}
%Genzel et al. 2008, 2011; Forster-Schreiber et al. 2009
and references therein).  The 
H$\alpha$ kinematics show a wide range of properties, from normal,
differentially rotating disks to disturbed disks, and a subset
with signatures of ongoing or recent mergers.  The disks tend
to be characterized by unusually high velocity dispersions, which
have been interpreted as reflecting a combination of dynamical
instability and possibly energy injection into the ISM from young
stars.  Further discussion of high-redshift galaxies is beyond
the scope of this article, but we shall return to the cold ISM
properties and SFRs of these galaxies in \S6.

% starbursts
%earlytype.tex
%080711 RCK (and many before)
%081111 NJE made a separate file from Rob's exgal section
%101011 RCK revision and trim
%101411 RCK update/trim references
%101811 RCK citation edits
%110411 NJE minor edits
%110611 RCK minor edits
%112311 RCK major slimdown
%121211 NJE added note about outer MW comparison
%121611 NJE minor edits
%050912 NJE update refs and minor edits

\subsection{The Low SFR and Low Density Regimes}
\label{earlytype}

One of the most important discoveries from the GALEX
mission was the detection of low levels of star formation
in environments which were often thought to
have been devoid of star formation.  These include
early-type galaxies,
dwarf galaxies, low surface brightness galaxies, and the extreme
outer disks of many normal galaxies.

Although star formation had been detected occasionally
in nearby elliptical and S0 galaxies (e.g., \citealt{1993AJ....106.1405P}),
% Pogge \& Eskridge (1993)
most of these galaxies have historically been regarded
as being ``red and dead" in terms of recent star formation.
Deep GALEX imaging of E/S0 galaxies (e.g., \citealt{2007ApJS..173..619K}),
however, has revealed that approximately 30\%\ of these 
%galaxies exhim of 
early-type galaxies exhibit near-ultraviolet emission in
excess of what could reasonably arise from an evolved stellar
population (e.g., \citealt{2007ApJS..173..619K}).   
%Kaviraj et al. 2007).  
Confirmation of the star formation has come from visible-wavelength
IFU spectroscopy, as exploited for example by the SAURON survey
(\citealt{2010MNRAS.402.2140S}).
%Shapiro et al. 2010).
The SFRs in these galaxies tend to be very low, 
with at most 1--3\%\ of the stellar mass formed over the past Gyr.
High-resolution UV imaging with HST often reveals extended 
star-forming rings or spiral arms in these galaxies 
\citep{2010ApJ...714L.290S}.
%Salim \& Rich (2010)
Follow-up CO observations of these galaxies reveals significant
detections of molecular gas in a large fraction of the galaxies
with detected star formation, 
with molecular gas masses of $10^7 - 10^{10}$\,\msun\
(\citealt{2007MNRAS.377.1795C, 2011MNRAS.410.1197C}),
%(Combes et al. 2007, Crocker et al. 2011), 
and with SFRs
roughly consistent with the Schmidt law seen in normal galaxies.

The GALEX images also reveal that star formation in more
gas-rich disk galaxies often extends much farther in radius
than had previously been appreciated.  In exceptional cases
(e.g., NGC\,5236=M83, NGC\,4625) the star formation extends
to 3--4 times the normal $R_{25}$ optical radius, 
to the edge of the HI disk \citep{2005ApJ...619L..79T, 2005ApJ...627L..29G}.
%(Thilker et al. 2005, Gil de Paz et al. 2005).
A systematic study by \citet{2007ApJS..173..538T}
%Thilker et al. (2007)
reveals that extended ``XUV disks" are found in $\sim$20\%\ of spiral galaxies,
with less distinct outer UV structures seen in another 
$\sim$10\% of disk galaxies.  
Follow-up deep \halpha\ imaging and/or spectroscopy reveals 
extended disks of HII regions that trace the UV emission in
most cases (e.g., \citealt{2007AJ....134..135Z, 2010MNRAS.405.2549C,
2010MNRAS.405.2791G}),  
%Zaritsky \& Christlein 2007, Christlein et al. 2010, Goddard et al. 2010), 
confirming the earlier detection of
HII regions at large radii (e.g., 
\citealt{1998ApJ...506L..19F, 2004AJ....127.1431R}).
%Ferguson et al. 1998, Ryan-Weber et al. 2004).
Individual examples of extended UV features without \halpha\
counterparts are sometimes found, however 
\citep{2010MNRAS.405.2791G}.
%(Goddard et al. 2010).
The GALEX images also have led to a breakthrough in measurements
of star formation in low surface brightness spiral galaxies (LSBs).
These provide a homogeneous body of deep
measurements of the star formation and the nature of the 
star formation law at low surface densities 
(\S6; \citealt{2009ApJ...696.1834W}).
%(Wyder et al. (2009). 
Comparison to the outer \mw\  (\S \ref{milkyway}) suggests that
deep searches for molecular gas will find few clouds, which will
nonetheless, be the sites of star formation.

The other low-density star formation regime can be
found in dwarf irregular galaxies.  Drawing general inferences
about the star formation properties of these galaxies requires
large samples with well-defined (ideally volume-limited) selection
criteria, and studies of this kind (based on UV, visible, and/or
\halpha\ observations) have been carried out by several
groups (e.g., 
\citealt{2004AJ....127.2031K, 2005ApJ...631..208B, 
2006ApJS..165..307M, 2008ApJS..178..247K,
2009ApJS..183...67D, 2010AJ....139..447H}).
%Karachentsev et al. 2004, Blanton et al. 2005, 
%Meurer et al. 2005, Gil de Paz et al. 2007, Kennicutt et al. 2008, Dalcanton et al. 2009,
%Hunter et al. 2010).  
One perhaps
surprising result is the near ubiquity of star formation
in the dwarf galaxies. \citet{2008ApJS..178..247K}  
%Kennicutt et al. (2008) 
observed or 
compiled H$\alpha$ luminosities for galaxies within the local
11 Mpc, which included $\sim$300 dwarf galaxies (M$_B > -17$).
Excluding a handful of dwarf spheroidal galaxies which have
no cold gas, only 10 of these ($\sim$3\%) were not detected in H$\alpha$,
meaning that star formation has taken place over the last 3--5 Myr
in the other 97\%\ of the systems.  Moreover many of the handful
of \halpha\ non-detections show knots of UV emission, demonstrating
that even fewer of the galaxies have failed to form stars over
the last $\sim$100 Myr \citep{2011ApJS..192....6L}.
%(Lee et al. 2011).  
Recent star formation is seen in
all of the galaxies with M$_B < -13$ and 
$M(HI) > 5 \times 10^7$\,\msun.  For less massive galaxies
it is possible for star formation to cease for timescales that
are longer than the ionization lifetime of an HII region
(up to 5 Myr), but examples of galaxies with extended periods
of no star formation are extremely rare.  

These data also provide a fresh look at the temporal properties
of the star formation in dwarf galaxies generally.  The
distribution of SSFRs shows a marked increase
in dispersion for galaxies with M$_B > -14.5$ or circular
velocity of 50\,km\,s$^{-1}$ 
\citep{2007ApJ...671L.113L, 2009MNRAS.400..154B}.
%(Lee et al. 2007, Bothwell et al. 2009).
The corresponding neutral gas fraction does not show
a similar increase in dispersion \citep{2009MNRAS.400..154B},
%(Bothwell et al. 2009), 
so this suggests
an increased short-term fluctuation of integrated SFRs in 
low-mass galaxies.  \citet{2009ApJ...692.1305L}
%Lee et al. (2009) 
analyzed the statistics
of the SFRs in more depth and confirmed a larger fraction of
stars formed in bursts in low-mass galaxies ($\sim$25\%),
but these still represent only a small fraction of the total
stars formed-- even in the dwarfs the majority of stars 
appear to form in extended events of duration longer than
$\sim$10 Myr.  Recent modelling of the statistics of \halpha\
and UV emission by 
\citet{2012ApJ...744...44W} and \citet{2011ApJ...741L..26F}
%Weisz et al. (2012) and Fumagalli et al. (2011)
as well as analysis of resolved stellar populations
(e.g., \citealt{2008ApJ...689..160W})
%Weisz et al. (2008)
confirm the importance of fluctuations in the SFR.

% earlytype

\section{STAR FORMATION RELATIONS}\label{sflobs}

%sflobs.tex
%081111 broken out of Rob's exgal.tex to make first sub of new section
%081111 NJE a few edits
% 082511 NJE moved discussion of Liu from spec.tex to here
% 083111 NJE added discussion of Schmidt law in MW
% 090811 NJE added note re dense gas to disc. of Daddi et al.
% 090911 NJE added note on SMC work
% 092311 RCK full edit
% 100711 NJE minor fixes
% 101211 RCK full scrub and trim
% 101411 RCK edit/trim references
% 101911 RCK add bibtex citations
% 101911 NJE fixed latex problems
% 102111 NJE changed Orion to refer to Chomiuk paper
% 102411 NJE added figure refs
% 110411 NJE minor edits
% 110611 RCK minor edits
% 121111 RCK edits/trims following ARAA review
% 121211 NJE minor edits, used macro for X(CO), need to say what value
% 121211 NJE dealt with some RG comments, including support for change in X
% 121911 NJE inserted some text from demographics.tex
% 012312 NJE edited after comments
% 012612 NJE updated refs back to MW section
% 012912 RCK edits in response to comments
% 020112 NJE further edits, minor fixes, dealing with dense gas issues
% 020612 RCK minor edits, addressing NJE comments
% 020712 NJE minor edits
% 021212 RCK added mention of new figure
% 021412 NJE worked in figure reference
% 021712 NJE minor edits
% 031212 RCK edits for revised figure and NJE suggestions
% 031212 NJE
% 050512 NJE updated references
% 050912 NJE updated references, minor edits
% 051412 NJE minor updates, HI, HII using ions

%\subsection{Star Formation Relations on Galactic Scales}\label{sflobs}

The immense dispersion in SFR properties seen in Figure \ref{demographicsfig}
collapses
to a remarkably tight scaling law when the SFR surface densities (\sigmasfr)
are plotted against mean gas surface densities (\sigmagas)
\citep{1998ApJ...498..541K}.  
This emergent order reflects the fact that gas is the input
driver for star formation.  
The concept of a power-law relation between
SFR density and gas density 
dates to \citet{1959ApJ...129..243S, 1963ApJ...137..758S},
%Schmidt (1959, 1963)
and relations of this kind are commonly referred to as 
``Schmidt laws."
%\footnote{The term ``Kennicutt-Schmidt law" or KS law
%is also in common use.}
On physical grounds we might expect the most fundamental relation between
the volume densities of star formation and gas, but since most observations
of external galaxies can only measure surface densities integrated along
the line of sight, the most commonly used relation, often called a
Kennicutt-Schmidt (KS) law, is in terms of 
surface densities.

\begin{equation}
\Sigma_{SFR} = A\, \Sigma{{^N}{_{gas}}} 
\end{equation}

\noindent
The precise form of this relation
depends on assumptions about how \sigmagas\ is derived from the
observations (\S \ref{gastracers}), 
but a strong correlation is clearly present.

\subsection{The Disk-Averaged Star Formation Law}\label{avgsfl}

\begin{figure}
\center
%\begin{center}
%\begin{tabular}[htp]{ll}
%\includegraphics*[totalheight=5.0cm, viewport=20 355 540 700]{fig_Schmidt_global.eps} &
%\includegraphics*[totalheight=5.0cm, viewport=20 355 540 700]{fig_gao.eps} \\
%\includegraphics*[scale=0.6]{fig_Schmidt_global.eps} &
%\includegraphics*[scale=0.4]{fig_gao.eps} \\
%\end{tabular}
%\end{center}
\includegraphics*[scale=0.6]{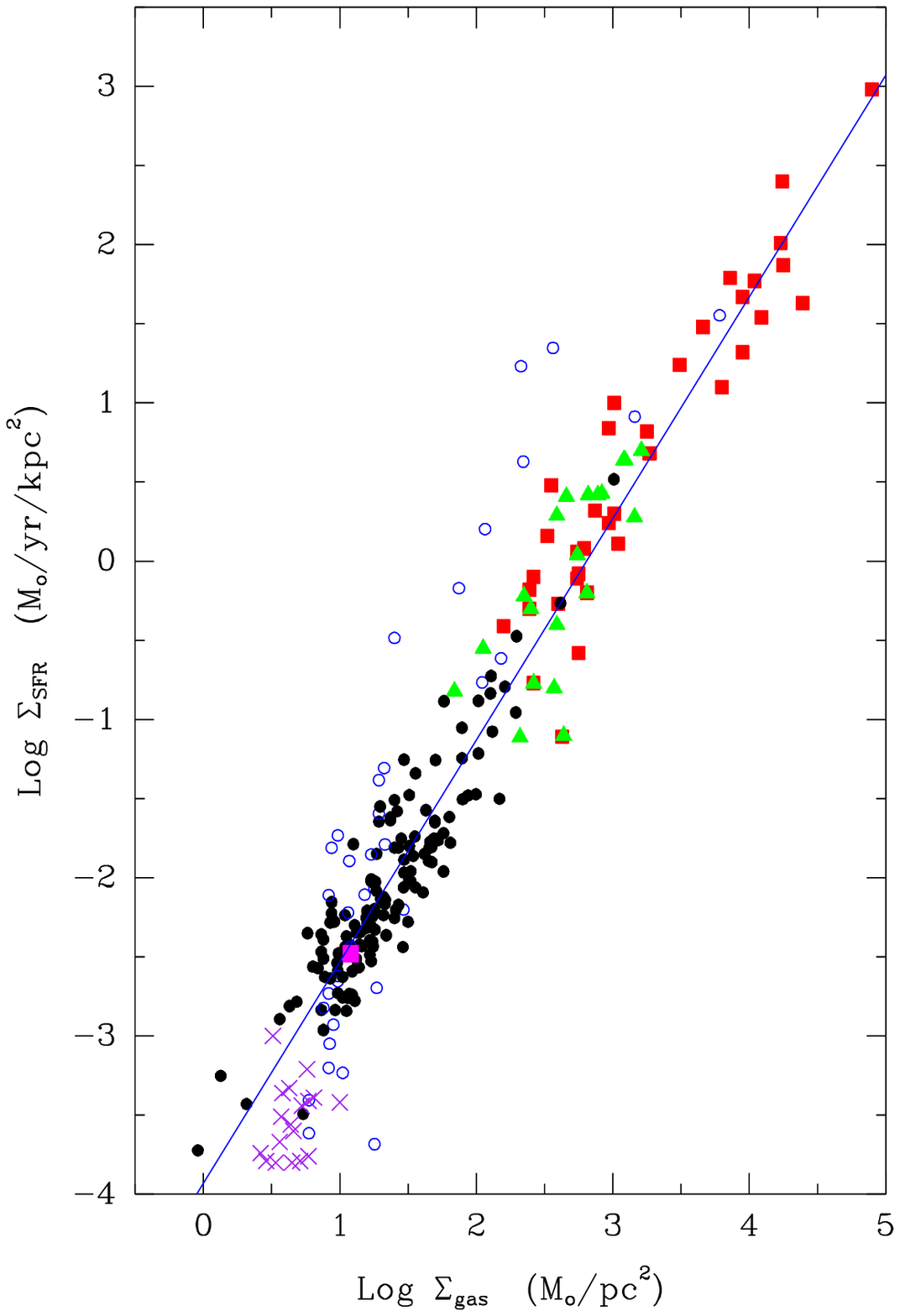} \\
\includegraphics*[scale=0.4]{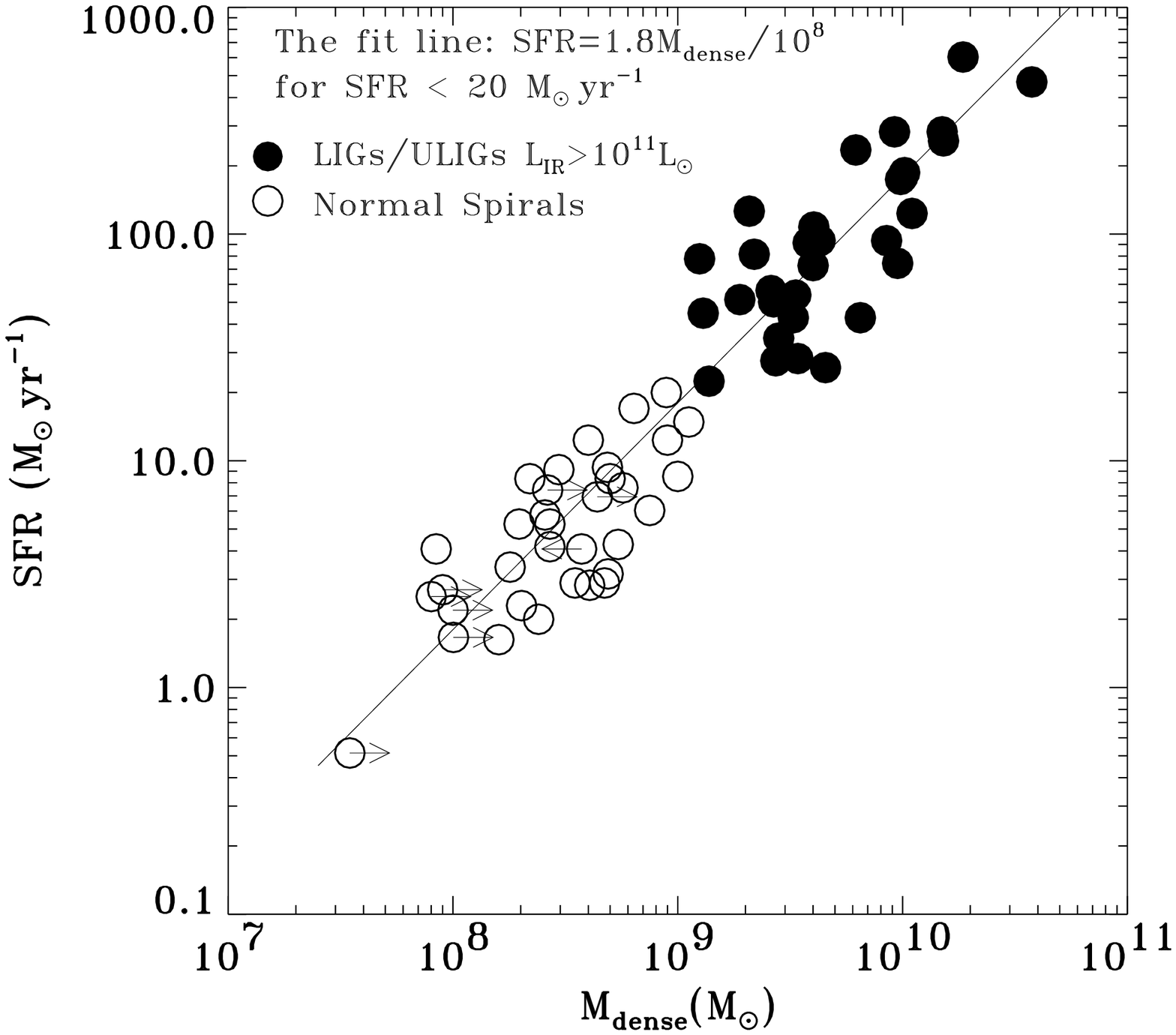} 

\caption{ 
%\footnotesize 
Upper:  Relationship between the disk-averaged surface 
densities of star formation and gas (atomic and molecular) for
different classes of star-forming galaxies.  Each point represents
an individual galaxy, with the SFRs and gas masses normalized to 
the radius of the main star-forming disk.  
The line shows the original $N$ = 1.4 fit from 
\citet{1998ApJ...498..541K}
%Kennicutt (1998b)
superimposed on the data.  Most of the galaxies form a 
tight relation (exceptions discussed below), and with the
improved dataset, even the normal galaxies (shown with black
points) follow a well-defined Schmidt law on their own.
The dispersion of the normal galaxies overall from the average relation
($\pm$0.30 dex rms) is considerably higher than can be attributed
to observational uncertainties, which suggests that
much of the dispersion is physical.  The Milky Way
(magenta square) fits well on the main trend seen for
other nearby normal galaxies.  The purple crosses show data for
low surface brightness galaxies.
%Colors are similar to Figure \ref{demographicsfig}, 
%with black points representing normal spiral and irregular
%galaxies, red points infrared-selected starburst galaxies (mostly
%LIRGs and ULIRGs, and green points denoting circumnuclear starbursts
%with SFRs measured from Pa\,$\alpha$ measurements. 
%The \mw\ appears as a magenta square. Purple crosses represent
%nearby low surface brightness galaxies, as described in the text.
%Open blue circles denote low-mass irregular and starburst galaxies  
%with estimated metal (oxygen) abundances less than 0.3\,$Z_\odot$.
%For this plot a constant $X(CO)$ factor was applied to all galaxies.
%The line shows a fiducial relation with slope $N = 1.4$ (not
%intended as a fit to these data).
The sample of galaxies has been enlarged
from that studied in \citet{1998ApJ...498..541K}, with many improved
measurements as described in the text.
Lower:  Corresponding relation between the total (absolute) SFR and
the mass of {\it dense} molecular gas as traced in HCN.  The line
is a linear fit, which contrasts with the non-linear fit in the upper
panel.  Figure adapted from \citet{2004ApJ...606..271G}. 
Reproduced by permission of the AAS.
}

\label{schmidt3}
\end{figure}

K98 presented a review of observations of the Schmidt law up to
the time, and nearly all of that work characterised the relation
between the disk-averaged SFR and gas surface densities in galaxies.
The upper panel of Figure \ref{schmidt3} 
presents an updated version of 
the global Schmidt
law in galaxies.  Each point is an individual galaxy (color
coded as explained in the caption), with
the surface density defined as the total gas mass (molecular
plus atomic) or SFR normalized to the radius of the main
star-forming disk, as measured from H$\alpha$, Pa$\alpha$,
or IR maps.  For simplicity, a constant \xco\ factor [$2.3\ee{20}$
cm$^{-2}$ (K km s$^{-1}$)$^{-1}$ with no correction for helium, 
see \S \ref{gastracers}]
has been applied to all of the galaxies; the consequences 
of a possible breakdown in this assumption are discussed later.
The sample of galaxies has been enlarged
from that studied in \citet{1998ApJ...498..541K},
%Kennicutt (1998b), 
and all of the 
H$\alpha$-based SFR measurements have been improved by
incorporating individual (IR-based) corrections for dust
attenuation and [\ion{N}{2}] contamination.  
%The line shows the original $N$ = 1.4 fit from 
%\citet{1998ApJ...498..541K}
%Kennicutt (1998b)
%superimposed on the data.  Most of the galaxies form a 
%tight relation (exceptions discussed below), and with the
%improved dataset, even the normal galaxies (shown with black
%points) follow a well-defined Schmidt law on their own.
%The dispersion of the normal galaxies overall from the average relation
%($\pm$0.30 dex rms) is considerably higher than can be attributed
%to observational uncertainties, which suggests that
%much of the dispersion is physical.  The Milky Way 
%(magenta square) fits well on the main trend seen for
%other nearby normal galaxies.  The purple crosses show data for 
%low surface brightness galaxies, as discussed below.

The form of this integrated Schmidt law appears to be surprisingly
insensitive to SFR environment and parameters such
as the atomic versus molecular fraction, but some metal-poor
galaxies (defined as $Z < 0.3 Z_\odot$) deviate
systematically from the main relation, as shown by the
blue open circles in Figure \ref{schmidt3} (upper panel).  
These deviations could 
arise from a physical change in the star formation
law itself, but are more likely to reflect a breakdown in 
the application of a constant \xco\ factor 
(\S2.4; \citealt{2011ApJ...737...12L}, 
%Leroy et al. 2011, 
and references therein).
Adopting higher values of \xco\ for metal-poor galaxies 
brings the 
galaxies much more into accord with the
main relation in Figure \ref{schmidt3}.

Recent observations of LSBs by \citet{2009ApJ...696.1834W} 
%Wyder et al. (2009) 
extend the measurements of the integrated
star formation law to even lower mean surface densities, 
as shown by the purple crosses in the upper panel of Figure \ref{schmidt3}.  A  
clear turnover is present, which is consistent with 
breaks seen in spatially-resolved observations of the star
formation law (\S6.2).

The slope of the integrated Schmidt law is 
non-linear, with $N \simeq$ 1.4--1.5 \citep{1998ApJ...498..541K}, 
%(Kennicutt 1998b)
when a constant \xco\ factor is applied.  This uncertainty
range does not include all possible systematic errors, arising
for example from changes in \xco\ or the IMF with 
increasing surface density or SFR.  Major systematic changes in either
of these could easily change the derived value of $N$ by as
much as 0.2--0.3.   For example, if \xco\ were five times
lower in the dense starburst galaxies (\S \ref{gastracers}),
the slope of the overall
Schmidt law would increase from 1.4--1.5 to 1.7--1.9 
\citep{2012MNRAS.421.3127N}.
% Narayanan et al. 2012
 
Usually the Schmidt law is parametrized in terms of the 
total (atomic plus molecular) gas surface density, but
one can also explore the dependences of the disk-averaged
SFR densities on the mean atomic and molecular surface densities
individually.  Among normal galaxies with relatively low
mean surface densities, the SFR density is not particularly well
correlated with either component, though variations in \xco\ 
could partly explain the poor correlation between SFR and derived
H$_2$ densities (e.g., \citealt{1998ApJ...498..541K}).
%Kennicutt 1998b).  
In starburst galaxies
with high gas surface densities, however, the gas is overwhelmingly
molecular, and a strong non-linear Schmidt law is observed
(upper panel of Figure \ref{schmidt3}). 
 
A similar non-linear dependence is observed for 
total SFR (as opposed to SFR surface density) versus total
molecular gas mass 
(e.g., \citealt{1988ApJ...334..613S, 2004ApJ...606..271G}),
%Solomon \& Sage 1988, Gao \& Solomon 2004), 
and presumably is another manifestation of the same underlying physical 
correlation.
The dependence of the SFR on {\it dense} molecular gas mass
is markedly different, however.
The lower panel of Figure \ref{schmidt3}, 
taken from \citet{2004ApJ...606..271G},
%Gao \& Solomon (2004),
shows the relation between the integrated SFRs and the
dense molecular gas masses, as derived from HCN \jj10\ measurements 
(\S \ref{gastracers}) for a sample of normal and starburst galaxies.  
In contrast to the correlation with total molecular mass from CO \jj10,
this relation is linear, implying
a strong coupling between the masses of dense molecular
clumps and stars formed, which is largely independent
of the galactic star-forming environment.   
\citet{2005ApJ...635L.173W}
%Wu et al. (2005) 
have subsequently shown that this linear  
relation extends down to the scales of individual star-forming
molecular clouds and dense clumps in the Galaxy. 
%Taken together, 
%these observations appear to  validate 
%NJE added something to fix syntax
Combined with the \mw\ studies (\S \ref{local}),
these dense gas relations for galaxies suggest that
dense clumps are plausible fundamental star-forming units.
If so, the mass fraction of the ISM (and 
fraction of the total molecular gas) residing in dense 
clumps must itself increase systematically with the SFR.

Because \lhcn\ in the \jj10\ line is only an approximate tracer for dense gas
(\S \ref{densegas})
and because the definition of a dense clump is by no means precise
(\S \ref{massfunctions}),
%RCK minor edit below
one should be careful not to overinterpret these trends.
Similar comparisons of the SFR with the emission from other
molecular tracers suggest that the slope of the SFR versus gas
mass relation changes continuously as one proceeds from lower-density
tracers such as CO \jj10\ to higher-excitation (and higher-density)
CO transitions, and further to high-density tracers such as HCN and
\hcop\ and to higher transitions of those molecules 
(e.g., \citealt{2009ApJ...707.1217J}).
%Juneau et al. 2009 and references therein).
% NJE removed as redundant
%This result may have a bearing on the interpretation of Schmidt laws
%measured with different rotational transitions of CO as well.
Multi-transition studies, along with realistic modeling will
help to refine the interpretation of line luminosities of
dense gas tracers (e.g., \citealt{2008A&A...479..703G, 2009ApJ...707.1217J}).
% Gracia-Carpio 2008, Juneau 2009

The data shown in Figure \ref{schmidt3} all come from observations
of nearby galaxies ($z < 0.03$), but recently a number of studies
have addressed the form of the molecular gas Schmidt law for 
starburst galaxies extending to redshifts $z \geq 2$ 
(e.g., \citealt{2007ApJ...671..303B, 2010ApJ...714L.118D, 2010MNRAS.407.2091G}).
%Bouche et al. 2007; Daddi et al. 2010; Genzel et al. 2010).
The interpretation of
these results is strongly dependent on the assumptions made about \xco\ 
in these systems.  When a Galactic \xco\ conversion
is applied, the high-redshift galaxies tend to fall roughly on 
the upper parts of the Schmidt law seen locally (e.g., Figure \ref{schmidt3}).
However if a lower \xco\ factor is applied to the
most compact starburst and submillimeter galaxies (SMGs), 
as  suggested by many independent analyses of \xco\ (\S 2.4),
the Schmidt relations shift leftwards  by the same factor, 
forming a parallel relation \citep{2010ApJ...714L.118D, 2010MNRAS.407.2091G}.
%(Daddi et al. 2010, Genzel et al. 2010).

Taken at face value, these results suggest the presence of two
distinct modes of star formation with different global efficiencies,
which separate the extended star-forming disks of normal galaxies 
from those in the densest circumnuclear
starbursts; this distinction is likely to be present both
in the present-day Universe and at early cosmic epochs. 
This inferred bimodality, however, is a direct consequence of
the assumption of two discrete values for \xco\ in the two
modes.  A change in the interpretation of CO emission is
certainly plausible when the derived molecular surface density
is similar to that of an individual cloud ($\sigmamol > 100$ \msunpc),
but the behavior  of \xco\ may be complex (\S \ref{gastracers}).
Variation of  \xco\  over a continuous range  would indicate
a steeper Schmidt law, rather than
a bimodal law \citep{2012MNRAS.421.3127N}.
% Narayanan et al. on smoothly varying X(CO)

\citet{2010ApJ...714L.118D}
%Daddi et al. (2010) 
also note
that the higher concentration of gas in the ULIRGs/SMGs
is consistent with an enhanced fraction of dense gas and
the dense gas relations of \citet{2004ApJ...606..271G}.
%Gao and Solomon (2004).
However, the star formation rate also appears to 
be larger for a given mass of dense gas in extreme starbursts 
(\citealt{2012A&A...539A...8G} and references therein),
%Garcia-Burillo et al. on HCN studies
especially when a lower value of $\alpha_{\rm HCN}$ is used.
As with CO, one can interpret these as bimodal relations
or as steeper than linear dependences on the gas tracer lines.
These results, together with evidence for {\it lower} SFR per
mass of dense gas in the CMZ of the \mw\ (\S \ref{milkyway})
warn against an overly simplistic picture of dense clumps
as the linear building blocks for massive star formation, with
no other variables in the picture.

The correlations between SFR and gas surface densities (and masses)
are not the only scaling laws that are
observed.  \citet{1998ApJ...498..541K}
%Kennicutt (1998b) 
pointed out that the 
SFR surface densities also correlate tightly with the ratio
of the gas surface density
to the local dynamical time, defined in that case to
be the average orbit time.  This prescription is 
especially useful for numerical
simulations and semi-analytical models of galaxy 
evolution.  Interestingly, 
\citet{2010ApJ...714L.118D}
%Daddi et al. (2010)
and \cite{2010MNRAS.407.2091G}
%Genzel et al. 2010
found that the bifurcation of Schmidt laws
between normal galaxies and ULIRGs/SMGs described above
does not arise in the dynamical form of this relation. 
 
\citet{2006ApJ...650..933B} 
%Blitz \& Rosolowsky (2006) 
have discovered another strong scaling
relation between the ratio of molecular to atomic
hydrogen in disks and the local hydrostatic pressure. 
The relation extends over nearly three orders of magnitude
in pressure and H$_2$/\hi\ ratio and is nearly linear
(slope = 0.92).  Technically speaking this scaling
relation only applies to the phase balance of cold gas rather than the
SFR, but it can be recast into a predicted
star formation law if assumptions are made about the
scaling between the SFR and the molecular
gas components (e.g., \citealt{2006ApJ...650..933B, 2008AJ....136.2782L}).
%Blitz \& Rosolowsky 2006; Leroy et al. 2008).   
Recent work by \citet{2010ApJ...721..975O} and \citet{2011ApJ...731...41O}
provides a theoretical explanation for these relations.
 
Finally, a number of workers have 
explored the scaling between SFR surface density
and a combination of gas and stellar surface densities.
For example \citet{1985ApJ...295L...5D} and \citet{1994ApJ...430..163D}
%Dopita (1985) and Dopita \& Ryder (1994) 
proposed a scaling between
the SFR density and the product of gaseous and stellar
surface densities; the latter scales with the disk 
hydrostatic pressure and hence bears some relation to
the picture of \citet{2006ApJ...650..933B}.
%Blitz \& Rosolowsky (2006).  
More recently \citet{2011ApJ...733...87S} 
%Shi et al. (2011) 
show that the scatter in the star formation law is minimized
with a relation of the form 
$\Sigma_{SFR} \propto \Sigma_{gas} \, \Sigma_{*}^{0.5}$.

Determining which of these different formulations
of the star formation law is physical and which
are mere consequences of a more fundamental
relation is difficult to determine from observations alone.
Some of the degeneracies between these various relations can be understood
if most gas disks lie near the limit of gravitational stability
($Q$\,$\sim$\,1; \citealt{1989ApJ...344..685K}), 
%Kennicutt 1989), 
and the critical column
densities for the formation of cold gas phase, molecule formation,
and gravitational instability lie close to each other 
(e.g., \citealt{1994ApJ...435L.121E, 2004ApJ...609..667S}).
%Elmegreen \& Parravano 1994, Schaye 2004).
In such conditions it can be especially difficult
to identify which physical process is most important
from observations.  We return to this
topic later.

We conclude this section by mentioning that there are other
SFR scaling laws that can be understood as arising from an
underlying Schmidt law.  The best known of these is a 
strong correlation between characteristic dust attenuation
in a star-forming galaxy and the SFR itself, with the consequence
that galaxies with the highest absolute SFRs are nearly all dusty 
infrared-luminous and ultraluminous galaxies 
(e.g., \citealt{1996ApJ...457..645W, 2005ApJ...619L..59M, 2011MNRAS.415.1815B}).
%Wang \& Heckman 1996; Martin et al. 2005b, Bothwell et al. 2011).
This opacity versus SFR relation is partly a manifestation of the Schmidt
law, because we now know that the most intense star formation in galaxies
takes place in regions with abnormally high gas surface densities,
and thus also in regions with abnormally high dust surface densities.
The other factor underlying the SFR versus opacity correlation is the prevalence of highly 
concentrated circumnuclear star formation in the most intense starbursts
observed in the present-day universe (\S \ref{starbursts}); this may
not necessarily be the case for starburst galaxies at early cosmic epochs.

%\subsection{Azimuthally-Resolved Observations}\label{radialsfl}
\subsection{Radial Distributions of Star Formation and Gas}\label{radialsfl}
%NJE My suggested title.

Over the past few years major progress has been made in characterizing
the spatially-resolved star
formation law within individual galaxies.  This work has  
been enabled by large multi-wavelength surveys of nearby galaxies
such as the Spitzer Infrared Nearby Galaxies Survey (SINGS)
\citep{2003PASP..115..928K}, 
%Kennicutt et al. 2003), 
the GALEX Nearby Galaxies Survey \citep{2007ApJS..173..185G},
%(Gil de Paz et al. 2007), 
the Spitzer/GALEX Local Volume
Legacy survey (LVL; \citealt{2009ApJ...703..517D, 2011ApJS..192....6L}), 
%Dale et al. 2009, Lee et al. 2011),
and the Herschel KINGFISH Survey (Kennicutt et al. 2011).
The resulting datasets provide the means to measure spatially-resolved 
and dust-corrected SFRs across a wide range of galaxy properties
(\S \ref{sfrdiagnostics}).
These surveys in turn have led to large spinoff surveys in 
\hi\ (e.g., THINGS, \citealt{2008AJ....136.2563W};
%Walter et al. 2008; 
FIGGS, \citealt{2008MNRAS.386.1667B}; 
%Begum et al. 2008; 
Local Volume \hi\ Survey; \citealt{2010ASPC..421..137K};
%Koribalski 2010; 
LITTLE THINGS, \citealt{2007AAS...211.9506H})
%Hunter et al. 2007) 
and in CO (e.g., BIMA-SONG, \citealt{2003ApJS..145..259H};
%Helfer et al. 2003; 
IRAM HERACLES, 
\citealt{2009AJ....137.4670L};
%Leroy et al. 2009; 
JCMT NGLS, \citealt{2009ApJ...693.1736W};
%Wilson et al. 2009; 
CARMA/NRO Survey, \citealt{2009AAS...21348504K};
%Koda et al. 2009, Momose 2011;
CARMA STING, \citealt{2011ApJ...730...72R, 2012ApJ...745..183R}).
%Rahman et al. 2011, 2012).  

The capabilities of these new datasets are illustrated in
Figure \ref{n6946optir}, which shows Spitzer 24\,$\mu$m, \halpha, 
VLA \hi, and CO (in this case from HERACLES) maps of NGC\,6946
(see Figure \ref{MWradial} for the corresponding radial
distributions of gas and star formation).
 
The first step in exploiting the spatial resolution of
the new observations is to analyze the azimuthally-averaged
radial profiles of the SFR and gas components
(as illustrated for example in Figure \ref{MWradial})
and the resulting SFR versus gas surface density correlations.
% RCK added reference to Fig 6
This approach has the
advantage of spatially averaging over large physical areas, which helps
to avoid the systematic effects that are introduced on smaller
spatial scales (\S3.9).  Radial profiles also have limitations arising
from the fact that a single radial point represents an
average over sub-regions with often wildly varying local gas and SFR densities, 
and changes in other radially varying physical 
parameters may be embedded in the derived SFR versus gas density relations.  

Large-scale analyses of the star formation law derived in this
way have been carried out by numerous authors 
(e.g., \citealt{1989ApJ...344..685K, 2001ApJ...555..301M, 2002ApJ...569..157W,
2003MNRAS.346.1215B, 2004ApJ...602..723H, 2005PASJ...57..733K, 
2007A&A...461..143S, 2008AJ....136.2782L, 2011AJ....142...37S,
2010A&A...522A...3G}). 
%Kennicutt (1989), Martin \& Kennicutt (2001), 
%Wong \& Blitz (2002), Boissier et al. (2003), Heyer et al. (2004), 
%Komugi et al. (2005), Schuster et al. (2007), 
%Leroy et al. (2008), Verley et al. (2010), 
%Schruba et al. (2011), Gratier et al. (2010), 
A strong correlation between SFR and gas surface density is
seen in nearly all cases, with a non-linear slope when plotted
in terms of total (atomic$+$molecular) density.  The best fitting
indices $N$ vary widely, however, ranging from 1.4--3.1 for the
dependence on total gas density and 
1.0--1.4 for the dependence on molecular gas surface
density alone (for constant \xco ).  
The measurements of the \mw\ range from $N = 1.2\pm0.2$, when
only molecular gas is used \citep{2006ApJ...641..938L},
%Luna et al. (2006), 
to $2.18\pm0.20$ when total (atomic and molecular) gas is used
\citep{2006A&A...459..113M}.
% Misiriotis

Some of the differences between these results can
be attributed to different schemes for treating dust attenuation, different
CO line tracers, different metallicities (which may
influence the choice of \xco), and different fitting methods.
The influence of a  metallicity-dependent \xco\ factor
is investigated explicitly by \citet{2003MNRAS.346.1215B}.
%Boissier et al. (2003).  
Since any variation in \xco\ is likely to flatten the radial molecular
density profiles, a variable conversion factor tends to steepen the
slope of the derived Schmidt law; for the prescription they use, the best
fitting slope can increase to as high as $N = 2.1 - 3.6$, underscoring
once again the critical role that assumptions about \xco\  play in
the empirical determination of the star formation law. 
Despite these differences in methodology, real physical variation
in the SFR versus gas density relation on these scales cannot be ruled out.

This body of work also confirms the presence of a turnover or threshold
in the star formation relation at low surface densities in many
galaxies.  Early
work on this problem (e.g., 
\citealt{1989ApJ...344..685K, 2001ApJ...555..301M, 2003MNRAS.346.1215B})
% Kennicutt 1989, Martin \& Kennicutt 2001, Boissier et al. 2003)
was based on radial profiles in \halpha, 
and some questions have been raised about whether these thresholds
resulted from breakdowns in the SFR versus \halpha\ calibration in low
surface brightness regimes, as opposed to real thresholds in the SFR.
Subsequent comparisons of UV and \halpha\ profiles appear to
confirm the presence of radial turnovers in the SFR in most disks,
however, even in cases where lower levels of star formation persist
to much larger radii (e.g., 
\citealt{2007ApJS..173..538T, 2010MNRAS.405.2549C, 2010MNRAS.405.2791G}).
%Thilker et al. 2007, Christlein et al. 2010, Goddard et al. 2010). 
% RCK new sentence
This extended star formation can be seen in Figure \ref{MWradial} 
($R > 7 - 8$ kpc).
Recent studies of nearby galaxies with unusually extended XUV disks
also show that the UV-based SFR falls off much more rapidly than the cold
gas surface density, with global star formation
efficiencies ($\epsilon^{\prime}$) an order of magnitude or more lower 
than those in the inner disks
\citep{2010AJ....140.1194B}.
% NJE just shortened a bit
% RCK corrected your correction, because it implied higher rather than lower 
% efficiencies in the outer disks 
%(Bigiel et al. 2010).  
%lower (higher) by at least a factor of ten relative
%to the inner disks 
Taken together, these recent
results  and studies of the outer \mw\ (\S \ref{milkyway})
confirm the presence of
a pronounced turnover in the Schmidt law for total gas at surface densities
of order a few \msun\,pc$^{-2}$ (Figure \ref{bigiel}).  

% RCK commented out paragraph below to remove repetition with
% discussion in spec.tex, as per our discussion

%An important outstanding question is whether the gas
%surface density at the star formation threshold varies systematically
%within galaxies.  Observational evidence for such variations 
%was presented by \citet{1989ApJ...344..685K} and \citet{2001ApJ...555..301M},
%% Kennicutt 1989, Martin \& Kennicutt 2001
%and they tentatively attributed the differences to variations in
%the critical densities for gravitational stability in the ISM.
%Their analyses were based on relatively poor CO observations by
%modern standards, however, and more recent work by \citet{2008AJ....136.2782L}
%% Leroy et al. 2008
%find less evidence for a threshold related to gravitational stability.
%% NJE my suggested more direct statement
%%has (quite properly) raised questions about the robustness of those conclusions.
%Further evidence for systematic departures from a constant
%threshold density comes from observations of low-mass (and lower metallicity) 
%galaxies such as the SMC, where the turnover in the large-scale SFR 
%is reported to occur at much higher (total) gas surface densities than in 
%massive spirals \citep{2011ApJ...741...12B}.
%% Bolatto et al. 2011
%Such measurements potentially can offer important clues
%to the physical mechanisms responsible for the thresholds, with
%due caution about effects of metallicity on \xco.
%% NJE added caution

\subsection{The Star Formation Law on Sub-Kiloparsec Scales}
\label{kpcsfl}

The same multi-wavelength data can be used
in principle to extend this approach to point-by-point 
studies of the star formation law.  The angular 
resolution of the currently available datasets 
(optical/UV, mid-far IR, \hi, CO)
allows for extending this analysis down to angular scales of $\sim$10\arcsec,
which corresponds to linear scales of 30--50\,pc
in the Local group and 200--1000\,pc for the galaxies in the 
local supercluster targeted by SINGS, THINGS, KINGFISH, 
and similar surveys.  Recent analyses have become available
(\citealt{2007ApJ...671..333K, 2008AJ....136.2846B, 2010AJ....140.1194B, 
2011ApJ...730L..13B, 2008AJ....136.2782L, 2009ApJ...704..842B, 
2010A&A...518L..62E, 2010A&A...510A..64V, 2011ApJ...735...63L,
2011ApJ...730...72R, 2012ApJ...745..183R, 2011AJ....142...37S}),
%Kennicutt et al. (2007), Bigiel et al. 2008, 2010, 2011;
%Leroy et al. (2008), Blanc et al. (2009), 
%Eales et al. (2010), Verley et al. (2010),
%Liu et al. (2011), Rahman et al. (2011, 2012), Schruba et al. (2011), and
%Momose (2011), 
and several other major studies are ongoing.

\begin{figure}
\center
\includegraphics[scale=0.8]{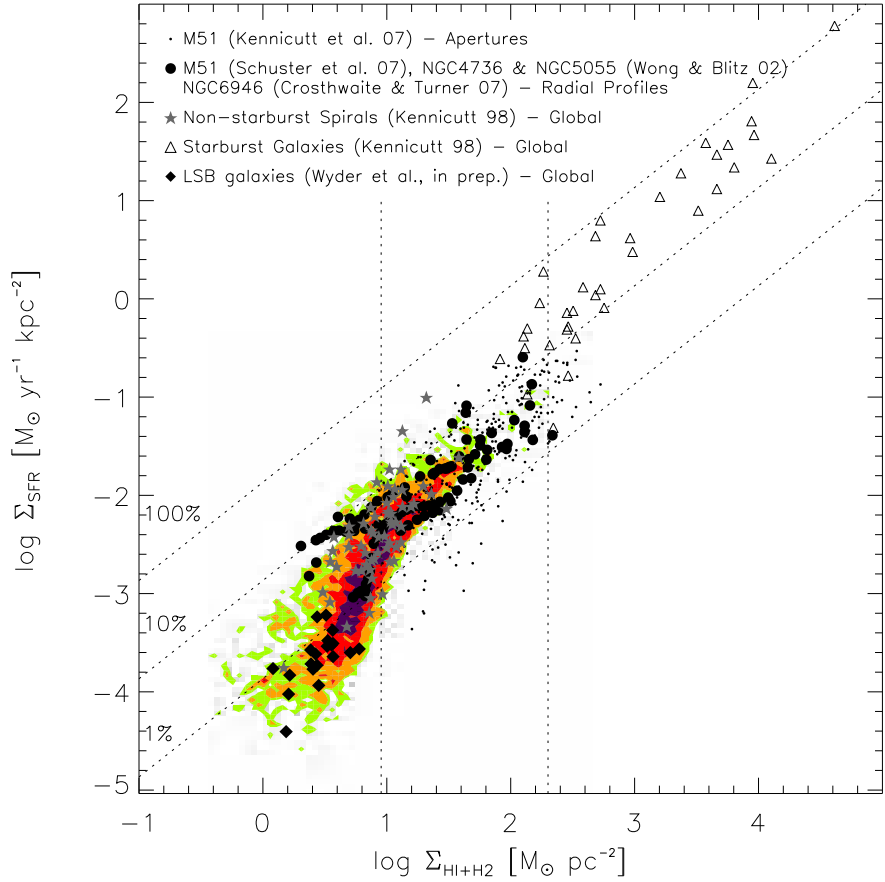}

\caption{Relation between SFR surface densities and total (atomic and molecular)
gas surface densities for various sets of measurements, from 
\citet{2008AJ....136.2846B}.  Colored contours show the distribution
of values from measurements of sub-regions of SINGS galaxies.  Overplotted 
as black dots are data from measurements in individual apertures in M51 
\citep{2007ApJ...671..333K}.  Data from radial profiles from M51 
\citep{2007A&A...461..143S}, NGC 4736, and NGC 5055 \citep{2002ApJ...569..157W},
and NGC 6946 \citep{2007AJ....134.1827C} are shown as black filled circles. 
The disk-averaged measurements from 61 normal spiral galaxies (filled gray stars) 
and 36 starburst galaxies (triangles) from \citet{1998ApJ...498..541K}
are also shown. The black filled 
diamonds show global measurements from 20 low surface brightness galaxies 
\citep{2009ApJ...696.1834W}.  In all cases, calibrations of IMF, $X(CO)$, etc.
were placed on a common scale.  The three lines extending from lower left to
upper right show lines of constant global star formation efficiency.  The
two vertical lines denote regimes which correspond roughly to those discussed
in \S \ref{sflobs} of this review.  
Figure taken from \citet{2008AJ....136.2846B}.
Reproduced by permission of the AAS.
}

\label{bigiel}
\end{figure}

The most comprehensive attack on this problem to date is based
on a combination of SINGS, THINGS, and HERACLES observations
(papers above by Bigiel, Leroy, and Schruba), and Figure \ref{bigiel},
taken from \citet{2008AJ....136.2846B}, 
%Bigiel et al. (2008), 
nicely encapsulates 
the main results from this series of studies.
The colored regions are the loci of individual sub-kiloparsec
measurements of SFR surface densities (measured from a
combination of FUV and 24\,$\mu$m infrared fluxes)
and total gas densities (\hi\ plus H$_2$ from CO(2--1) with a constant
\xco\ factor), for 18 galaxies in the SINGS/THINGS
sample.  Other data (including \citealt{1998ApJ...498..541K})
%Kennicutt 1998b) 
are overplotted as
described in the figure legend and caption.  The measurements
show at least two distinct regimes, a low-density sub-threshold
regime where the dependence of the SFR density on gas
density is very steep (or uncorrelated with gas density), and a higher density
regime where the SFR is strongly correlated with the
gas density.   The general character of this relation, with
a high-density power law and a low-density threshold confirms
most of the results presented earlier.

One can also examine separately the dependence of the SFR density
on the atomic and molecular surface densities and the 
two relations are entirely different.  
The local SFR density is virtually uncorrelated with \hi\ density,
in part because the range of \hi\ column densities is truncated
above $\sigmahi \sim 10$\,\msun\,pc$^{-2}$ or 
$N(\hi) \sim 10^{21}$\,cm$^{-2}$.
This upper limit to the local \hi\ densities corresponds to the 
column density where the \hi\ efficiently converts to molecular form.
This phase transition also roughly coincides with the turnover 
in the SFRs in Figure \ref{bigiel}, which hints that 
the SFR threshold itself may be driven in part, if not entirely,
by an atomic--molecular phase transition (see \S \ref{synthesis}).

In contrast to the \hi, the SFR surface density is tightly correlated
with the H$_2$ surface density, as inferred from CO.  
A recent stacking analysis
of the HERACLES CO maps made it possible to 
statistically extend this comparison to low H$_2$
column densities, and it shows a tight correlation between
SFR and CO surface brightness that extends into the \hi-dominated
(sub-threshold) regime \citep{2011AJ....142...37S}, reminiscent
of studies in the outer \mw\ (Fig. \ref{MWradial}, \S \ref{milkyway}).  
%A similar relation is found in the very metal-poor 
%($Z/Z_\odot = 0.2$) SMC when
%dust emission, rather than CO emission, is used to trace molecular gas
%\citep{2011ApJ...741...12B}.
 
The strong local correlation between \sigmasfr\ and \sigmamol\ 
is seen in all of the recent studies which probe linear scales of
$\sim$200\,pc and larger.  This is hardly surprising, given
the strong coupling of star formation to dense molecular gas in local
clouds (\S4, \S6.4).  However, on smaller linear scales the measurements 
show increased scatter for reasons discussed in \S \ref{challenge}.
% shortened by referring to earlier discussion
%will
%begin to sample individual molecular clouds (or small ensembles of clouds),
%%and for scales below $\sim$50\,pc, the observations will begin to probe
%substructure within individual clouds 
%and their associated star clusters, \hii\ regions, and dust clouds.
On these scales we expect the scaling laws to break down, as the
stars and gas may arise from separate regions; this effect
has been directly observed in high-resolution observations of M33
\citep{2010ApJ...722L.127O}.
%(Onodera et al. 2010).  

The main results described above -- the presence of a power-law
SFR relation with a low-density threshold, the lack of correlation of
the local SFR with \hi\ column density, the strong correlation with
H$_2$ column density, and the rapidly increasing scatter in the
Schmidt law on linear scales below 100--200\,pc -- are seen consistently 
across most, if not all, recent spatially-resolved studies.
Most of the recent studies also derive a mildly non-linear slope to
the SFR density versus total gas density Schmidt law in the high-density
regime, with indices $N$ falling in the range 1.2--1.6.  
% RCK Reworded next sentence.
Less clear from the recent work is the linearity and
slope of the Schmidt law on small scales, especially the dependence of SFR
surface density on molecular gas surface density.
The results from the HERACLES/THINGS studies have consistently 
shown a roughly linear $\Sigma_{SFR}$ versus \sigmamol\ relation 
($N$ = 1.0--1.1), and similar results have been reported by 
\citet{2009ApJ...704..842B}, \citet{2010A&A...518L..62E}, and
\citet{2012ApJ...745..183R}. 
%Eales et al. (2010), Rahman et al. (2012). Blanc et al. (2009)
%These slopes are systematically lower than the
%non-linear relations derived for disk-averaged measurements (\S6.1),
%and would signal an important change in the form of the Schmidt
%law between local and global scales.  
Other authors however have reported steeper dependences
($N$ = 1.2--1.7), closer to those seen in integrated measurements
(e.g., \citealt{2007ApJ...671..333K, 2010A&A...510A..64V, 2011ApJ...730...72R,
2011ApJ...735...63L, 2012momose}).
%Kennicutt et al. 2007, Verley et al. 2010,Rahman et al. 2011a,Liu et al. 2011
%Momose 2012, PhD thesis, called in separate thesis.bib file made by hand 
The discrepancies between these results are much larger than the estimated
fitting errors, and probably arise in part from systematic
differences in the observations and in the way the data are analyzed
(see discussion in e.g., \citealt{2009ApJ...704..842B}).
% Blanc et al. 2009
One effect which appears to be quite important is the way in
which background diffuse emission is treated in measuring the local
SFRs (\S3.9).  \citet{2011ApJ...735...63L} show that they can
produce either a linear or non-linear local molecular SFR law
depending on whether or not the diffuse emission is removed.
Another CARMA-based study by \citet{2011ApJ...730...72R}, however,
%Rahman et al. (2012) 
suggests that the effects of diffuse emission may not be sufficient
to account for all of the differences between different analyses.
Another important factor may be 
the excitation of the CO tracer used (e.g., \citealt{2009ApJ...707.1217J}).
%Juneau et al. 2008),
%The profusion recent papers 
%testifies to the dynamic (and fluid) nature of this subject, 
%and with such a concentration of observations and effort
%we expect these issues to be clarified in the near future.  In the 
%meantime readers should be aware of the systematic uncertainties
%which remain to be addressed, and to take them into consideration
%when interpreting the observations.  

It is important to bear in mind that nearly all of our
empirical knowledge of the form of the local star formation
law is based on observations of massive gas-rich spiral
galaxies with near-solar gas-phase metal abundances.
Most studies of the star formation relations in metal-poor
dwarf galaxies have been limited to \hi\ data or marginal
detections in CO at best (e.g., \citealt{2008AJ....136.2846B}).
%Bigiel et al. 2008
A recent study of the SMC ($\sim1/5\,Z_\odot$) by \citet{2011ApJ...741...12B}
%Bolatto et al. 2011
offers clues to how these results may change in low-metallicity
environments.  They find that the atomic-dominated
threshold regime extends up to surface densities that are an
order of magnitude higher than in spirals.

% sflobs
% sflaw has subsubsections with labels avgsfl, radialsfl, kpcsfl
% localtests.tex
% formerly synthesisobs.tex (Neal's version)
% 060711 NJE
% 060911 NJE
% 070411 NJE
% 070511 NJE
% 070811 NJE
% 072711 NJE
% 080911 NJE
% 081111 NJE moved to localtests and put in new SFRelations section
% 081111 NJE, edited to fit with new structure and remove redundancies
% 082211 NJE added stuff from Rosie Chen work on LMC
% 090611-090811 NJE considerable editing
% 092511 RCK editing
% 100711 NJE minor edits
% 102111 NJE clean-up, added Chomiuk ref and followed consequences
% 110411 NJE minor edits
% 100611 RCK minor edits
% 121211 NJE minor edits
% 121611 NJE minor edits and rearranged text for better logical flow
% 122311 NJE edits after comments
% 020212 NJE further edits, added ref re eff in clumps, re MK20 comment
% 020612 NJE minor clean up
% 020812 NJE added refs to obs and theory re threshold
% 021712 NJE minor edits
% 031212 NJE minor edits
% 050912 NJE minor edits
% 051412 NJE minor edits

\subsection{Local Measurements of Star Formation Relations} \label{localtests}

With the ongoing large-scale surveys of the Galaxy
and the Magellanic Clouds in recent years it is becoming possible
to investigate star formation rate indicators, the scaling laws,
 and other star formation relations for local samples.  
These can provide valuable external checks on
the methods applied on larger scales to external galaxies and 
probe the star formation relations on physical scales that are
not yet accessible for other galaxies.  Local studies of the conversion of CO
observations into column density or mass were discussed in \S \ref{gastracers}.

Comparison of various star formation rate indicators
for the \mw\ (\S \ref{milkyway}), as would be used by observers 
in another galaxy, against more direct measures,
such as young star counts, suggests that star formation rates 
based on the usual prescriptions (\mir\ and radio continuum)
may be underestimating absolute \sfr\ by factors of 2-3
\citep{2011AJ....142..197C}. 
%Chomiuk and Povich 2011
The authors suggest that changes to the intermediate mass IMF,
timescale issues, models for O stars, and stochastic sampling of the
upper IMF can contribute to the discrepancy.
As discussed earlier, measures of star formation rate that use
ionizing photons (\halpha\ or radio continuum) require regions with
a fully populated IMF to be reliable, and they systematically
underestimate the SFR in small star-forming clouds (\S \ref{challenge}).  
For example, star formation in clouds
near the Sun would be totally invisible to these measures
and the star formation rate would be badly underestimated in
the Orion cluster (\S \ref{challenge}). 
Tests of other star formation rate tracers, such as 24 \micron\
emission, within the Milky Way would be useful.  Changes in 
the IMF as large as those proposed by \citet{2011AJ....142..197C}
should also be readily observable in more evolved Galactic star
clusters, if not in the field star IMF itself.

Using a YSO-counting method for five GMCs in the N159 and N44 regions
in the LMC,  \citet{2010ApJ...721.1206C}
%Chen et al. (2009), 
were able to reach stellar masses of about 8 \msun, below 
which they needed to extrapolate. The resulting ratios of 
\sfr(YSOs) to \sfr(H$\alpha\ + 24 \micron$) ranged from 0.37 to 11.6, 
with a mean of 3.5.

Recent studies have begun to probe the form of the star formation
relations on scales of clouds and clumps.  
Using the YSO star counting method to get \sfr\ and extinction maps
to get mean mass surface density of individual, nearby clouds, 
\citet{2009ApJS..181..321E}
% Evans et al. (2009) 
found that the local clouds all lay well above
the K98 relation. Taken in aggregate, using the mean \sigmagas,
they lay a factor of 20 above the Kennicutt et al. (1998b) 
relation and even farther above the relation of Bigiel et al. (2008).

\begin{figure}%4	% Example of Figure pull
\center
\includegraphics[scale=1.0]{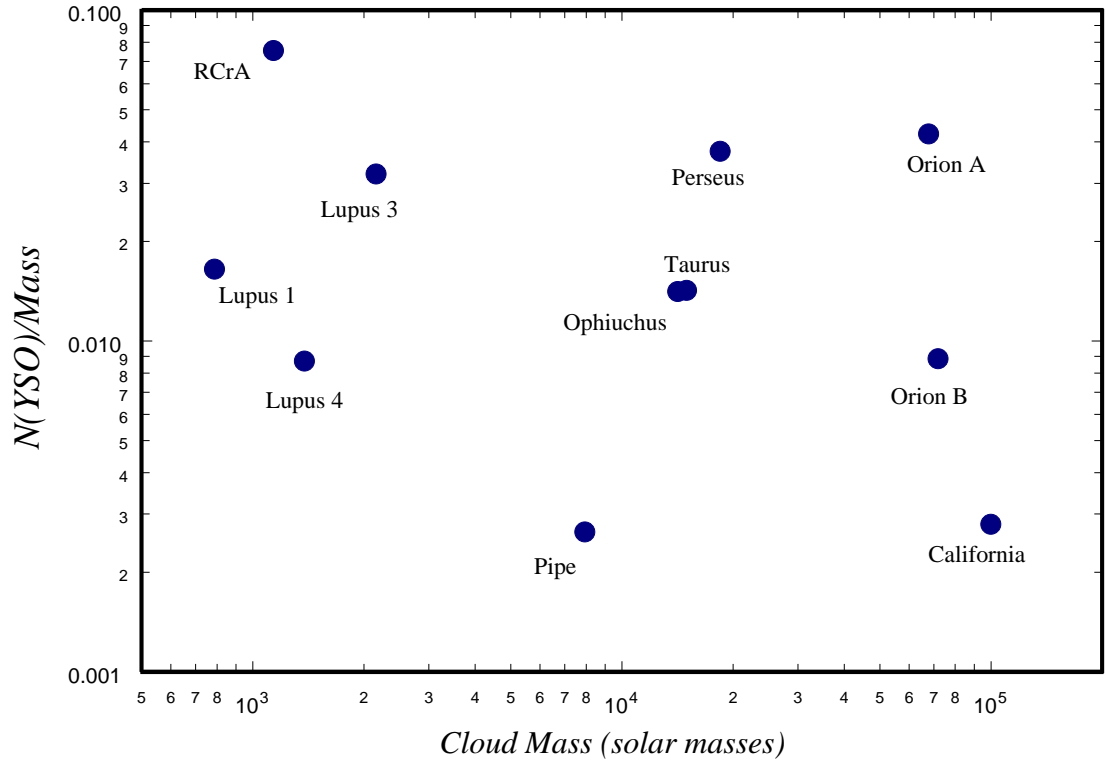}
\caption{
Plot of the ratio of the total YSO content of a cloud to the 
total cloud mass vs. total cloud mass. This is equivalent to a measure of the 
star formation efficiency as a function of cloud mass for the local sample. 
It is also equivalent to the measure of the SFR per unit cloud mass as a 
function of the cloud mass. The plot shows large variations in the efficiency 
and thus the SFR per unit mass for the local cloud sample.
Taken from \citet{2010ApJ...724..687L}.
%Lada et al. 2010.
Reproduced by permission of the AAS.
}
\label{ladafig2}
\end{figure}

\begin{figure}%4	% Example of Figure pull
\center
\includegraphics[scale=1.0]{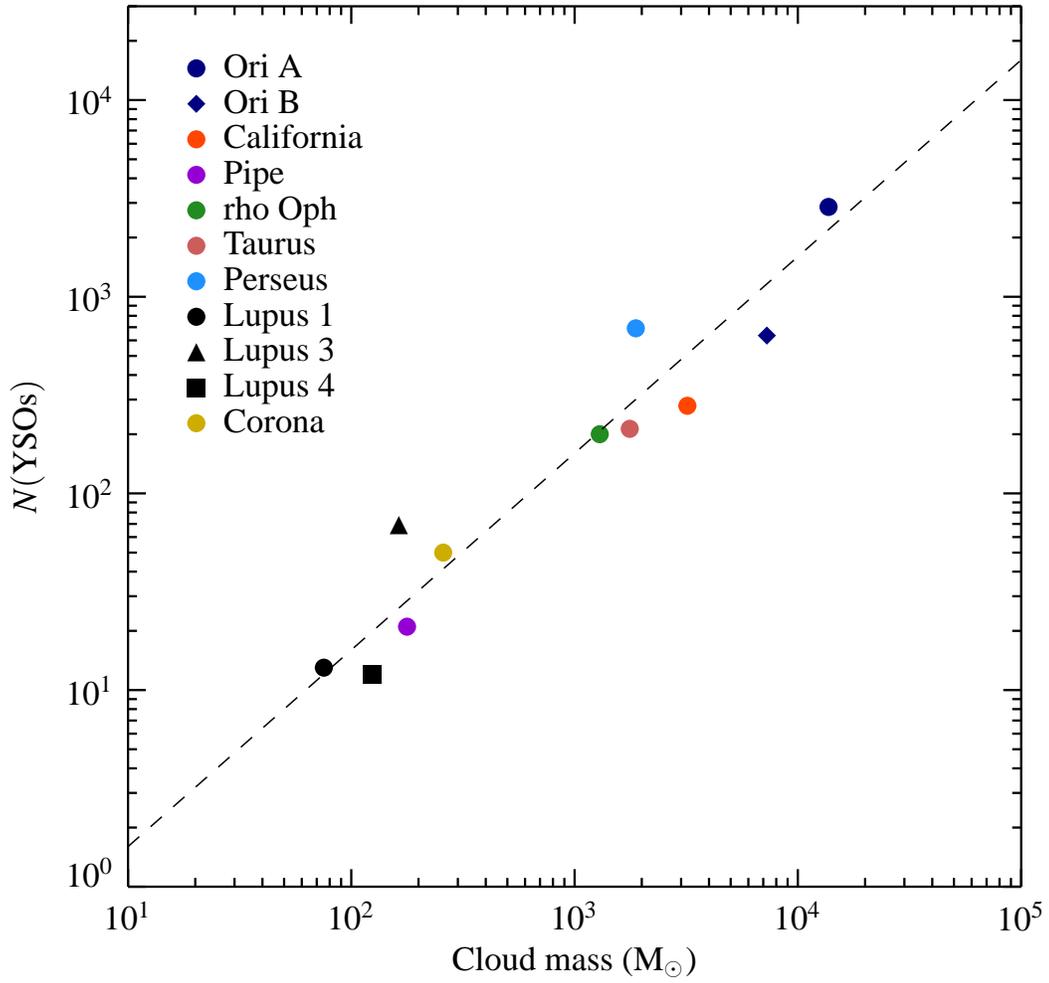}
\caption{
Taken from \citet{2010ApJ...724..687L}.
%Lada et al. 2010.
Relation between N(YSOs), the number of YSOs in a cloud, and $M_{0.8}$, 
the integrated cloud mass above the threshold extinction of $A_{K0} = 0.8$ mag. 
For these clouds, the SFR is directly proportional to N(YSOs), and thus 
this graph also represents the relation between the SFR and the mass of 
highly extincted and dense cloud material. A line representing the best-fit 
linear relation is also plotted for comparison. There appears to be a 
strong linear correlation between N(YSOs) (or SFR) and $M_{0.8}$, the cloud 
mass at high extinction and density.
Taken from \citet{2010ApJ...724..687L}.
%Lada et al. 2010.
Reproduced by permission of the AAS.
}
\label{ladafig4}
\end{figure}

\begin{figure}%4	% Example of Figure pull
%\epsfscale1200		% Figure enlarged to 120%
\begin{center}
\includegraphics[scale=0.6]{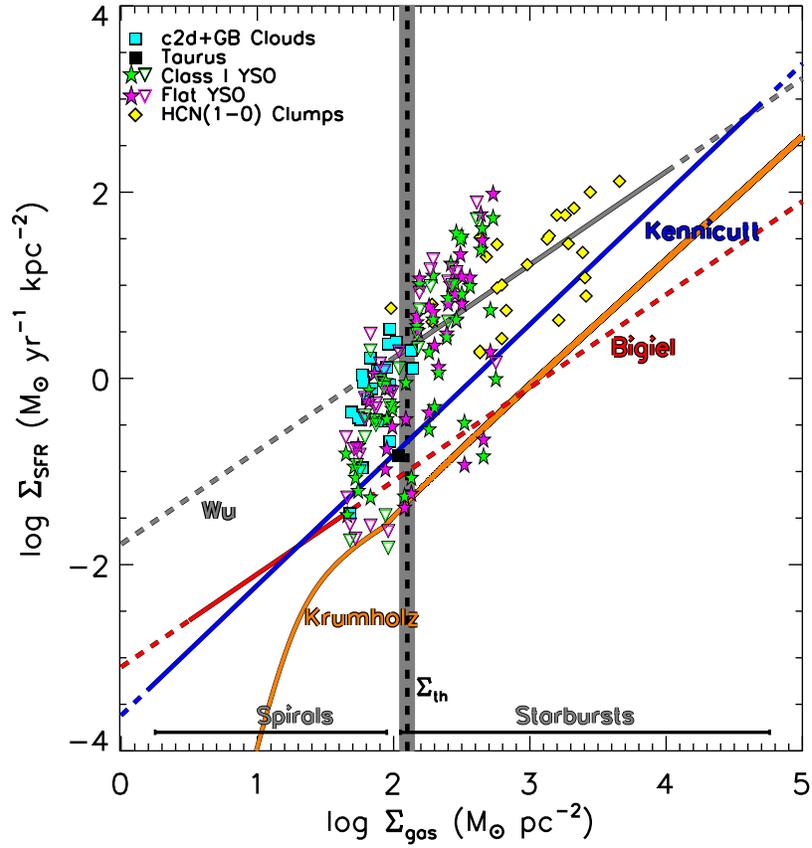}
\caption{Comparison of Galactic total c2d and GB clouds, YSOs, and massive 
clumps to extragalactic relations. SFR and gas surfaces densities for the 
total c2d and GB clouds (cyan squares), c2d Class I and Flat SED YSOs 
(green and magenta stars), and $\lir > 10^{4.5}$ \lsun\  
massive clumps (yellow diamonds) are shown. 
The range of gas surface densities for the spirals and 
circumnuclear starburst galaxies in the \citet{1998ApJ...498..541K} sample is 
denoted by the gray horizontal lines. The gray shaded region denotes the 
range for $\Sigma_{th} =  129 \pm 14$ \msunpc.
Taken from \citet{2010ApJ...723.1019H}. 
Reproduced by permission of the AAS.
}
\label{Heidermanfig9}
\end{center}
\end{figure}

Subsequent studies show evidence for both a surface density threshold for
star-forming clumps and a \sigmasfr-\sigmamol\  relation in
this high-density sub-cloud regime.
\citet{2010ApJ...724..687L}
% Lada et al. (2010) 
found  a threshold surface density
for efficient star formation in nearby clouds; the star formation rate
per cloud mass scatters widely (Fig. \ref{ladafig2}), but
is linearly proportional to the cloud mass above a surface density
contour of $116\pm 28$ \msunpc\ (Fig. \ref{ladafig4}), 
and the coefficient agrees well with the
dense gas relations of \citet{2004ApJ...606..271G}.
% Gao and Solomon  
\citet{2010ApJ...723.1019H}
% Heiderman et al. (2010) 
studied the behavior of
\sigmasfr\ on smaller scales using contours of extinction. They limited
the YSOs to Class I and Flat SED objects to ensure that they were still 
closely related
to their surrounding gas and  found a steep increase in \sigmasfr\
with increasing \sigmagas\ up to about 130 \msunpc, above which
a turnover was suggested. By adding the dense clumps data from 
\citet{2010ApJS..188..313W},
% Wu et al.  (2010), 
they identified a turnover at $129\pm 14$ \msunpc, where the
$\Sigma$(SFR) began to match the dense gas relation, 
but was far above the K98 relation for total gas
(Figure \ref{Heidermanfig9}). 
The agreement between these two independent approaches is encouraging,
suggesting that a contour of about 125 \msunpc\ is a reasonable
defining level for a star-forming clump (\S \ref{names}).
Other studies have found similar thresholds for efficient star
formation or the presence of dense cores
(\citealt{1998ApJ...502..296O, 2007ApJ...666..982E, 2004ApJ...611L..45J,
2010A&A...518L.102A, 1997ApJ...488..277L, 1992ApJ...393L..25L}).
% Onishi et al. 1998, Enoch 2007, Johnstone 2004, Andre(Herschel) 2010,
% Li et al. 1997, Lada et al. 1992
Theoretical explanations for such thresholds can be found by considering
magnetic support \citep{1976ApJ...210..326M}
% Mouschovias and Spitzer
or regulation by photo-ionization \citep{1989ApJ...345..782M}.
% McKee 1989
Alternatively, this threshold may simply correspond to the part of
the cloud that is gravitationally bound (cf. \S \ref{sftheory}).

The star formation rate density is even higher {\it within} clumps.
\citet{2011ApJ...739...84G}
% Gutermuth et al. 2011  
find a continuation of the steep increase in \sigmasfr\
to higher \sigmagas\ in a study including embedded clusters.
They find that $\sigmasfr \propto \sigmagas^2$ up to several 100
\msunpc, with no evidence of a threshold.  
Near the centers of some centrally condensed clumps, 
\sigmagas\ reaches 1 gm cm$^{-2}$, or about 4800 \msunpc\
\citep{2010ApJS..188..313W},
% Wu et al.
a threshold for the formation of massive stars suggested by theoretical
analysis \citep{2008Natur.451.1082K}.
% Krumholz and McKee

Other studies have identified possible scaling relations in which
SFR surface densities 
are not simply proportional to surface densities of dense gas.
One analysis found that clouds forming stars with masses
over 10 \msun\ satisfied the following relation:
$m(r) \ge 870 \msun (r/pc)^{1.33}$
(\citealt{2010ApJ...716..433K, 2010ApJ...723L...7K}),
where $m(r)$ is the enclosed mass as a function of radius.
This is essentially a criterion they call ``compactness"; it
can be thought of as requiring a central condensation, with
$\rho(r) \propto r^p$, with $p \ge 1.67$.

\citet{2011ApJ...741..110D}
% Dunham  M K on ammonia survey
have compared a subset of clumps in the BGPS catalog
(\citealt{2011ApJS..192....4A,2010ApJS..188..123R})
% Aguirre et al. , Rosolowsky et al.
with kinematic distances to the Heiderman-Lada (HL) criterion
for efficient star formation and to the Kauffman-Pillai (KP)
criterion for massive star formation. About half the clumps
satisfy both criteria. Interestingly, \citet{2011ApJ...731...90D}
%Dunham, Robitaille et al.
found that about half the BGPS sample contained at least one
mid-infrared source from the GLIMPSE survey. For the clumps
most securely identified with star formation, 70\% to 80\%
satisfy the HL or KP criteria.

All these studies consistently show that \sigmasfr, especially for massive
stars, is strongly localized to dense gas, and is much higher 
(for $\Sigma_{gas} > 100$ \msunpc) than in the extragalactic Schmidt
relations (\citealt{1998ApJ...498..541K, 2008AJ....136.2846B}).
% Kennicutt ApJ,  Bigiel 2008
This offset between the extragalactic and dense clump ``Schmidt laws"
can be straightforwardly understood as reflecting the mass fraction 
(or surface filling factor) of dense clumps in the
star-forming ISM.  If most or all star formation 
takes place in dense clumps, but the clumps contain only a small
fraction of the total molecular gas, we would expect the characteristic
star formation \tdep\ measured for clumps to be proportionally
shorter than \tdep\ of the total cloud mass and the efficiency
($\epsilon$) to be higher.
% Changed to refer to depletion time re Mark's comment number 20 NJE
Using a threshold of \cooo\ emission to define clumps, 
\citet{2009ApJ...705..468H}
% Higuchi et al. 2009
found that the star formation efficiency ($\epsilon$) in clumps
varies widely but averages 10\%, about 4 times that in clouds as a whole
(\S \ref{lowmass}).
If this interpretation is correct, we would also expect the relations
between SFR and dense gas mass for galaxies (Gao-Solomon relation)
to be similar for galaxies and the clumps, and 
they indeed appear to be roughly consistent \citep{2005ApJ...635L.173W}.
% Wu et al.  2005

This conclusion is subject to some caveats, however.  
Various authors
(\citealt{2007ApJ...669..289K, 2008ApJ...684..996N, 2009ApJ...707.1217J})
% Krumholz and Thompson, Narayanan, Juneau
have pointed out that lines
like those of HCN \jj10\ are not thermalized at lower densities, so
can result in a linear relation even if the underlying star
formation relation is
a local version of a non-linear KS law that extends to much lower densities.
The SFR vs dense gas relation is based on a correlation
between total FIR and HCN luminosities, and \lfir\ is likely to 
underestimate the SFR in young clusters 
(\citealt{2010ApJ...710.1343U,2011ApJ...739...84G,2007ApJ...669..289K}),
% Urban et al. 2010, Gutermuth et al. 2011 Krumholz and Thompson 2007
perhaps by up to factors of 3--30.  
However, 
\citet{2012ApJ...745..190L}
%Lada et al. (2011, preprint)
find a similar continuity between the Gao-Solomon starbursts
and SFRs measured by counting YSOs in gas above a threshold
surface density.

% localtests

\section{TOWARD A SYNTHESIS}\label{synthesis} 

%cluesobs.tex
%081111 NJE was synthesis_rob.tex
%081111 NJE some edits
%082611 NJE total rewrite to include MW studies for each regime, ...
%082911 NJE minor edit to refer correctly to milkyway section
%090811 NJE more substantial edits
%092511 RCK substantial edits
%100411 RCK more edits
%100511 RCK yet more edits (some backwards!)
%100711 NJE minor edits and some questions to Rob
%101411 RCK edits to address NJE questions and reference updates
%101911 RCK citation links
%110411 NJE minor edits
%110611 RCK minor edits
%110911 NJE changed tdep to "as short as 10 Myr"
%121311 NJE various edits, added subsubsecs
%121611 NJE minor edits
%121911 NJE changed title of last subsubsec
%020212 NJE edited to make efficiency consistent and shortened
%020812 NJE minor edits, added refs for prestellar core eff and life
%021712 NJE minor edits
%031212 NJE minor edits
%050512 NJE updated references
%050912 NJE insert ref to Table 3
%051012 NJE insert ref to Table 3
%051412 NJE edited efficiency para to agree with Rob, used ion

\subsection{Summary: Clues from Observations}\label{cluesobs}

Before we embark on an interpretation of the observed
star formation law, it is useful to collect the main
conclusions which can be drawn from the observations
of star formation both outside and inside the Galaxy.

Beginning on the galactic scale ($>1$ kpc), 
we can identify at least two and probably
three distinct star formation regimes (Table \ref{regimetab}).
%, separated by characteristic surface densities of order 10 \msunpc\
%and 100-300 \msunpc. 
Whole galaxies may lie in one of these
regimes, but a single galaxy may include two or three regimes.  

\subsubsection{The Low-density Regime}

The lowest-density regime (sometimes referred to
as the sub-threshold regime), is most
readily observed in the outer disks of spiral galaxies, but
it also can be found in the interarm regions of some spiral
galaxies and throughout the disks of some gas-poor galaxies.
The solar neighborhood lies near the upper end of this 
low-density regime and the outer disk of the MW is clearly in this regime
(\S \ref{milkyway}).

The cold gas in these regions is predominantly atomic,
though local concentrations of molecular gas are often found.
Star formation is highly dispersed, with young
clusters and \hii\ regions only observed in regions of
unusually high cold gas densities.
The global ``efficiency" of star formation,
$\epsilon^{\prime}$ (defined in \S \ref{definitions}),
or \sigmasfr/\sigmagas, 
%the ratio of SFR surface density to gas surface density) 
is very low, and it is uncorrelated with \sigmagas.
%the total gas surface density.  
The steep, nearly vertical ``Schmidt" relation 
seen in this regime (e.g., Fig. \ref{bigiel}) mainly reflects lack of 
correlation over a region where the SFR has a much
larger dynamic range than the local gas density;
any apparent correlation is not physical.  Recent stacking analysis
of CO maps and studies of the outer \mw\ suggest, however, 
that there may be a strong  
correlation  with molecular surface density (\S \ref{sflobs}, 
\citealt{2011AJ....142...37S}).
% (Schruba as discussed in section on obs of sfl).  

\subsubsection{The Intermediate-density Regime}

The next, intermediate-density regime is roughly characterized by
average gas surface densities of $\sigmagas > 10$ \msunpc, 
corresponding roughly to $ N({\rm H}) \sim 10^{21}$\,cm$^{-2}$, 
or $\av \sim 1$ mag, for solar metallicity.
The upper limit to this regime is about $\sigmagas \sim100 - 300$ \msunpc,
as discussed in the next section.
The intermediate regime applies within the main optical
radii (R$_{25}$) of most gas-rich, late-type spiral and irregular galaxies.
The ``Galactic Ring" region of the Milky Way lies at the low end of 
this regime, while the CMZ may lie near the high end (\S \ref{milkyway}).

The transition from the low-density 
regime roughly corresponds to the transition between \hi-dominated
and H$_2$-dominated ISMs.
Above $\sigmagas = 10$ \msunpc, both the phase balance of the ISM 
and the form of the star formation law begin to change.
The intermediate range features an increasing filling factor of 
molecular clouds, and star formation becomes more pervasive.
The SFR surface density is strongly
and tightly correlated with the cold gas surface density, whether expressed
in terms of the total (atomic plus molecular) or only the
molecular surface density.  In most massive spiral galaxies,
the cold gas in this regime is molecular-dominated.  In low-mass
and irregular galaxies, atomic gas can dominate, though this is
somewhat dependent on the value of \xco\ that is assumed.
The characteristic 
%``efficiency" ($\epsilon^{\prime}$, \S \ref{definitions})
% NJE changed to definition, but then commented since we talk depletion time
depletion time (\tdep\ \S \ref{definitions})
for the interstellar gas is 
%of order 5--10\%\ per 100 Myr or 100\%\ per 
% NJE just shortening
1--2 Gyr
(e.g., \citealt{2011ApJ...730L..13B}).

When expressed in terms of $\epsilon^{\prime}$ (\S \ref{definitions}), 
the star formation rate per unit total gas mass, nearly all studies
suggest that $\epsilon^{\prime}$ increases with gas surface density,
with an exponent of $0.2 - 0.5$, (the Schmidt law exponent, $N-1$).  
When measured against
molecular mass (or surface density),
some studies suggest $\epsilon^{\prime}$ is constant, 
but others suggest $\epsilon^{\prime}$
increasing systematically with surface density.

\subsubsection{The High-density Regime}

The two regimes discussed above were able to reproduce all of the early
observations of the large-scale star formation law by 
\citet{1989ApJ...344..685K, 1998ApJ...498..541K} and 
%Kennicutt (1989, 1998b),
\citet{2001ApJ...555..301M}.
%Martin \& Kennicutt (2001).  
However a number of recent
observations suggest the presence of a third regime (and second
transition) around  $\sigmagas >100 - 300$\,\msunpc,
into what one might call the high-density, or starburst, regime.
Around this value of \sigmagas, the interpretation of \ico\ is likely to change
(\S \ref{gastracers}), and studies of local \mw\ clouds indicate
that a similar \sigmagas\ value may correspond to the theoretical
notion of a cluster-forming clump, in which the SFR is much higher
than in the rest of the cloud (\S \ref{localtests}).
In the most extreme starburst galaxy environments,
if standard values of \xco\ are used,
the average surface densities of gas (virtually all molecular)
reach 1000 to \eten4\ \msunpc,
and the {\it volume} filling factor of clumps
could reach unity \citep{2009ApJ...707..988W}.
% Wu et al. high-z SHARC paper
While the interpretation of molecular emission
in these conditions warrants skepticism
(\S \ref{gastracers}, \citealt{2012A&A...539A...8G}), the highest inferred
surface density also corresponds to
the densest parts of cluster forming clumps and the
theoretical threshold of 1 gm cm$^{-2}$ for efficient
formation of massive stars (\S \ref{sftheory}).

Some recent observations strongly hint at the existence of
such a transition.  As discussed in \S \ref{sflobs}, luminous and
ultraluminous starburst galaxies (and high-redshift SMGs) have characteristic 
ratios of $L_{IR}/L(CO)$ that are as much as 1--2 orders of magnitude
higher than in normal galaxies, implying that 
at some point the SFR per molecular mass
must increase dramatically.  This could be explained 
by a break in the slope of the Schmidt law at high densities, 
a continuous non-linear Schmidt law slope extending
from the intermediate to high-density regimes, or a
second mode of star formation in extreme starbursts with much higher
``efficiency" ($\epsilon^{\prime}$).  
As discussed in \S \ref{kpcsfl}, the constancy of the molecular
$\epsilon^{\prime}$ at intermediate surface densities is uncertain.
% NJE since we just discussed this, I think we can shorten
%, with
%some studies pointing to a constant efficiency and others to
%a nonlinear dependence.  
Observations of CO in high-redshift 
galaxies have been interpreted in terms of just such a bimodal Schmidt law
\citep{2010ApJ...714L.118D, 2010MNRAS.407.2091G}.
%(Daddi et al. 2010, Genzel et al. 2010).  
This interpretation
rests on the assumption of a bimodality in \xco\ (\S \ref{avgsfl}), 
and a change in $\epsilon^{\prime}$ that follows these
changes in \xco; this is not entirely implausible, because the
same physical changes in the ISM environment in the densest starbursts
could affect both the CO conversion factor and $\epsilon^{\prime}$ if
an increasing fraction of the gas is in dense clumps.
% Doesn't need to be all
%the star formation
%efficiency, if virtually all of the gas were in dense clumps.

Both 
%the interpretation of 
a Galactic value of \xco\ in the starbursts
and a much lower value have their discomforting aspects.  Applying
a Galactic conversion factor produces total molecular masses
which often exceed dynamical mass limits for the regions, whereas adopting
values of \xco\ which are factors of several lower produces 
gas consumption times as short as 10 Myr
\citep{2010ApJ...714L.118D, 2010MNRAS.407.2091G},
%(Daddi et al. 2010, Genzel et al. 2010)  
with implications for
the triggering and duty cycles of these massive starbursts.

The relations between SFR and ISM properties summarized above
strictly refer to the correlations with the total cold gas
surface density or in some cases the total \hi\ and total H$_2$
densities; these are the relations of most interest for applications
to galaxy modelling and cosmology.  However the observational picture
is quite different when we correlate the SFR with the supply of
dense gas, as traced by HCN \jj10\ and other dense clump tracers.
Here there seems to be a single linear relation which extends
across all of the SFR regimes described above, and which even
extends to star-forming clouds in the Milky Way 
(\citealt{2004ApJ...606..271G, 2005ApJ...635L.173W, 2007ApJ...660L..93G,
2012ApJ...745..190L}),
%(Gao \& Solomon 2004, Wu et al. 2005, Gao et al. 2007).
% and Lada et al. new paper
However, there is still some evidence of bimodality even if tracers of
dense gas are used \citep{2012A&A...539A...8G}, 
% Garcia-Burillo et al.
and physically based models, rather than simple conversion factors, are needed.

\subsubsection{Clues from Studies of Molecular Clouds}

The biggest hurdle one confronts when attempting to interpret
these observations of galaxies is the severe influences of 
spatial averaging, both across the sky and along the line of
sight.  The star formation law relates surface densities of
young stars and interstellar gas-- already smoothed along the
line of sight-- averaged over linear dimensions ranging from
order 100 pc to 50 kpc.  These are 2--4 orders of magnitude
larger than the sizes of the dense clump regions in which most
stars form, and 4--8 orders of magnitude larger in terms of the surface
areas being measured.  This difference in scales means that the
``surface densities" measured in extragalactic studies are 
really characterizing the filling factors of gas clumps and star-forming
regions, rather than any measure of physical densities.  Understanding
how this ``active component" of the star-forming ISM works is 
essential to even an empirical understanding of large-scale star
formation, much less understanding its underlying physics.
Detailed studies within the \mw\ provide a way to ``zoom in"
further than is possible for other galaxies.

A common observed feature across all of the density regimes
observed in galaxies is that star formation takes place in 
molecular clouds.  In the lowest-density regimes, clouds are
rare and widely separated, but within individual molecular
clouds in the \mw, which can be probed in detail, the star 
formation seems to be indistinguishable from that in regions of somewhat
higher average surface density (\S \ref{milkyway}).

These observations also reveal that nearly all
star formation within molecular clouds is highly localized, taking
place in clumps, roughly defined by $\sigmamol > 125$ \msunpc, or 
$n > \eten4$ \cmv\ (\S \ref{localtests}).
The clumps host young stars, YSOs, and pre-stellar cores, the sites
of individual star formation.  
This scale is as close to a deterministic environment for star
formation as can be found.  Once a pre-stellar core reaches
substantial central condensation,
about one-third  of its mass will subsequently turn into young stars
within a few Myr (\citealt{2007A&A...462L..17A, 2008ApJ...684.1240E}).
% Alves et al. 2007, Enoch et al. 2008
The relatively low global efficiencies (both $\epsilon$ and
$\epsilon^{\prime}$) within GMCs are largely a reflection
of the low mass fraction in clumps and cores (\S \ref{lowmass}).  
The star formation rate surface density in the clumps is 20-40
times that predicted for the mass surface density from the extragalactic
Schmidt relation (\S \ref{localtests}), 
and this likewise can be largely understood as 
reflecting the low mass fraction of molecular gas in clumps and cores. 

One might be tempted to identify the dense clumps within molecular
clouds as a possible fourth density regime, in addition to the 
three regimes already discussed from observations of galaxies.
However this would be very misleading, because so far as is currently
known the formation of most stars in dense clumps is a common
feature of star formation across all of the ISM environments in
galaxies, extending from low-density sub-threshold disks to the
most intense starbursts.  The dense clump may well be the 
fundamental unit of massive, clustered star formation 
\citep{2005ApJ...635L.173W}.  
% Wu et al. 2005
If this picture is correct, then the three regimes identified in galaxies
and the order-of-magnitude increases in gas-to-star formation 
conversion rate  across them must reflect changes in the fraction of
the ISM that is converted to dense clumps, approaching 100\%\ in
the densest and most intense starbursts.

% cluesobs
% Rob's version
%spec.tex
% 081111 NJE made from bits from NJE's synthesis section
% meant to follow cluesobs.tex
% 082311 Added some refs and started to smooth out, still needs work.
% 082411 corrected numbers in local example (which may move to "challenge")
% 082511 began major rework, moved a lot to sflobs and rewrote the rest
% 082611 adapted to new version of cluesobs
% 090911 NJE further adapted and added some refs from deleted sfltheory
% 092611 RCK
% 092611 NJE, minor fixes
% 100511 RCK, ditto
% 100711 NJE, minor fixes
% 101411  RCK, reference updates
% 110411 NJE, minor edits
% 120811 NJE added comments about Hennebelle, Krumholz, Lada
% 120911 NJE further edits to balance this
% 121611 NJE minor edits
% 012312 NJE edits after comments, fixed missing refs
% 012412 NJE further small edits
% 012912 RCK added the missing Silk 1997 reference
% 020212 NJE minor edits, added some refs re similar efficiencies
% 020612 RCK minor edits, removed NJE flags where I agree with wording
% 020712 NJE very minor edit
% 021412 NJE added issue of bound gas, def of crossing time re Pringle comments
% 021712 NJE minor edits
% 031212 RCK minor edit on grav instabilities
% 031212 NJE minor edits
% 050912 NJE minor edit
% 051412 NJE minor edits and ion 

\subsection{Some Speculations}\label{spec}

The observations summarized in this review have stimulated a
rich literature of theoretical ideas, models, and simulations
aimed at explaining the observed star formation relations and
constructing a coherent picture of galactic-scale star formation.
Unfortunately we have neither the space nor the expertise to 
review that large body of theoretical work here.  A review of
many of the theoretical ideas can be found in \citet{2007ARA&A..45..565M},
% McKee \& Ostriker ARAA 
and an informative summary of the main models in the literature
up to 2008 can be found in \citet{2008AJ....136.2782L}.
%  Leroy et al. (2008). 
Here we offer some speculations 
aimed toward explaining the
observations and connecting the extragalactic and MW studies,
and identifying directions for future work.  Many of these
speculations reflect ideas being discussed in the literature,
and we make no claims for originality in the underlying concepts.

As mentioned in the introduction, the formation of stars represents
the endpoint of a chain of physical processes that begins with
cooling and infall of gas from the intergalactic medium onto disks, 
followed by the formation of a cool atomic phase, contraction to
gravitationally bound clouds, the formation of molecules and molecular
clouds, the formation of dense clumps within those molecular clouds,
and ultimately the formation of pre-stellar cores, stars, and star
clusters.  Although the beginning and end points of this process are
relatively clear, the sequence of intermediate steps, in particular
the respective roles of forming cool atomic gas, molecular gas, and
bound clouds is unclear, and it is possible that different processes
dominate in different galactic environments.  By the same token, a
variety of astrophysical time scales may be relevant: e.g., the free-fall
time of the gas within clouds, the crossing time for a cloud,
the free-fall time of the gas layer,
or the dynamical time scales for the 
disc and spiral arm passages. Any of these timescales may be invoked for
setting the time scale for star formation and the form of the
star formation law.  However one can construct two scenarios
to explain the observations outlined in \S \ref{cluesobs},
which illustrate the boundaries of fully locally-driven versus
globally-driven approaches.  In some sense,
these approaches are complementary, but they need to be brought
together.

The first approach, which might be called a bottom-up picture, assumes that
star formation is controlled locally within molecular clouds, 
(e.g., \citealt{2005ApJ...630..250K}),
% Krumholz and McKee 2005 
building on what we observe in  well-studied local regions of star formation.
In this picture, one can identify three distinct regimes (\S \ref{cluesobs}),
and associate the transitions between them to the crossing of two
physically significant thresholds, the threshold for
conversion of atomic to molecular gas, and the threshold
for efficient star formation in a molecular cloud, identified
with the theoretical notion of a clump (\S \ref{names}). 
In the purest form of this picture the SFR is driven completely by
the amount and structure of the molecular gas.  Current evidence
from studies of nearby clouds indicates that the star formation 
rate scales linearly with the amount of dense gas in clumps
\citep{2010ApJ...724..687L}, 
%Lada et al. 2010 
and this relation extends to starburst galaxies if the HCN emission
is used as a proxy for the dense gas 
(\citealt{2005ApJ...635L.173W, 2012ApJ...745..190L}). 
%Wu et al. 2005, Lada et al. published
In this picture, the non-linear slope ($N \sim 1.5$) of the global 
Schmidt relation would arise from either a decrease
in the characteristic timescale \citep{2012ApJ...745...69K}
% Krumholz, Dekel, McKee 2012
or by an increase 
in the fraction of gas above the clump threshold (\fdense), 
from its typical value in local clouds 
[$\fdense \sim 0.1$ \citep{2012ApJ...745..190L}]
% Lada et al., 2012
with $\fdense \propto \sigmagas^{0.5}$. 

The second approach, which we could call a top-down picture,
assumes that star formation is largely controlled by global dynamical phenomena,
such as disk instabilities (e.g., \citealt{1997ApJ...481..703S}),
and the dynamical timescales in the parent galaxy.
In this picture the transition between the low-density and higher-density
SFR regimes is mainly driven by gravitational instabilities in the
disk rather than by cooling or molecular formation thresholds,
and the non-linear increase in the SFR relative to gas density
above this threshold reflects shorter self-gravitational timescales
at higher density or the shorter dynamical timescales.  In this
picture there is no particular physical distinction between the 
intermediate-density and high-density regimes; in principle the
same dynamical processes can regulate the SFR continuously across
this wide surface density regime.  Likewise it is the total surface
density of gas, whether it be atomic or molecular, that drives the
%SFR.  This picture is partly inspired by the observation
%of a tight correlation between \sigmasfr\ and $\sigmagas/\tau_{dyn}$
%and the preservation of this relation across normal
%disks and starbursts (\S \ref{sflobs}, Kennicutt 1998b, 
%\citealt{2010ApJ...714L.118D}).
% Daddi et al. 2010
SFR.  The asymptotic form of this picture is a self-regulated star
formation model:  the disk adjusts to an equilibrium in which
feedback from massive star formation acts to balance the hydrostatic
pressure of the disk or to produce an equilibrium porosity of
the ISM (e.g., \citealt{1981ApJ...245..534C, 1985ApJ...295L...5D,
1997ApJ...481..703S, 2010ApJ...721..975O}).
%Cox 1981, Dopita 1985, Silk 1997, Ostriker et al. 2010
 
When comparing these pictures to the observations, each has
its particular set of attractions and challenges.  
The bottom-up picture has the attraction of simplicity, associating
nearly all of the relevant SFR physics with the formation of 
molecular gas and molecular cloud clumps.  It naturally fits with 
a wide range of observations including the tight correlation of 
of the SFR and molecular gas surface densities 
(e.g., \citealt{2011AJ....142...37S} and references therein), 
%Schruba  et al. 2011 
the concentration of star formation within clouds in regions of 
dense gas (\S \ref{local}), and the observation of a 
linear relation between the dense gas traced by HCN emission
and the total SFR in galaxies (\S \ref{sflobs}).  
If the resolved star formation
relation at intermediate surface densities is linear (\S \ref{sflobs},
\citealt{2008AJ....136.2846B}),
%Bigiel et al.
\fdense\ would be constant in that regime, while increasing
monotonically with \sigmagas\ in the higher-density starburst regime
with the transition occuring where \sigmagas\ derived from CO
is similar to the threshold for dense gas. 
This picture does not explain why a particular
galaxy or part of a galaxy lies in one of these regimes, a question
perhaps best answered by the top-down picture.

Aspects of the top-down picture date back to early dynamical 
models of the ISM and the first observations of 
star formation thresholds in disks (e.g., \citealt{1973ApJ...179...69Q,
1987sbge.proc..467L, 1988Afz....29..190Z, 1989ApJ...344..685K, 
1991ApJ...378..139E, 1997ApJ...481..703S}).
%Quirk \& Tinsley 1972, 
%Larson 1987, Zasov \& Simikov 1988, Kennicutt 1989, Elmegreen 1991,
%Silk 1997).  
There is some observational evidence for associating
the observed low-density thresholds in disks with gravitational instabilities
in the disk (e.g., $Q$ instabilities), rather than with atomic or molecular phase transitions
(e.g., \citealt{1989ApJ...344..685K, 2001ApJ...555..301M}),
%(e.g., Kennicutt 1989, Martin \& Kennicutt 2001), 
but recent observations
and theoretical analyses have raised questions about this interpretation 
(e.g., \citealt{2004ApJ...609..667S, 2008AJ....136.2782L}).
%(e.g., Schaye 2004, Leroy et al. 2008).  
At higher surface densities 
the global relation between \sigmasfr\ and the ratio of gas density
to local dynamical time  
($\sigmagas/\tau_{dyn}$) shows a correlation that is nearly as tight
as the conventional Schmidt law \citep{1998ApJ...498..541K}, 
% Kennicutt 1998b, ApJ article
and it also removes the double sequence of disks and starbursts
that results if \xco\ is systematically lower in the starbursts
(\S \ref{sflobs}, \citealt{2010ApJ...714L.118D, 2010MNRAS.407.2091G}).
% Daddi et al. 2010, Genzel et al. 2010, MNRAS
In this picture, the higher star formation rate for a given gas
surface density in mergers is caused by the compaction of the gas
and the resulting shorter rotation period. However, the efficiency
per orbital period is not clearly explained in this picture.
%Normal galaxies require a low efficiency
%per rotation time, but merger driven starbursts do not. 
Theories of feedback-regulated star formation
show promise in explaining the low
efficiency per orbit in normal galaxies
\citep{2011ApJ...743...25K}, and
%Kim, Kim, Ostriker
they may also explain the less effective role of negative feedback in
merger-driven starbursts
(e.g., \citealt{2011ApJ...731...41O}).
% Ostriker and Shetty on starburst regulation
This model does not extend to cloud-level star formation. If feedback
from massive stars is fundamental, the similarity between the efficiency
in local clouds forming only low mass stars and
regions with strong feedback from massive stars 
(cf. \citealt{2009ApJS..181..321E}
%Evans et al. c2d results
and 
\citealt{2011ApJ...729..133M})
% Murray measured for \hii\ regions
remains a mystery.

Both of these scenarios can claim some successes, 
but neither provides a complete
explanation that extends over all scales and environments. 
%RCK new sentences added below
As but one example, the large changes in the present-day SFRs
and past star formation histories of galaxies as functions of
galaxy mass and type may well be dictated mainly by external
influences such as the accretion history of cold gas from 
the cosmic web and intergalactic medium.  Neither the 
bottom-up or top-down pictures as articulated above incorporate
these important physical processes.   On smaller scales a complete
model for star formation may combine features of both scenarios. 
%It is interesting to
%seek to combine their best features.
The dynamical picture is 
quite attractive on the scales of galaxies, and it would be interesting
to extend it to smaller scales. In doing so, the question of what to
use for $\tau_{dyn}$ arises. The galaxy rotation period cannot
control star formation in individual clouds, so a more local
dynamical time is needed.  
One option is the cloud or clump crossing time
(essentially the size over the velocity dispersion) 
\citep{2000ApJ...530..277E},
% Elmegreen 2000, "star formation in a crossing time"
which may be particularly relevant in regions of triggered star
formation.  Since clumps are the star forming units, their crossing 
times may be the more relevant quantities.

A popular option is the free-fall time
(e.g., \citealt{2005ApJ...630..250K}).
Since $\tff \propto \rho^{-0.5}$, a volumetric star formation law,
$\rho_{\rm SFR} \propto \rho/\tff \propto \rho^{1.5}$, where
$\rho$ is the gas density, is a tempting rule.
With this rule, the roughly 1.5 power of the KS relation
appears to be explained if we ignore the difference between
volume and surface density.
Indeed, \citet{2012ApJ...745...69K} 
% Krumholz, Dekel, McKee 2012
argue that such a volumetric law
reproduces observations from the scale of nearby clouds to starburst
galaxies. However, the definition of \tff\ changes
for compact starbursts (see eq. 9 in their paper), essentially at the
boundary between the intermediate-density and high-density regimes discussed in
\S \ref{cluesobs}. Including the atomic-molecular threshold, as treated in
\citet{2009ApJ...699..850K}, clarifies that this ``scale-free" picture
implicitly recognizes the same three
regimes discussed in \S \ref{cluesobs}.

Theories that rely on \tff\ face two problems.
First, no evidence has been found in well-studied
molecular clouds for collapse at \tff\
(\S \ref{sftheory}, e.g., \citealt{1974ApJ...192L.149Z}).
To match observations, an ``efficiency" of about 0.01 must be
inserted. Why the star formation density should remain
proportional to \tff, while being slowed by a factor of 100,
is a challenging theoretical question.  Studies of the
role of turbulence show some promise in explaining this
paradox  (\citealt{2005ApJ...630..250K, 2011ApJ...743L..29H}),
%Krumhholz and McKee 2005, Hennebelle and Chabrier 2010
and magnetic fields may yet play a role.

The second problem is how to calculate a relevant \tff\ in a cloud,
much less a galaxy, with variations in $\rho$ by many orders of magnitude,
and much of the gas unlikely to be gravitationally bound.
For example, \citet{2012ApJ...745...69K} calculate \tff\ 
from the  mean density of the whole cloud ($\tff \propto 1/\mean{\rho}$). 
A computation of \mean{\tff} from \mean{1/\rho} would emphasize
the densest gas, where star formation is observed to occur.
In a model that accounts for this, \citet{2011ApJ...743L..29H}
% Hennebelle and Chabrier 2011
reproduce at some level the observations of surface density thresholds
for efficient star formation observed in nearby clouds
(\citealt{2010ApJ...724..687L, 2010ApJ...723.1019H}).
An alternative view to the models involving \tff\ and emphasizing
instead the critical role of the dense gas threshold can be found
in \citet{2012ApJ...745..190L}. 
% Lada et al. on connection to exgal
Because both models claim to apply to cloud-level star formation,
observations of \mw\ clouds should be able to test them.

% spec
%future.tex
%061411
%070811 NJE
%080911 NJE
%081111 NJE made into subsection in synthesis
%090911 NJE some clean up
%100611 RCK revision incorporating questions
%100611 NJE added some more things
%100711 NJE minor fixes
%100911 RCK minor additions, addressed and removed NJE and RCK comments
%101911 RCK reference and citation link updates
%110411 NJE minor edits
%110711 NJE small edits to condense
%121311 NJE minor edits
%121611 NJE minor edits
%012412 NJE edits after comments
%012912 RCK minor edits after comments
%020212 NJE minor edits, changed Galaxy to \mw\ for consistency
%020612 RCK added draft final paragraph
%020712 NJE minor edits, final para.
%031212 NJE
 
\subsection{Future Prospects}\label{future}

We hope that this review has conveyed the 
%major activity that has typified this subject over the past decade, and the 
tremendous progress over the last decade
%that has been made 
in understanding the systematic behavior of 
%massive 
star formation, both in the \mw\  and in other galaxies.  
As often happens
when major observational advances are made, observational
pictures that once seemed simple and certain have proven
to be more complex and uncertain.  We have attempted to inject
a dose of skepticism about some well-accepted truisms
and to highlight important questions where even the observations
do not present a completely consistent picture.

As we return to the key questions outstanding in this
subject (\S \ref{questions}), a few clear themes emerge,
%(text box near the first section)
many of which lie at the cusp between \mw\  and extragalactic
observations, and between theory and simulation on the sub-cloud
scale on the one hand, and the galactic and cosmological scales
on the other.  The recent observations within the \mw\ of a near-universal
association of stars with dense molecular clumps (\S6.4)
offers the potential key of a fundamental sub-unit of high-efficiency
star formation in all galactic environments, from the low-density
and quiescent environments of outer disks and dwarf galaxies to
the most intense starbursts.  However this hypothesis needs
to be validated observationally in a wider range of environments.  
Restated from another perspective,
we need much better information on the structure (physical structure,
substructure) and dynamics of star-forming clouds across the
full range of star-forming environments found in galaxies today.

Fortunately we are on the brink of major progress 
on multiple fronts.  One approach is to assemble more 
complete and ``zoomed-out" views of the \mw, while preserving
the unparalleled spatial resolution and sensitivity of \mw\
observations to fully characterise the statistical trends in 
cloud structure, kinematics, mass spectra, and associated star
formation for complete, unbiased, and physically diverse samples.
% to combine with
%more sensitive and ``zoomed-in" views of other galaxies.
In the near-term, {\it Herschel} surveys will deliver images of nearby
clouds, the plane of the MW, and many galaxies in bands from 60 to 500
\micron. When the MW plane data are combined with higher resolution surveys 
of the MW at 0.87 to 1.1 mm
from ground-based telescopes, spectroscopic follow-up, and improved distances
from ongoing VLBA studies, we will have a much improved and
nuanced picture of the gas in the \mw, which can then provide a more
useful template for understanding similar galaxies.
Recent surveys have doubled the number of known \hii\ regions in the MW
\citep{2011ApJS..194...32A},
% Anderson et al.
allowing study of a wider range of star formation outcomes. Tests of
the limitations on star formation rate tracers used in extragalactic
work will be possible, as will comparisons to KS relations on
a variety of spatial scales.  The Magellanic Clouds also offer
great potential for extending this approach to two sets of environments
with different metallicities, ISM environments, and star formation
properties (exploiting for example the 30 Doradus region).

This expanded information from \mw\  surveys will then need to
be combined with more sensitive and ``zoomed-in" observations of other galaxies.
For such studies the weakest link currently is the
knowledge of the molecular gas. We can trace \hi\ and SFRs to much lower
levels than we can detect CO, even with stacking of the CO maps.
When fully commissioned, {\it ALMA} will have a best resolution of 
about 13 mas at 300 GHz (0.5 pc at the distance of M51), 
and about 8 times the total collecting
area of any existing millimeter facility, allowing us to trace molecular
emission to deeper levels and/or to obtain resolution at least 10 times better 
than the best current studies. 
It may be possible to study dense structures within molecular clouds
in other galaxies
for comparison to dense clumps in the \mw\ clouds.
To be most effective, these programs will need 
to expand beyond the typical surveys in one or two CO rotational
transitions to include the high-density molecular tracers and ideally
a ladder of tracers with increasing excitation and/or critical density.
Continuum observations will gain even more because of large bandwidths
and atmospheric stability. Maps of mm-wave dust emission may become the
preferred method to trace the gas in other galaxies, as they are becoming
already in the \mw, bypassing the issues of \xco.
Observations of radio recombination lines may also provide alternative
tracers of star formation rate in highly obscured regions.
It is difficult to exaggerate the potential transformational power
of {\it ALMA} for this subject.
%, though observing time will be so
%precious that experiments will have to be carefully designed and
%targeted.

For molecular line observations of nearby galaxies, large single dishes
(the Nobeyama 45-m, the IRAM 30-m, and the future CCAT), along
with the smaller millimeter arrays, IRAM PdB and CARMA, will provide
complementary characterization of the molecular and dense gas on larger scales
than can be efficiently surveyed with ALMA.  A substantial expansion 
of the IRAM interferometer (NOEMA) 
will provide complementary advances in sensitivity in the northern hemisphere. 
This subject will also advance with a substantial expansion of \hi\ 
mapping (by e.g., the EVLA) to expand the range of environments 
probed and ideally to attain spatial resolutions comparable to the 
rapidly improving molecular line observations.  
Although it is likely that star formation on the
local scale concentrates in molecular-dominated regions, the formation of
this molecular gas from atomic gas remains a critical 
(and possibly controlling) step
in the entire chain that leads from gas to stars.

Future opportunities also await for improving our measurements of
SFRs in galaxies, though the truly transformational phase of that
subject may already be passing with the end of the ISO, Spitzer,
GALEX, and Herschel eras.  
HST continues to break new ground, especially in direct mapping of
young stars and their ages via resolved color-magnitude diagrams.
Herschel surveys of nearby galaxies will provide \fir\
data, complementary to the other wavelength regions that trace star
formation (\S \ref{sfrdiagnostics}), and multiband imaging with 
the EVLA will allow better separation of non-thermal from
the free-free continuum emission,
which directly traces the ionization rate and massive SFR.
SOFIA offers the opportunity for mapping
the brightest regions of star formation with higher spatial 
and spectral resolution than Spitzer.
Further into the future, major potential lies with JWST,
which will vastly improve our capabilities for tracing star formation
via infrared radiation, both on smaller scales in nearby galaxies and
in very distant galaxies.  SPICA may provide complementary capability
at longer wavelengths. 
As highlighted in \S \ref{questions},
%the questions box, 
key unknowns in this
subject include robust constraints on the ages and lifetimes of star-forming
clouds; observations of resolved star clusters, both
in the \mw\ and nearby galaxies, offer potential for considerable
inroads in this problem.

A number of the questions we have listed also can be attacked with
groundbased OIR observations.  As discussed in \S \ref{challenge}, despite the
availability now of spatially-resolved multi-wavelength observations
of nearby galaxies, we still do not have an absolutely reliable way
to measure dust-corrected SFRs, especially from 
short-lived tracers that are critical for exploring the local 
Schmidt law.   For observations of normal galaxies where local
attenuations are modest, integral-field mapping of galaxies offers
the means to produce high-quality H$\alpha$ maps corrected for dust
using the Balmer decrement (e.g., \citealt{2009ApJ...704..842B,
2010ASPC..432..180B, 2012A&A...538A...8S}).
% Blanc et al. 2009, 2010; Sanchez et al. 2012
A ``gold standard" for such measurements  also is available in
the hydrogen recombination lines of the Paschen and Brackett series
in the near-infrared.  The emissivities of these lines are directly
related to ionizing luminosities in the same way as are the more
widely applied Balmer lines (albeit with somewhat stronger density
and temperature dependences), and these lines suffer much lower
dust attenuation (readily calibrated by comparing to H$\alpha$ or
other shorter wavelength lines).  These lines are much fainter,
however, and are subject
to strong interference from telluric OH emission (and thermal emission
at longer wavelengths).  Continuing advances in near-infrared detectors
have now brought many of these lines into the accessible range,
and soon we should begin to see large-scale surveys that will 
provide high-resolution emission maps and robust ``SFR maps", at 
least on spatial scales larger than individual \hii\ regions.  These
maps in turn can be used to test and hopefully recalibrate other
dust-free tracers such as combinations of infrared dust emission
with UV and optical emission-line maps.  Looking further ahead,
spectral imaging in  the near-infrared (and with \halpha\ in the visible)
with massively parallel integral-field spectrometers will provide 
more precise and full views of star formation.
%to complement the high-resolution maps of molecular gas and dust
%(e.g., \citealt{2010ASPC..432..180B}).
%Blanc et al. on VENGA
%The Herschel surveys of nearby galaxies will provide \fir\
%data, complementary to the other wavelength regions that trace star
%formation (\S \ref{sfrdiagnostics}), and the EVLA will allow much
%deeper images of radio continuum emission.
 
The current, pioneering studies of star formation at $z \sim 1-3$ will
become much easier with new instrumentation.
Advances in submillimeter array technology (e.g., SCUBA-2 on the
JCMT) and larger dishes at very high altitude, such as CCAT,
will allow deep and wide searches for
dust continuum emission from distant galaxies. Many of the biases in
current samples can be alleviated and a more complete picture of
star formation through cosmic time can be constructed.
Huge surveys of Ly$\alpha$ emitting galaxies
will be undertaken to constrain dark energy, providing as a by-product,
nearly a million star-forming galaxies at $1.9< z < 3.5$ and
a large number of [OII] emitting galaxies for $z<0.5$
(e.g., \citealt{2008ASPC..399..115H}).

Since the theoretical side of the subject lies outside 
the scope of this review, we comment only briefly on this area. 
Numerical simulations are poised to
make major contributions to the subject over the next several years.
Simulations with higher resolution and more sophisticated treatments
of the heating, cooling, phase balance, and feedback are providing
deeper insights into the physical nature of the star formation scaling
laws and thresholds and the life cycles of molecular clouds
(e.g., \citealt{2008ApJ...680.1083R, 2008MNRAS.391..844D, 
2010MNRAS.409..396D, 2009ApJ...700..358T, 2011ApJ...730...11T, 
2011ApJ...728...51B}).
%Robertson \& Kravtsov 2008, Dobbs 2008, Dobbs \& Pringle 2009,
%Tasker \& Tan 2009, Tasker 2011).  
%Brooks et al. 2011
%. Improving resolution to 100 pc and increasing
%the star formation threshold density from 0.1 \cmv\ to 100 \cmv\ in simulations
%of evolving galaxies has already resulted in more realistic disks 
Inclusion of realistic stellar feedback may obviate the need for 
artificial constraints on efficiency (e.g., \citealt{2011MNRAS.417..950H}).
% Hopkins, Quataert, and Murray 2011

Analytical models which incorporate the wide range of relevant physical
processes on the scales of both molecular clouds and galactic disks
are also leading to deeper insights into the triggering and regulation
of star formation on the galactic scale (e.g., 
\citealt{2010ApJ...721..975O}, 
%Ostriker et al. 2010,
and papers cited in \S \ref{spec}).
Exploring  different prescriptions for
cloud-scale star formation in galaxy evolution models
will illuminate the effects of the cloud-scale prescription on
galaxy-scale evolution. 
At the ``micro-scale" of molecular clouds,
inclusion of radiative and mechanical feedback from star formation is
producing more realistic models, and perhaps the relative importance
of feeding from the local core and from the greater clump may be
quantified (e.g., references in \S \ref{sftheory}).

%NJE some wordsmithing
%This review 
We conclude with a hope and a prediction.  From the 
start, this review was designed to 
% bring together expertise from the communities studying 
% NJE build precedence for "this exchange"
exchange ideas between those studying
star formation in the Milky Way and
those studying star formation in other galaxies.  
The authors have benefited 
immensely from this exchange, and we hope for more of this
cross-talk between our two communities: people studying star formation
within the Milky Way can put their results in a larger context; those
studying other galaxies can appreciate that our home galaxy offers 
unique advantages for understanding galaxies.  A common theme
across both arenas has been the rapid pace of advance, both in
observations and theory.  When the next review of this  subject is written,
we predict
%NJE to make parallel to "hope and prediction"
%anticipate 
that it will focus on observational
tests of detailed physical theories and simulations rather than on
empirical star formation laws.  We conclude with heartfelt thanks
to all of those who are working toward this common goal.

% future

\acknowledgments
%070111
%070811
%091311
%100311 NJE
%101811 RCK
%102011 NJE minor edit
%012412 NJE added names of folks who read and commented
% 061912 NJE Added B. Catinella for reading

We would like to acknowledge the many people who have supplied
information and preprints in advance of publication. In addition,
we thank
Alberto Bolatto,
Leonardo Bronfman,
Peter Kaberla,
Jin Koda, 
Adam Leroy,
Yancy Shirley, 
and
Linda Tacconi
for useful and stimulating discussions. 
Henrik Beuther,
Guillermo Blanc,
Daniela Calzetti,
Barbara Catinella,
Reinhard Genzel, 
Rob Gutermuth,
Mark Krumholz,
Charles Lada,
Fred Lo,
Steve Longmore,
Eve Ostriker,
and
Jim Pringle
provided valuable comments on an early draft.
We both would like to express
special thanks to our current and former students and postdocs, with 
whom we have shared countless valuable discussions.
NJE thanks the Institute of Astronomy, Cambridge,
 and the European Southern Observatory, Santiago, 
for hospitality during extended visits, during which much of his
work on the review was done. NJE also acknowledges support from
NSF Grant AST-1109116 to the University of Texas at Austin.

\bibliographystyle{aa}

\bibliography{robk1,robk2,njebib,thesis}

\clearpage
% Tables
%\section{Tables}
% table1.tex
% was table2, but reordered 050912
% 031512 NJE edited to work with AAS latex
%\begin{document}
\begin{table}
\begin{center}
\caption{Star Formation Rate Calibrations \label{sfrequations}}
\vspace{0.5cm}
\footnotesize
\begin{tabular}{l|c|l|c|c|l}
%\hline
Band & Age Range (Myr)\tablenotemark{a} & $L_x$ Units & $\log C_x$ & \mbox{$\dot M_{\star}$}/\mbox{$\dot M_{\star}$}(K98) & References \\
\hline
FUV & $0-10-100$ & ergs\,s$^{-1}$ ($\nu L_\nu$) & 43.35 & 0.63 & 1, 2 \\
\hline
NUV & $0-10-200$ & ergs\,s$^{-1}$ ($\nu L_\nu$) & 43.17 & 0.64 & 1, 2 \\
\hline
H$\alpha$ & $0-3-10$ & ergs\,s$^{-1}$ & 41.27 & 0.68 & 1, 2 \\
\hline
TIR & $0-5-100$\tablenotemark{b} & ergs\,s$^{-1}$ (3$-$1100\,$\mu$m) & 43.41 & 0.86 & 1, 2 \\
\hline
24\,$\mu$m & $0-5-100$\tablenotemark{b} & ergs\,s$^{-1}$ ($\nu L_\nu$) & 42.69 & & 3 \\
\hline
70\,$\mu$m & $0-5-100$\tablenotemark{b} & ergs\,s$^{-1}$ ($\nu L_\nu$) & 43.23 & & 4 \\
\hline
1.4 GHz & $0-100:$ & ergs\,s$^{-1}$\,Hz$^{-1}$ & 28.20 & & 1 \\
\hline
%Thermal Radio & $0-3-10$ & ergs\,s$^{-1}$\,Hz$^{-1}$ & $27.34 + 0.45 log{(T_e \over 10^4)} + 0.1 \nu_{GHz}$ & & 1 \\
%\hline
2$-$10 keV & $0-100:$ & ergs\,s$^{-1}$ & 39.77 & 0.86 & 5 \\
\hline
\end{tabular}
\tablenotetext{a}{Second number gives mean age of stellar population 
contributing to emission, third number gives age below which 90\%\ of 
emission is contributed.} 
\tablenotetext{b}{Numbers are sensitive to star formation history, 
those given are for continuous star formation over 0--100 Myr.  
For more quiescent regions (e.g., disks of normal galaxies) the 
maximum age will be considerably longer.}
\tablecomments{
References: (1) \citet{2011ApJ...737...67M};
%Murphy et al. 2011; 
(2) \citet{2011ApJ...741..124H};
%Hao et al. 2011; 
(3) \citet{2009ApJ...692..556R};
%Rieke et al.  (2009); 
(4) \citet{2010ApJ...719L.158C}; \\
%Calzetti et al. 2010; 
(5) \citet{2003A&A...399...39R}
%Ranalli et al. (2003)
} 
\end{center}
\end{table}
%\end{document}

% table2.tex
% was table 1, but reordered 050912
% 031512 NJE edited to work with AAS latex
\begin{table}
\begin{center}
\caption{Multi-Wavelength Dust-Corrections \label{multitab}}
\vspace{0.5cm}
\begin{tabular}{l|c}
\hline
Composite Tracer & Reference \\
\hline
$L(FUV)_{corr} = L(FUV)_{obs} + 0.46 L(TIR)$ & 1 \\
\hline
$L(FUV)_{corr} = L(FUV)_{obs} + 3.89 L(25\micron)$ & 1 \\
\hline
$L(FUV)_{corr} = L(FUV)_{obs} + 7.2\ee{14} L(1.4 {\rm GHz})$\tablenotemark{a} & 1 \\
\hline \hline
$L(NUV)_{corr} = L(NUV)_{obs} + 0.27 L(TIR)$ & 1 \\
\hline
$L(NUV)_{corr} = L(NUV)_{obs} + 2.26  L(25\micron)$ & 1 \\
\hline
$L(NUV)_{corr} = L(NUV)_{obs} + 4.2\ee{14} L(1.4 {\rm GHz})$\tablenotemark{a} & 1 \\
\hline \hline
$L(H\alpha)_{corr} = L(H\alpha)_{obs} + 0.0024 L(TIR)$ & 2 \\
\hline
$L(H\alpha)_{corr} = L(H\alpha)_{obs} + 0.020 L(25\micron)$ & 2 \\
\hline
$L(H\alpha)_{corr} = L(H\alpha)_{obs} + 0.011 L(8\micron)$ & 2 \\
\hline
$L(H\alpha)_{corr} = L(H\alpha)_{obs} + 0.39\ee{13} L(1.4 {\rm GHz})$\tablenotemark{a} & 2 \\
\hline
\end{tabular}

%$^a$ Radio luminosity in units of ergs\,s$^{-1}$\,Hz$^{-1}$ \\
\tablenotetext{a}{Radio luminosity in units of ergs\,s$^{-1}$\,Hz$^{-1}$} 
\tablecomments{References: (1) \citet{2011ApJ...741..124H}; 
%Hao et al. 2011
(2) \citet{2009ApJ...703.1672K}
% Kennicutt et al. (2009)
}
\end{center}
\end{table}
%\end{document}

% 122311 initiated by NJE
% 020312 NJE edited to fit better
% 031512 NJE edited to work with AAS latex
\begin{table}
\begin{center}
\caption{Star Formation Regimes \label{regimetab}}
\vspace{0.5cm}
\footnotesize
\begin{tabular}{l|l|c|c|l}
\hline
Name & \sigmagas & Gas Properties  & Star Formation & Examples \\
     & (\msunpc) &                 &                &           \\
\hline
Low         & $< 10$   & Mostly atomic   & low, sparse   & outer disks \\
Density     &     &  & & early type galaxies \\
            &     &  & & LSB galaxies \\
            &     &  & & dwarf galaxies \\
            &     &  & & solar nbd in MW \\
Intermediate &  $10- \Sigma_{dense}$  & atomic$\rightarrow$ molecular  & moderate   & normal disks \\
Density     &     &  & & inner MW \\
            &     &  & & molecular clouds \\
High & $> \Sigma_{dense}$   & molecular$\rightarrow$ dense   & high, concentrated   & some nuclear regions \\
Density     &     &  & & starbursts \\
            &     &  & & mergers \\
            &     &  & & clumps and cores \\
\hline
\end{tabular}
\end{center}
\tablecomments{The dividing line between Intermediate and High-density regimes
($\Sigma_{dense}$)  ranges from 100 to 300 \msunpc.}
\end{table}

%Figures
%\section{Figures}
%figtacconi.tex
% 120711 NJE added as first figure
% 020312 NJE fixed exponents
% 031612 NJE improved for emulateapj
% Tacconi figure

% fig_ngc6946_ir.tex
% 102411 NJE added figure call and refs
% 021212 RCK updated citation and added link
% 031612 NJE Adjusting size for emulateapj

% caption there, but not showing fig because it is very large
%this file causes problems now, fig too big?
% figdustcorr.tex
% 102411 NJE added fig call and label
% 110711 NJE adapted for new figure
% 031612 NJE adapted for emulateapj and added copyright line

% figure call in and caption for Heiderman figure 2

% figure call in and caption for Churchwell MW figure
% 031612 NJE added copyright line

%fig_n6946optircap.tex
% 021412 NJE renamed and labeled for consistency
% 021412 NJE cap version for caption, no figure yet
% 031212 NJE added call to figure
% 040212 NJE added permissions statement
% 040512 NJE removed permissions, since all original

% fig_radial.tex
% 012012 NJE copied from another and mod to do radial plots
% 021412 NJE corrected ref to Kalberla and Dedes 2008 for HI, and He issues
% 021512 NJE minor edits
% 031612 NJE adjusted size for emulateapj
% 031612 NJE added copyright info

% fig_galex.tex
%102411 NJE added figure call and label
% 031612 NJE adjusted size for emulateapj
% 031612 NJE added copyright
% 051412 NJE fixed to have only right panel

% fig_demographics.tex
% 102411 NJE added figure call and label
% 021212 RCK shortened caption, references now in main text
% 031612 NJE adjusted size for emulateapj
% 051412 NJE moved some text here from main body

% fig_bothwell.tex
% 102411 NJE added figure call and label
% 031512 NJE changed to call .eps file
% 040512 NJE updated ref to figure since original

% fig_Schmidt_gao.tex
% 102411 NJE added figure calls and labels
% Need to add two other figs, but need eps files
% 020312 Added ref to MW point, which needs to be added
% 031212 point has been added
% 031612 NJE fixing errors in emulateapj
% 051412 NJE moved text to caption from main text

%fig_bigiel.tex
% 102411 NJE added figure call and label
% 031612 NJE adjusted size for emulateapj
% 031612 NJE added copyright

% figlada.tex
% figure call in and caption for Lada figures
% 031612 NJE adjusted size for emulateapj
% 031612 NJE added copyright

% figure call in and caption for Heiderman figure 9
% 031612 NJE added copyright

\end{document}